\documentclass[11pt]{article}
\usepackage[top=1in,bottom=1in,left=1in,right=1in]{geometry}

\usepackage[noEucal]{main}
\usepackage{aas_macros}
\usepackage{graphicx}
\usepackage[cal=boondoxo]{mathalfa}
 
\renewcommand{\tr}{\text{tr}}

\def\ds{{\slashed{\d}}}
\def\ci{{\cal I}}
\def\mr{\mathring}
\def\mrx{\mr{X}}

\def\mru{{\mr{u}}}

\newcommand{\ttt}[1]{\text{($#1$)}}

\def\m{\text{$(-1)$}}
\def\0{\text{(0)}}
\def\1{\text{(1)}}
\def\2{\text{(2)}}
\def\3{\text{(3)}}
\def\4{\text{(4)}}
\def\5{\text{(5)}}
\def\6{\text{(6)}}
\def\7{\text{(7)}}

\def\tf{\text{\scalebox{0.8}{TF}}}

\begin{document}
\begin{titlepage}
\unitlength = 1mm
~
\vskip 3cm
\begin{center}

{\LARGE{\textsc{Phase Space Renormalization and Finite BMS\\Charges in Six Dimensions}}}

\vspace{0.8cm}

Federico Capone$^a$, Prahar Mitra$^b$, Aaron Poole$^c$, Bilyana Tomova$^b$

\vspace{1cm}

{\it $^a$Theoretisch-Physikalisches Institut, Friedrich-Schiller-Universit\"at Jena,\\Max-Wien-Platz 1, D-07743 Jena, Germany}\\
{\it $^b$Department of Applied Mathematics and Theoretical Physics, University of Cambridge, Wilberforce Road, Cambridge, CB3 0WA, UK}\\
{\it $^c$Department of Physics and Research Institute of Basic Science, Kyung Hee University, \\26 Kyungheedae-ro, Seoul 02447, Republic of Korea}\\

\vspace{0.8cm}

\begin{abstract}\noindent
We perform a complete and systematic analysis of the solution space of six-dimensional Einstein gravity. We show that a particular subclass of solutions -- those that are analytic near $\ci^+$ -- admit a non-trivial action of the generalised Bondi-Metzner-van der Burg-Sachs (GBMS) group which contains \emph{infinite-dimensional} supertranslations and superrotations. The latter consists of all smooth volume-preserving Diff$\times$Weyl transformations of the celestial $S^4$. Using the covariant phase space formalism and a new technique which we develop in this paper (phase space renormalization), we are able to renormalize the symplectic potential using counterterms which are \emph{local} and \emph{covariant}. The Hamiltonian charges corresponding to GBMS diffeomorphisms are non-integrable. We show that the integrable part of these charges faithfully represent the GBMS algebra and in doing so, settle a long-standing open question regarding the existence of infinite-dimensional asymptotic symmetries in higher even dimensional non-linear gravity. Finally, we show that the semi-classical Ward identities for supertranslations and superrotations are precisely the leading and subleading soft-graviton theorems respectively.

\end{abstract}

\vspace{1.0cm}
\end{center}
\end{titlepage}
\pagestyle{empty}
\pagestyle{plain}
\pagenumbering{arabic}

\tableofcontents

\section{Introduction}
\label{sec:intro}

The asymptotic symmetry group (ASG) of a geometry $(M,G)$ is defined as the set of all transformations which preserve the boundary and gauge conditions that define the metric $G$ and which act \emph{non-trivially} on the asymptotic boundary data. This is often described by a simple equation
\begin{equation}
\begin{split}
\text{ASG} = \frac{ \text{Allowed Transformations} }{ \text{Trivial Transformations} } .
\end{split}
\end{equation}
For four-dimensional asymptotically flat spacetimes, the asymptotic symmetry group was first studied by Bondi, Metzner, van der Burg \cite{Bondi:1962px} and Sachs \cite{Sachs:1962wk} in the early 60s. A surprising outcome of their analysis was that the ASG for asymptotically flat spacetimes is an infinite-dimensional extension of the Poincar\'e group which today we refer to as the BMS group. This group consists of the usual 4D Lorentz transformations and an infinite-dimensional extension of translations known as \emph{supertranslations} (no relation to supersymmetry). This group has received much attention over the past decade due to the seminal work by Strominger \cite{Strominger:2013jfa} who proved that the semi-classical Ward identity for supertranslation symmetry \cite{He:2014laa} is the leading soft-graviton theorem \cite{Weinberg:1965nx}. It has also been suggested \cite{deBoer:2003vf,Banks:2003vp, Barnich:2011ct} that the analysis of BMS can be generalised to include an infinite-dimensional generalisation of Lorentz transformations, aptly named \emph{superrotations}. Superrotations generalise Lorentz transformations -- which are isomorphic to global conformal transformations of the celestial $S^2$ (i.e. the sphere at infinity) -- to the local Virasoro transformations. It was shown in \cite{Kapec:2014opa} that the Ward identity of this symmetry is a new, hitherto unknown \emph{subleading} soft-graviton theorem \cite{Cachazo:2014fwa}.

These asymptotic symmetries, the corresponding soft theorems together with the associated gravitational memory effects \cite{Strominger:2014pwa, Pasterski:2015tva}, form what is now known as \emph{infrared triangles} (see \cite{Strominger:2017zoo} for an early review). These triangles -- which are now known to be ubiquitously present in many theories in asymptotically flat spacetimes -- thread together three previously unrelated fields of research -- formal relativity (asymptotic symmetries), perturbative QFT (infrared effects) and observational/experimental relativity (memory effects). This remarkable discovery has unveiled many interesting features of gauge and gravitational theories and has revitalised research in flat holography. In particular, the infrared triangle is a universal feature that any microscopic/holographic formulation of quantum gravity in asymptotically flat spacetimes will have to exhibit, independently of the fine-grained details of the model. Indeed, over the past decade, two separate (but perhaps, equivalent) formulations of flat holography have emerged -- \emph{Celestial Holography} and \emph{Carrollian Holography} \cite{Pasterski:2021raf, Bagchi:2022emh, Donnay:2022wvx, Saha:2023hsl} both of which manifest the infrared triangle in slightly different ways!

Exploration of the rich structure described above has largely been limited to four spacetime dimensions. Efforts to analyse and study this structure in dimensions greater than four are sparse \cite{Hollands:2003ie, Hollands:2003xp, Tanabe:2011es, Godazgar:2012zq, Kapec:2015vwa, Kapec:2017gsg, Pate:2017fgt, Aggarwal:2018ilg, He:2019jjk, He:2019pll, He:2019ywq, Capone:2019aiy, Capone:2021ouo, Kapec:2021eug, Chowdhury:2022nus, Chowdhury:2022gib}. There are several reasons for this. A BMS-like analysis reveals that the asymptotic symmetry group of asymptotically flat gravity is simply the finite-dimensional Poincar\'e group \cite{Hollands:2003xp}, i.e. there is no infinite-dimensional extension! It is also well-established that while four-dimensional theories with massless particles (e.g. graviton) are plagued with infrared divergences, there are no such divergences in higher dimensions. Finally, it was shown in \cite{Hollands:2016oma} that gravitational memory effects are also not present in higher dimensions. Altogether, these results seem to suggest something quite obvious -- namely that the infrared sector of higher dimensional gravity is trivial. There is simply not enough structure to admit the richness we see in four dimensions!

The gaping outliers in all of this are the soft theorems. The leading and subleading soft-graviton theorems exist and have the exact same qualitative structure in all dimensions \cite{Laddha:2017ygw}. This fact alone has forced us and others to revisit the aforementioned results and to question if it is indeed true that BMS symmetries are absent in higher dimensions. The first of this ``new-wave'' of results was obtained in \cite{Kapec:2015vwa} where it was shown that by weakening the definition of asymptotic flatness employed in \cite{Hollands:2003xp}, the asymptotic symmetry group in higher even dimensions can include supertranslations. Working in linearised gravity, it was then shown these supertranslations are related to the leading soft-graviton theorem, exactly as in four dimensions! Following this, \cite{Pate:2017fgt} showed that there are memory effects in higher dimensions, albeit at a subleading order (which is why they were not observed in \cite{Hollands:2016oma}) \emph{and} that they are related to supertranslations and soft theorems in the usual way. These results were partially generalised to non-linear gravity in \cite{Chowdhury:2022nus}. Despite the progress, no coherent analysis of the inevitable divergences of such supertranslation-compatible phase spaces has yet been given.

Having gathered enough evidence about the existence of supertranslations in higher dimensions, the obvious next step is to question the existence of superrotations. As mentioned before, the subleading soft theorem exists in all dimensions. \emph{Does this not imply that superrotations should also exist in all dimensions?} At this stage, we encounter our first complication. Superrotations in four dimensions were obtained as a Virasoro generalization of the global conformal transformations, $SL(2,\mcc) \to \text{Vir}\times\overline{\text{Vir}}$. Such an extension can be realised on the $S^2$ because of the holomorphicity of two-dimensional conformal transformations. In higher dimensions, we lose this holomorphic structure and there is no ``local conformal group''. The conformal group on $S^d$ is the \emph{finite-dimensional} group $SO(1,d+1)$ which cannot reproduce the infinitely-many subleading soft theorems (one for each point on $S^d$).

This problem actually has a relatively straightforward resolution. To understand this, we revisit the four-dimensional problem. One of the issues with the Virasoro superrotations is that they are only defined locally on the celestial sphere. The Hamiltonian charges that generate superrotation transformations on the phase space involve integrals over the entire $S^2$ and are therefore ill-defined. This issue was observed by Campiglia and Laddha in  \cite{Campiglia:2015yka} and they proposed an alternative definition of superrotations which are parameterized by smooth vector fields on $S^2$. In doing so, they resolved some of the technical issues that lie at the heart of the equivalence between soft theorems and asymptotic symmetries while still preserving the infrared triangle. For our purposes, what is particularly interesting about \emph{this} definition of superrotations is that it naturally generalises to higher dimensions! This natural expectation prompted the analyses of \cite{Colferai:2020rte,Campoleoni:2020ejn, Capone:2021ouo,Chowdhury:2022gib}. Specifically, closely following the linearised gravity analysis of \cite{Kapec:2015vwa}, the early work \cite{Colferai:2020rte} concocts the Poisson bracket of a pair of asymptotic fields and partially derives the subleading soft theorem. The work \cite{Campoleoni:2020ejn} explores superrotation charges within the context of linear higher-spin theories. Later, in full non-linear general relativity, the authors of \cite{Chowdhury:2022gib}, adapting \cite{Chowdhury:2022nus} and \cite{Capone:2021ouo} (about which we will say more shortly),
provide a partial construction of superrotation charges linearising around a BMS frame and the matching with the subleading soft theorem in the case of scalar external particles.

Having identified the right set of transformations that we are after, we must next understand what boundary conditions one should use to define asymptotic flatness such that these transformations would be allowed, but not trivial. The goal of this paper is to answer this question and more generally, to formalize the rather scattered discussion of the BMS group in higher dimensions. In the process of formalization, we shall settle several questions whose answers are either completely unknown or only partially known, e.g.
\begin{enumerate}

\item What is the smallest solution space that admits supertranslations \emph{and} superrotations as asymptotic symmetries? What are the boundary conditions that define this space?

\item What is the pre-symplectic form on the solution space? Is it finite? Is it invertible?

\item If not finite, how do we regulate and renormalize the divergences in a local and covariant manner?

\item If not invertible, how should we reduce the solution space to make the pre-symplectic form invertible? How does this affect the asymptotic symmetries?

\item Are supertranslations and superrotations canonical transformations?

\item What is the algebra of the supertranslations and superrotation charges? Does the charge algebra have a central extension?

\item How are these symmetries related to the leading and subleading soft-graviton theorems?

\end{enumerate}
In this paper, we will answer every single question listed above and provide ``first-principles'' derivations of all our results. The starting point is the general analysis of boundary conditions in dimensions greater or equal to four performed in \cite{Capone:2021ouo}. Along the way, we will encounter many of the usual complications associated with superrotations and several new ones that are unique to higher dimensions. The first, and most obvious complication is the dramatic increase in the difficulty of calculations. To simplify our work, we restrict ourselves to six spacetime dimensions. The results of this paper can be easily generalised to all higher even dimensions, although the explicit calculations are more and more tedious as we go higher up in dimension.\footnote{The structure of odd-dimensional gravity is qualitatively very different essentially owing to the fact that the massless retarded/advanced Green's functions are supported in the entire interior of the light-cone whereas in even dimensions they are supported on the light-cone. This smearing of information in odd spacetime dimensions implies that the asymptotic expansion near $\ci^+$ is actually non-local. These issues have been studied in the case of gauge theories in \cite{He:2019jjk,He:2019pll,He:2019ywq} but the generalization to gravity is as yet unknown.} To facilitate our calculations, we have extensively used {\tt Mathematica} and the package {\tt diffgeo.m}.\footnote{This package can be found at \href{https://people.brandeis.edu/~headrick/Mathematica/}{https://people.brandeis.edu/headrick/Mathematica/}.} The second, far more important, complication that we encounter are the large volume divergences which arise in the construction of the generalised phase space. To deal with this issue, we develop a novel formalism which we refer to as ``phase space renormalization'' which allows us to regulate and renormalize all the divergences in a local and covariant way. Once this hurdle is crossed and we have a finite symplectic potential, the rest of the construction follows in a relatively straightforward manner.

We conclude the introduction with a brief comment on the philosophy that has guided our work, namely AdS/CFT. There, in order not to over-constrain the CFT and to properly develop the holographic dictionary via holographic renormalization \cite{deHaro:2000vlm, Skenderis:2002wp}, the gravitational solution with the most general boundary conditions is first constructed. Any additional restriction on the bulk solution often implies extra constraints on the CFT side which changes the physics of the theory. In the same vein, we proceed in this paper by first studying the most general solution space that is consistent with Einstein's equations and imposing additional restrictions \emph{only} when absolutely necessary! Another point where the AdS/CFT line of reasoning plays a role is our approach toward the renormalization of the phase space. We insist on finding local and covariant counterterms, as is done in the holographic renormalization of the bulk AdS action. However, in contrast to AdS/CFT, we do not attempt to renormalize the action. Instead, we only consider a renormalization of the phase space.\footnote{Counterterm actions for asymptotically flat geometries has been previously studied \cite{Kraus:1999di,deHaro:2000wj} and the results are confusing and still not properly understood. In particular, unlike in AdS/CFT, the counterterm action in the flat case is non-local.}

\subsection{Outline of Paper and Results}
\label{sec:summary}

Since this paper is rather lengthy and the discussions across sections are intertwined, we present in this section a detailed outline of the paper along with the important results.

\subsubsection*{Constructive Definition of Analytic Solutions from Boundary Conditions at $\ci^+_-$}

We begin this paper with Section \ref{sec:afs}, which describes three different classes of solutions of Ricci-flat gravity in dimensions greater than four.

In Section \ref{sec:gensol}, we present the results from the previous work of one of us \cite{Capone:2021ouo}.  The metric is given in Bondi-Sachs gauge \eqref{Bondi-metric} (this gauge is used throughout this paper as well), where every component is a function of all the coordinates and is asymptotically expanded in terms of the radial coordinate $r$. The procedure put forward in \cite{Capone:2021ouo} shows that generic solutions are polyhomogeneous, containing logarithmic terms in the expansion \eqref{metric-exp}, and allows to identify all such independent terms and their descendants. The qualitative structure is presented in the main text and the explicit (and gory) details of the solutions are in Appendix \ref{app:genlargeexp}.

In Section \ref{sec:analyticsol}, we construct a smaller solution space by imposing the vanishing of the Riemann tensor at $\ci^+ _-$ in an orthonormal frame
\begin{equation}
\label{vanishingRiemann1}
r^4 e^\mu_{\hat \mu} e^\nu_{\hat \nu} e^\rho_{\hat \rho} e^\s_{\hat \s} R_{\mu\nu\rho\s}[G] |_{\ci^+_-} = 0 .
\end{equation}
This condition, along with an appropriate definition of the frame fields, reduces the general solution space to an analytic solution space $\CS_\text{analytic}$ (discussded in \ref{sec:analyticsol}) without logarithmic terms in the radial expansion (this is proved in Appendix \ref{sec:solution-nolog}). This is the solution space that we focus on in the majority of the paper. An important physical implication of this condition is that gravitational radiation vanishes in the far past of $\ci^+$, i.e. near $\ci^+_-$ (see Section \ref{sec:summary1} for discussion). The specific fall-off near $\ci^+_-$ is derived in Section \ref{sec:largeu-lingrav-1} using a linear approximation around Minkowski spacetime (this is done in Section \ref{sec:lin-grav}) and is further supported in Section \ref{sec:BMSconst} by the requirement that the large $u$ fall-offs must be consistent with the action of the subsequently derived diffeomorphism group.

Metrics in $\CS_\text{analytic}$ are  characterized by the following data on $\ci^+$ -- radiation ${\wt h}_{\2AB}(u,x)$, mass aspect $M(x)$, angular momentum aspect $N_A(x)$, four-dimensional diffeomorphisms $\chi^A(x)$, Weyl factor $\Phi(x)$, and a scalar function $C(x)$:
\begin{equation}
\begin{split}
\CS_\text{analytic} |_{\ci^+} = \left\{ {\wt h}_{\2AB}(u,x) , \Phi(x) , C(x) , \chi^A(x) , M(x) , N_A(x) \right\} .
\end{split}
\end{equation}
Here, $u$ is the null generator of $\ci^+$ and $x^A$ are generalised coordinates on a transverse cut. $\Phi(x)$ and $\chi^A(x)$ are related to the induced metric and $C(x)$ is related to the extrinsic curvature of $\ci^+$. On the other hand $M$ and (part of) $N_A$ are the boundary values of certain components of the Riemann tensor. The explicit expressions that relate the data to metric components are given in \eqref{eq:q-form}, \eqref{sec:h1} and \eqref{sec:nolog-U3W5tilde-def}. Explicit details of this solution are in Appendix \ref{sec:solution-nolog}.

The two solutions spaces described above are novel. In Section \ref{sec:mrS0-explicit-details}, we discuss the standard or canonical solution space $\CS_\text{canonical}$. This is the solution that is discussed in almost all previous literature on higher dimensional gravity. The solution space $\CS_\text{analytic}$ is related to $\CS_\text{canonical}$ by fixing $C=\Phi=0$ and $\chi^A = x^A$. Explicit details of this solution are in Appendix \ref{app:mrS0-explicit-details}. In Appendix \ref{sec:news-tensor}, we discuss various notions of the news tensor which have appeared in literature.

\subsubsection*{Computational Trick: Large Diffeomorphisms to single out Canonical d.o.f.}

In Section \ref{sec:BMS-transform}, we study the asymptotic symmetry group of the three solution spaces described above. In particular, we show that the `general solutions' and `analytic solutions' both admit an infinite-dimensional Weyl-BMS symmetry whereas the `canonical solutions' admit only the finite-dimensional Poincar\'e group.

A crucial and central result of this paper comes from the observation that the spaces $\CS_\text{canonical}$ and $\CS_\text{analytic}$ are related to each other via BMS transformations. This also implies that $C$, $\Phi$ and $\chi^A$ are pure large gauge modes on $\CS_\text{analytic}$. More precisely, we show in Section \ref{sec:finite-BMS} that for every metric $G_{\mu\nu}(X)$ in the analytic space, there exists a unique metric $ \mr{G}_{\rho\s}(\mr{X}(X))$ in the canonical solution space, and a diffeomorphism $X^\mu \to \mr{X}^\mu(X)$ such that
\begin{equation} \label{eq: finite_BMS_transformation}
G_{\mu\nu}(X) = \p_\mu \mr{X}^\rho (X) \p_\nu \mr{X}^\s(X) \mr{G}_{\rho\s}(\mr{X}(X)).
\end{equation}
Explicit details of this map can be found in Appendix \ref{app:BMStransform}. It has several important consequences. Firstly, it highlights the appropriate variables to work with, as it clearly differentiates between canonical degrees of freedom and large gauge degrees of freedom. Furthermore, it helps in treating the divergences in the pre-symplectic potential, as we will discuss in the next paragraph.

\subsubsection*{All Divergences in the Pre-Symplectic Potential are Boundary Terms}

The immediate issue one faces when trying to construct the charges associated to the Weyl BMS diffeomorphism group is that the symplectic form computed on $\CS_\text{analytic}$ has large $r$ divergences. To proceed, we must then regulate and renormalize all these divergences.

The first step towards this goal is the realization that \emph{all} large $r$ divergences of the pre-symplectic potential, for a metric in $\CS_\text{analytic}$, reduce to a boundary term. The way to prove this is by using \eqref{eq: finite_BMS_transformation}, which implies a splitting of the pre-symplectic potential in the following way
\begin{equation}
\begin{split}
\Theta_\S (\d ) = \int _\S\t (\d) = \int _\S \t_\text{can} (\d) + \oint _{\p\S} & Q_{\ds \mr{X}} , \qquad \ds \mr{X}^\mu = \d X^\nu \p_\nu X^\mu , 
\end{split}
\end{equation}
where
\begin{equation}
\begin{split}
 [ \star \t_\text{can} (\d) ]^\mu = G^{\mu\nu}  \p_\nu \mr{ X}^\rho \left( [ \star \mr{\t} (\d) ]_\rho |_{  X \to \mr{ X} } \right)& , \qquad [ \star Q_\xi ]^{\mu\nu} = 2 \n^{[\mu} \xi^{\nu]}  . 
\end{split}
\end{equation}
Given that $\mr{\t} (\d)$ is the symplectic potential density for the canonical metric $ \mr{G}$, it is straightforward to see that the bulk part $\t _\text{can}$ encodes the variation of the canonical degrees of freedom, and the boundary part $  Q_{\ds \mr{ X}}$ captures the variations of the large gauge fields - $\d C$, $\d \chi ^A$ and $\d \Phi$. In Section \ref{sec:symppot-rad}, we show that the bulk part is finite and that all the divergences are in the boundary contribution which is derived in Section \ref{sec:symppot-bdy}. This result relies crucially on the boundary conditions we chose, highlighting again their importance. It is not clear whether relaxing them would preserve it.

\subsubsection*{Renormalization via Local and Covariant Counterterms}

At this point, anyone who is familiar with covariant phase space formalism can think that the divergence issue is immediately resolved following \cite{Compere:2018ylh,Chandrasekaran:2021vyu}. The symplectic potential density is a $(d-1)$--form, whose definition is ambiguous, up to the addition of an exact form (or in other words a boundary term). Using this fact, such divergences can be naively cancelled by simply adding to $\theta$ the divergent pieces with opposite sign (see the discussion around \eqref{counterterm-prescription}). This is however not enough to provide a sensible notion of renormalization as it appears to us to be too arbitrary. Without any sort of first principles prescription for the renormalization, pieces can be added at will to $\theta$ at any order and hence also the finite part can be rendered highly ambiguous, if not cancelled altogether!

In this paper, we propose a prescription for \emph{local} and \emph{covariant} counterterms (in an appropriate sense, specifically on the transverse directions). These two principles reduce the freedom in the choice of the counterterms by constraining them to depend only on the induced metric $\cal g$ of $\p\S$, its extrinsic curvature $\cal k$ with respect to one of the null normals, and linearly on the variation of these quantities
\begin{equation}
\begin{split}
\int_{\p\S} Y(\d) = \int_{\p\S} \dt^4 x \sqrt{{\cal g}} \CY( \cal g , \cal k, \d \cal g, \d \cal k) . 
\end{split}
\end{equation}
On the other hand, $\cal g$ and $\cal k$ themselves depend only on the large gauge fields $\chi ^A$, $\Phi$, and $C$. This has an important consequence - it implies that we can never add a counterterm that depends on the mass and angular momentum aspects of the metric. They will necessarily enter the symplectic form and subsequently the charges. Two independent, but equivalent, procedures for the construction of this counterterm are described in Section \ref{sec:counterterm}.  

\subsubsection*{Invertibility of the Symplectic Form and Resolution of Large $u$ Divergences}

The renormalization procedure described above cancels all the divergences at large $r$ producing a symplectic potential \eqref{Theta-final-1} that is finite as we take $r \to \infty$. However, this still leaves behind large $u$ divergences that we cannot remove with this new technique (this is discussed in Section \ref{sec:large-u-divergences-symp-pot}). These large $u$ divergences are of two types -- there is an explicit divergence which is linear in $u$ and an implicit divergence which arises from the dependence of the extrinsic curvature on $u$. 

The explicit divergence is automatically removed when we fix the completely separate problem of invertibility. When computing the pre-symplectic form, we observe that it is degenerate. This can be understood intuitively by noticing (see equation \eqref{Theta-final}) that while the supertranslation mode $C(x)$ is conjugate to the mass aspect $M(x)$, and the superrotations mode $\chi^A(x)$ is conjugate to angular momentum aspect $N_A(x)$, other independent degrees of freedom that could potentially be conjugate to $\Phi$ are simply not available! We are thus forced to freeze \eqref{Phi-final-form} this degree of freedom on the phase space.\footnote{This does not seem to be a dimension-dependent statement and we expect it to be true in four dimensions as well. However, this is in stark contrast with the conclusions of \cite{Freidel:2021fxf} which are in contradiction to our statements.} We do this by freezing the volume form on $\ci^+$. It turns out that this condition also gets rid of the explicit linear-in-$u$ divergence in the symplectic potential. With this restriction, we construct the non-degenerate symplectic form in Section \ref{sec:sympform}. We then invert this and construct all the Poisson/Dirac brackets of the theory explicitly (see equations \eqref{h2-brackets}--\eqref{N-brackets}).

The implicit large $u$ divergence cannot be renormalized and in fact plays a physical role in the charge algebra, discussed in detail in section \ref{sec:BMS-charge-algebra}.

\subsubsection*{GBMS Diffeomorphisms are NOT Canonical Transformations!}

In Section \ref{sec:BMS-charge-algebra}, we attempt to construct the Hamiltonian charges which generate GBMS diffeomorphisms. Unsurprisingly (based on our four-dimensional intuition), we find that the charge is in fact, not integrable! More precisely, there is no Hamiltonian function $H_{\xi }$ associated to a GBMS vector field $\xi$ \eqref{WBMSvector} on the phase space, that satisfies
\begin{equation}
    \d H _{\xi } = \Omega (\d, \d _{\xi} ).
\end{equation}
It follows that GBMS diffeomorphisms are not canonical transformations! The obstructions come precisely from the non-covariant pieces in the symplectic potential that are responsible for implicit large $u$ divergences. It might be possible to solve this integrability issue by choosing field-dependent parameters for the GBMS diffeomorphisms, but this requires solving a complicated functional derivative equation, which is beyond the scope of this project. The usual way of handling this non-integrability is to work with local non-integrable charges. Here we take a different route.

Noting that we cannot construct charges associated to GBMS diffeomorphisms, we do the next best thing -- we extract the ``integrable part'' of $\Omega (\d, \d _{\xi})$ and use that to construct the corresponding Hamiltonian charges. By construction, these functions would be the closest thing that we have to GBMS charges. This is done in Section \ref{sec:BMS-charge-algebra} and the corresponding charges are
\begin{equation}
\begin{split}
T_f = \frac{1}{4\pi G} \int \dt^4 x \sqrt{q} f M , \qquad J_Y = \frac{1}{16\pi G} \int \dt^4 x \sqrt{q}\, Y^A N_A , 
\end{split}
\end{equation}
where $f$ and $Y^A$ are the field-independent supertranslation and superrotation parameters. Using the brackets \eqref{h2-brackets}--\eqref{N-brackets}, we find that they faithfully represent the GBMS algebra
\begin{equation}
\begin{split}
\{ T_f , T_{f'} \} &= 0 , \qquad \{ J_Y , T_f \}  = T_{( Y \cdot \p  - \frac{1}{4} D \cdot Y ) f  } , \qquad \{ J_Y , J_{Y'} \} = J_{[Y,Y']} , 
\end{split}
\end{equation}
and the transformations they generate are almost identical to the variations from GBMS diffeomorphisms (denoted in the equations below with $\d _{\xi}$)
\begin{equation}
\begin{split}
\{ T_f + J_Y , {\wt h}_{\2AB} \} &= -  \left( f \p_u + \CL_Y + \frac{1}{4} D \cdot  Y u \p_u \right) {\wt h}_{\2AB}  = - \d _{\xi }  {\wt h}_{\2AB}, \\ 
\{ T_f + J_Y , C \} &= - \CL_Y C -  e^{-\Phi} f  = - \d _{\xi} C,  \\
\{ T_f + J_Y , M \} &= - \left( \CL_Y  + \frac{5}{4} D \cdot  Y \right) M  = - \d _{\xi } M  , \\
\{ T_f + J_Y , \chi^A \} &= - Y^B  \p_B \chi^A = - \d _{\xi } \chi ^A , \\
\{ T_f + J_Y , N_A \} &=  - ( \CL_Y + D \cdot Y ) N_A - ( f \p_A  + 5 \p_A  f ) M  \neq -  \d _{\xi } N_A. \\ 
\end{split}
\end{equation}
The only difference is in their action on the angular momentum aspect. Under GBMS diffeomorphisms $N_A$  has a more complicated non-covariant transformation.
We call these types of transformations GBMS transformations and distinguish them from GBMS diffeomorphisms.

\subsubsection*{GBMS Ward Identities and Soft Theorems}

In Section \ref{sec:HilbertSpace}, we quantize the classical phase space constructed in the previous section (using canonical quantization) and in doing so, we construct the asymptotic Hilbert space of the theory. This has an infinite-dimensional moduli space of vacua which are spanned by the zero mode operators $C(x)$, $M(x)$ and $\chi^A(x)$, $N_A(x)$. We next define creation and annihilation operators from the radiation field ${\wt h}_\2$ (essentially by taking a Fourier transform in $u\to \o$) and construct the rest of the Hilbert space as a Fock space. From this point, we follow the standard procedure to construct the Ward identities. In Section \ref{sec:matchingcond}, we impose a set of antipodal matching conditions \eqref{eq:matchingcond} between fields on $\ci^+_-$ and $\ci^-_+$ on the large gauge modes which immediately implies equality of $\ci^+$ and $\ci^-$ GMBS charges \eqref{GBMScharges}. This equality can be massaged into a ``soft theorem'' \eqref{soft-exp-2} by using Einstein's equations to recast the GBMS charges as integrals over all of $\ci^+$ \eqref{GBMScharges-uint} which then allows us to separate out a soft part which is linear in the radiation field and a hard part which is quadratic \eqref{GBMScharges_sh}. We then explicitly verify in Section \ref{sec:soft-theorems} that the Ward identities \eqref{Ward-Identity} constructed in this way are exactly the same as the soft-graviton theorems \eqref{soft-exp-2}, including the non-trivial generalization to spinning fields.

\section{Asymptotically Flat Spacetimes Near \texorpdfstring{$\ci^+$}{ip}}
\label{sec:afs}

In this section, we define and classify the space of geometries that we study in this paper.

\subsection{Definition}
\label{sec:afs-def}

\begin{figure}[ht!]
\label{fig:Penrose}
\begin{center}
\includegraphics[scale=1]{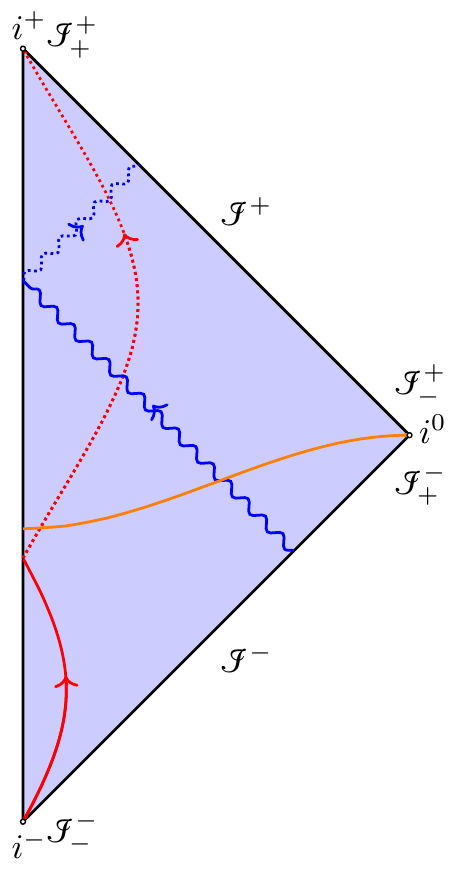}
\end{center}
\caption{\textbf{Penrose Diagram for Globally Asymptotically Flat Spacetimes} (1) Timelike future and past infinity $i^+$ and $i^-$ are spacelike boundaries and timelike geodesics (red) start on $i^-$ and end on $i^+$. (2) Spatial infinity $i^0$ is a timelike boundary on which spacelike geodesics (orange) end. (3) Lightlike past and future infinity $\ci^-$ and $\ci^+$ are null boundaries and null geodesics (blue) start on $\ci^-$ and end on $\ci^+$. $\ci^+$ ($\ci^-$) itself has future and past boundaries which denoted by $\ci^+_+$ ($\ci^-_+$) and $\ci^+_-$ ($\ci^-_-$) respectively.}
\label{fig:penrose}
\end{figure}

The Penrose diagram of a globally asymptotically flat spacetime is shown in Figure \ref{fig:Penrose}.

 In this paper, we will be interested in studying the structure of asymptotically flat spacetimes at asymptotic future null infinity $\ci^+$.\footnote{Formally, $\ci^+$ is defined by first conformally compactifying $(M,G)$ to an unphysical spacetime $(M,{\wt G})$ and then affixing $\ci^+$ as a \emph{finite} null boundary of $(M,{\wt G})$. In this paper, we do not discuss this compactification and $\ci^+$ is simply taken to be an \emph{asymptotic} null boundary of $(M,G)$.} To describe these regions, it is convenient to work in Bondi-Sachs coordinates $X^\mu = (u,r,x^A)$ where the metric takes the form
\begin{equation}
\begin{split}\label{Bondi-metric}
G_{\mu\nu} \dt X^\mu \dt X^\nu = -  e^{2\b} U \dt u^2 -  2 e^{2\b}   \dt u   \dt r  + r^2 h_{AB} ( \dt x^A - W^A \dt u ) ( \dt x^B - W^B \dt u ) , 
\end{split}
\end{equation} 
$h_{AB}$ satisfies the so-called Bondi determinant condition
\begin{equation}
\begin{split}\label{Bondi-detcond}
\p_r \det ( h_{AB} ) = 0 . 
\end{split}
\end{equation}
The metric functions $U$, $\b$, $W^A$ and $h_{AB}$ are all functions of $u$, $r$ and $x^A$.

In asymptotically flat spacetimes, for a fixed coordinate value $(u,x^A)$, the coordinate $r$ extends to infinity and the conformal boundary at $r \to \infty$ is $\ci^+$. Near $\ci^+$ (i.e. at large $r$), the metric functions behave as\footnote{More general boundary conditions, which we will not work with,  compatible with the requirement of a null boundary and Einstein equations are  $U = \CO(r)$, $\b = \CO(r^0)$, $W^A = \CO(r^{-1})$ and $h_{AB} = q_{AB}(u,x) + \CO(r^{-1})$. See \cite{Capone:2021ouo} for the discussion of the solution space with such conditions.}
\begin{equation}
\begin{split}\label{afs-falloff}
U = \CO(1) , \qquad \b = \CO(r^{-2}) , \qquad W^A = \CO(r^{-2}) , \qquad h_{AB} = q_{AB}(x) + \CO(r^{-1}) .
\end{split}
\end{equation}
These conditions ensure that $\ci^+$ is null. The past and future boundaries of $\ci^+$ are located at $u=-\infty$ and $u=+\infty$ and denoted by $\ci^+_-$ and $\ci^+_+$ respectively.

\subsection{Classification of Ricci-Flat Metrics}

\label{sec:afs-Einstein}

We now consider asymptotically flat geometries in Einstein gravity. These geometries are Ricci-flat,
\begin{equation}
\begin{split}\label{Einstein-EoM}
R_{\mu\nu}[G] = 0 . 
\end{split}
\end{equation}
In this section, we will study the behaviour of these geometries near $\ci^+$, i.e. near $r \to \infty$. The complete analysis of the asymptotic structure in \emph{any} dimension $D \geq 4$ can be found in \cite{Capone:2021ouo}. Here, we report the complete result for $D=6$.

\subsubsection{General (Polyhomogeneous) Solution Space: \texorpdfstring{$\CS_\text{general}$}{Sgen}}
\label{sec:gensol}
The large $r$ expansion for generic Ricci-flat metrics in $D=6$ is given by
\begin{equation}
\begin{split}\label{metric-exp}
\b &= \frac{\b_\2}{r^2} + \frac{\b_\3 + \b_\ttt{3,1} \ln  r }{r^3} + \frac{\b_\4 + \b_\ttt{4,1} \ln  r + \b_\ttt{4,2} \ln^2 r }{r^4}  + \CO(r^{-5}) ,  \\
U &= U_\0 + \frac{U_\1}{r} + \frac{U_\2 + U_\ttt{2,1} \ln  r }{r^2} + \frac{U_\3 + U_\ttt{3,1} \ln  r + U_\ttt{3,2} \ln^2 r  }{r^3} + \CO(r^{-4}),\\
W^A &= \frac{W^A_\2}{r^2} + \frac{W^A_\3 + W^A_\ttt{3,1} \ln  r }{r^3} + \frac{W^A_\4 + W^A_\ttt{4,1} \ln  r   }{r^4} \\
&\qquad \qquad \qquad \qquad \qquad + \frac{W^A_\5 + W^A_\ttt{5,1} \ln  r + W^A_\ttt{5,2} \ln^2 r + W^A_\ttt{5,3} \ln^3 r }{r^5}  + \CO(r^{-6}) ,   \\
h_{AB} &= q_{AB} + \frac{h_{\1AB}}{r} + \frac{h_{\2AB} + h_{\ttt{2,1}AB} \ln  r}{r^2} + \frac{h_{\3AB} + h_{\ttt{3,1}AB} \ln  r }{r^3} \\
&\qquad \qquad \qquad \qquad  \qquad \qquad \qquad + \frac{h_{\4AB} + h_{\ttt{4,1}AB} \ln  r + h_{\ttt{4,2}AB} \ln^2 r }{r^4}  + \CO(r^{-5}) .
\end{split}
\end{equation}
All the coefficients are functions of both $u$ and $x^A$ -- except $q_{AB}$ which is just a function of $x$. The Bondi determinant condition \eqref{Bondi-detcond} completely fixes the trace of $h_{AB}$ (w.r.t. $q_{AB}$) in terms of the trace-free part $h_{\{AB\}}$.\footnote{We will use this curly bracket notation around spacetime indices to denote the \emph{symmetric trace-free} part of a tensor $X_{\{A B\}} \equiv X_{(A B)} - \frac{1}{4} q_{AB}\tr [X] $.} This trace condition can be imposed order-by-order in large $r$. 

Einstein's equations \eqref{Einstein-EoM} determine the expansion and its coefficients as follows:
\begin{itemize}
\item $R_{rr}[G] = 0$ fixes $\b$ in terms of $h_{\{AB\}}$. 
\item $R_{rA}[G] = 0$ fixes $W^A$ in terms of $h_{\{AB\}}$  and $W_\5^A$. 
\item $G^{AB} R_{AB}[G] = 0$ fixes $U$ in terms of $h_{\{AB\}}$, $W_\5^A$ and $U_\3$. 
\item $R_{AB}[G] =0$ fixes $h_{\{AB\}}$ in terms of $h_{\2\{AB\}}$, $W_\5^A$ and $U_\3$ and $h_{\{AB\}}|_\G$ where $\G$ is the null surface defined by $u=-\infty$. Note that $\ci^+ \cap \G = \ci^+_-$.
\item  $R_{uu}[G] = 0$ fixes $U_\3$ in terms of $h_{\2\{AB\}}$, $U_\3|_{\ci^+_-}$ and $h_{\{AB\}}|_\G$. 
\item $R_{uA}[G] = 0$ fixes $W^A_\5$ in terms of $h_{\2\{AB\}}$, $W_\5^A|_{\ci^+_-}$, $U_\3|_{\ci^+_-}$ and $h_{\{AB\}}|_\G$. 
\item $R_{ur}[G] = 0$ does not impose any new constraints. 
\end{itemize}
Note that the first three sets of equations are algebraic in nature so they fix the corresponding functions entirely whereas the last three sets of equations are differential (in $u$) so they fix their corresponding functions up to $u$-independent integration constants. Here, we have chosen the values of the fields at a fixed null time $u=-\infty$ as the integration constants.\footnote{Some of the fields are actually divergent as $u \to -\infty$. In this case, what we really mean by $f|_{\ci^+_-}$ is the $\CO(1)$ term in the Taylor expansion of $f$ around $u=-\infty$.} Altogether, the full solution is determined entirely by the following data:
\begin{equation}
\begin{split}\label{S-coord}
\CS_\text{general} = \left\{ h_{\2\{AB\}} , q_{AB} , U_\3|_{\ci^+_-} , W_\5^A|_{\ci^+_-}, h_{\{AB\}}|_\G \right\}  .
\end{split}
\end{equation}
The complete details of the solution including explicit formulas for each of the large $r$ coefficients shown in \eqref{metric-exp} can be found in Appendix \ref{app:genlargeexp}. The interpretation of each of the fields in \eqref{S-coord} is as follows:
\begin{itemize}
\item $h_{\2\{AB\}}$ describes the gravitational flux across $\ci^+$.
\item $q_{AB}$ is the transverse induced metric on $\ci^+$.
\item $U_\3|_{\ci^+_-}$ describes the mass distribution in the system.
\item $W_\5^A|_{\ci^+_-}$ describes the angular momentum distribution in the system.
\item $h_{\{AB\}}|_\G$ describes the gravitational flux across $\G$. 
\end{itemize}
To conclude, let us point out that \eqref{metric-exp} is constructed following a principle of ``naturality''. We have only included those log terms that are \emph{forced} on us by the equations of motion. In this case, the independent log terms are $h_{\ttt{2,1}AB}$, $U_{\ttt{3,1}}$ and $W^A_{\ttt{5,1}}$. Every other log term in the asymptotic expansion is sourced by these three modes. These three ``fundamental'' log terms are themselves sourced by the non-log terms in the expansion, e.g. $h_\ttt{2,1}$ is sourced by $h_\1$ so we cannot set these to zero without imposing additional constraints on the solution. This procedure naturally produces maximally polyhomogeneous expansions starting from the radiation order (for example in the case of four-dimensional spacetimes it shows the generic existence of a time-independent term $h_\ttt{1,1}$).  Notice that this approach is different from that pursued in \cite{Chrusciel:1993hx, Godazgar:2020peu}, where the solution space is defined starting from an arbitrary choice of polyhomogeneous initial characteristic data.

\subsubsection{Analytic Solution Space: \texorpdfstring{$\CS_\text{analytic}$}{San}}
\label{sec:analyticsol}

Due to several large $r$ divergences, it is very difficult (though perhaps not impossible) to elevate the polyhomogeneous solution space of the previous section to a phase space. The main culprit turns out to be the log terms in \eqref{metric-exp}. As we will see in Section \ref{sec:symppot}, in the absence of such terms it \emph{is} possible to construct the phase space and even though the aforementioned large $r$ divergences are still present, it is possible to regulate and renormalize them.

The log terms can be removed by imposing the following diffeomorphism invariant constraint
\begin{equation}
\begin{split}\label{vanishingRiemann}
r^4 e^\mu_{\hat \mu} e^\nu_{\hat \nu} e^\rho_{\hat \rho} e^\s_{\hat \s} R_{\mu\nu\rho\s}[G] |_{\ci^+_-} = 0 . 
\end{split}
\end{equation}
Here $e^\mu_{\hat \mu}$ is an orthonormal basis of vielbein. Physically, we interpret this condition to imply the absence of gravitational flux through $\G$ near $\ci^+_-$. In Appendix \ref{sec:solution-nolog}, we present a detailed analysis of this constraint and show that it does, in fact, remove \emph{all} the log terms. In addition to this, it also imposes a few constraints on the non-log coefficients which we now discuss.

The first non-trivial constraint is the vanishing of the Weyl tensor of $q_{AB}$ which implies that it is conformally flat,
\begin{equation}
\begin{split}
\label{eq:q-form}
q_{AB}(x) = e^{2\Phi(x)} {\hat q}_{AB}(x) , \qquad {\hat q}_{AB}(x) = \p_A \chi^C(x) \p_B \chi^D(x) \d_{CD} . 
\end{split}
\end{equation}
$\Phi(x)$ is the Weyl factor and $\chi^A(x)$ is the finite diffeomorphism which maps the flat metric ${\hat q}_{AB}$ to the flat Cartesian metric $\d_{AB}$. From the bulk point of view, $\chi^A$ is a \emph{large} diffeomorphism (as we will see in Section \ref{sec:GMBSalgebra}) so it is \emph{not} trivial!

Next, \eqref{vanishingRiemann} imposes a constraint on $h_\1$,
\begin{equation}
\begin{split}\label{Omega-zero}
\o_{ABC} \equiv  D_{[A} h_{\1B]C} + \frac{1}{3} ( D \cdot h_\1 )_{[A} q_{B]C}  = 0 ,
\end{split}
\end{equation}
where $D_A$ is a covariant derivative w.r.t. $q_{AB}$. Here and in the rest of this paper, we employ vector and matrix notation where all repeated indices are contracted w.r.t. $q_{AB}$, e.g.
\begin{equation}
\begin{split}
( h_\1^2 )_{AB} \equiv h_{\1AC} q^{CD} h_{\1DB} , \qquad \tr [ h_\1^3 ] \equiv h_{\1AB} q^{BC} h_{\1CD} q^{DE} h_{\1EF} q^{FA} .
\end{split}
\end{equation}
Derivative indices are similarly contracted
\begin{equation}
\begin{split}
( D h_\1 )_{ABC} \equiv  D_A h_{\1BC} , \qquad ( D \cdot h_\1 )_A = q^{BC} D_B h_{\1CA} . 
\end{split}
\end{equation}
Together with conformal flatness of $q$ {\eqref{eq:q-form}} and the equation of motion $R_{AB}[G] = 0$, \eqref{Omega-zero} solves to
\begin{equation}\label{sec:h1}
\begin{split}
h_{\1AB} = - u R_{\{AB\}} - ( 2 D_{\{A} D_{B\}} + R_{\{AB\}} ) ( e^\Phi C  ) . 
\end{split}
\end{equation}
In other words, $h_\1$ is determined entirely in terms of a scalar $C(x)$.\footnote{$h_\1$ is defined in terms of $e^\Phi C$ instead of $C$ as this simplifies many of the equations derived in this paper.}

It is important to note that $q_{AB}$ and $h_{\1AB}$ are the actual relevant degrees of freedom in the geometry, so $\Phi(x)$, $\chi^A(x)$ and $C(x)$ are not independently uniquely defined. Rather, the trio $\big(\chi^A(x),\Phi(x),C(x) \big)$ is defined up to the identification
\begin{equation}
\begin{split}\label{chi-phi-identification}
\bigg( \chi^A(x) , \, \Phi(x) , \, C(x)  \bigg) \sim \bigg( {\cal x}_g^A(\chi(x)) , \, \Phi(x) - \ln \O_g(\chi(x))  , \, \O_g(\chi(x)) [C(x) + {\cal c}(x)] \bigg) ,
\end{split}
\end{equation}
where
\begin{equation}
\begin{split}
\pd{{\cal x}_g^C(\chi)}{\chi^A} \pd{{\cal x}_g^D(\chi)}{\chi^B} \d_{CD} = \O_g^2(\chi) \d_{AB}  , \qquad ( 2 D_{\{A} D_{B\}} + R_{\{AB\}} ) [ e^{\Phi(x)} {\cal c}(x) ]  = 0 . 
\end{split}
\end{equation}
In other words, ${\cal x}_g$ is a conformal transformation and $\O_g$ is the corresponding Weyl factor. It is clear that such a redefinition leaves $q$ and $h_\1$ invariant.

To describe the final set of constraints imposed by \eqref{vanishingRiemann}, it is useful to first define the following fields
\begin{equation}
\begin{split}
\label{sec:htilde-def}
{\wt h}_\2 &\equiv  \bigg( h_\2 - \frac{1}{4} h_\1^2  \bigg)^\tf , \\
{\wt h}_\3 &\equiv  \bigg( h_\3 -  h_\1 {\wt h}_\2 + 2 \b_\2 h_\1 \bigg)^\tf , \\
{\wt h}_\4 &\equiv \bigg( h_\4 - \frac{1}{2} {\wt h}_\2^2 - \frac{1}{2} h_\1 {\wt h}_\3  -  \frac{1}{4} h_\1 {\wt h}_\2 h_\1 + 4 \b_\2 {\wt h}_\2 +  \frac{1}{96} h_\1  \tr [ h_\1^3 ]  \bigg)^\tf .
\end{split}
\end{equation}
where the superscript $\tf$ denotes the trace-free part w.r.t. $q$. The constraints can now be written as\footnote{The latter two guarantee the vanishing of $U_{\ttt{3,1}}$ and $W^A_{\ttt{5,1}}$. See the comments after (5.31) in \cite{Capone:2021ouo} for some more considerations about them and the comparison with the case of four-dimensional spacetimes.}
\begin{equation}
\begin{split}\label{hh-div-constraints}
{\wt h}_{\2AB} |_{\ci^+_-} &=  {\wt h}_{\3AB} |_{\ci^+_-}  =  {\wt h}_{\4AB} |_{\ci^+_-}  = 0, \\
D \cdot D \cdot {\wt h}_\3  &= - \frac{1}{2} \tr [ {\wt h}_\3 R ]  , \\
D \cdot \left( {\wt h}_\4 - \frac{1}{2} h_\1 {\wt h}_\3 \right) &= - \frac{1}{16} \tr [  3 h_\1 D {\wt h}_\3 - {\wt h}_\3 D h_\1 ] . 
\end{split}
\end{equation}
With this, we are done! Once all the constraints described above are imposed, all the log terms vanish and we have an analytic solution.

Before ending this section, we note that while \eqref{vanishingRiemann} forces the tilded fields \eqref{sec:htilde-def} to vanish on $\ci^+_-$, it does \emph{not} tell us how fast those fields fall-off in a neighbourhood. However, it is possible to determine this using other considerations which are outlined in Section \ref{sec:largeufalloff}. For completeness and readability, we retroactively present those results in this section. First, it is noted that the tilded fields \eqref{sec:htilde-def} satisfy
\begin{equation}
\begin{split}\label{sec:htilde-largeu}
{\wt h}_\2 = \CO(u^{-3}) , \qquad  {\wt h}_\3 = \CO(u^{-2}) , \qquad {\wt h}_\4 = \CO(u^{-1}) \quad \text{as $u \to \pm\infty$.}
\end{split}
\end{equation}
Using this, we can determine the large $u$ behaviour of $U_\3$ and $W_\5$ as well. To do this, we first define
\begin{equation}
\begin{split}\label{sec:nolog-U3W5tilde-def}
{\wt U}_\3 &\equiv U_\3 + 8 \p_u  ( \b_\2^2 ) + W_\2 h_\1 W_\2  + 4 W_\2 D \b_\2  + \frac{R}{576}  \tr [ h_\1^3 ]  - \frac{1}{128} \p_u  \tr [    h_\1^4  ] , \\
{\wt W}_\5 &\equiv W_\5 + 10 D ( \b_\2^2 )  - \frac{1}{256}  D \tr [ h_\1^4 ]    - \frac{1}{48} \tr[ h_\1^3 ] W_\2  + 6 \b_\2 h_\1 W_\2  + \frac{1}{2} h_\1^3 W_\2 \\
&\qquad \qquad \qquad \qquad \qquad - \frac{1}{96} h_\1 D \tr[h_\1^3] + \frac{3}{2} h_\1^2 D \b_\2 .
\end{split}
\end{equation}
These fields satisfy the $u$-evolution equations \eqref{app:nolog-U3du} and \eqref{app:nolog-W5du} (which are too long to present here). They give an appropriate definition of the mass and angular momentum aspect. Using \eqref{sec:htilde-largeu}, we can immediately determine the large $u$ behaviour of these fields
\begin{equation}
\begin{split}\label{sec:MNAdef}
{\wt U}_\3(u,x) &~\xrightarrow{u\to-\infty}~ - M (x) + \CO(u^{-2}) , \qquad \qquad \qquad \qquad~~ {\wt U}_\3(u,x) ~\xrightarrow{u\to+\infty}~ \CO(u^{-2}) , \\
{\wt W}_{\5A}(u,x) &~\xrightarrow{u\to-\infty}~ - \frac{u}{5} \p_A M (x) - \frac{1}{5} {\cal N}_A(x) + \CO(u^{-1}) , \qquad {\wt W}_{\5} (u,x) ~\xrightarrow{u\to+\infty}~ \CO(u^{-1}) . 
\end{split}
\end{equation}
The behaviour at large positive $u$ is fixed by the requirement that the system reverts to the vacuum in the far future of $\ci^+_+$.\footnote{This condition is modified if there exist massive matter fields in the theory which contribute a non-vanishing flux at $\ci^+_+$. In this case, both ${\wt U}_\3$ and ${\wt W}_\5$ are non-vanishing at $\ci^+_+$.} $M(x)$ and $\CN_A(x)$ are the integration constants that are obtained upon integrating the $u$-evolution equations \eqref{app:nolog-U3du} and \eqref{app:nolog-W5du}.
With all of this, we can now summarise the data that defines the analytic solution space
\begin{equation}
\begin{split}\label{S-analytic-data}
\CS_\text{analytic} = \left\{ {\wt h}_{\2AB} , \, \Phi , \, C , \,\chi^A ,\, M ,\, {\cal N}_A ,\, h_{\{AB\}}|_\G \right\} .
\end{split}
\end{equation}
 The detailed asymptotic expansion for solutions in $\CS_\text{analytic}$ is given in Appendix \ref{sec:solution-nolog}.

\subsubsection{Canonical Solution Space: \texorpdfstring{$\CS_\text{canonical}$}{Scan}}
\label{sec:mrS0-explicit-details}

The two solution spaces described so far are novel and apart from \cite{Capone:2021ouo}, have \emph{not} been previously discussed in the literature. The standard or ``canonical'' solution space, which has been extensively studied, e.g. in \cite{Hollands:2003ie,Tanabe:2009va,Tanabe:2010rm,Tanabe:2011es,Hollands:2016oma}, involves a much \emph{stronger} definition for asymptotically flat spacetimes which is motivated by the asymptotic behaviour of linearised gravitational radiation, namely\footnote{This is not entirely correct. The definition that is actually used is $U=1+\CO(r^{-2})$, $\b = \CO(r^{-4})$, $W^A=\CO(r^{-3})$ and $h_{AB} = \g_{AB} + \CO(r^{-2})$ where $\g_{AB}$ is the round metric on $S^4$. Here, we use a slightly different but equivalent definition. This choice simplifies our calculations since we do not have to keep track of curvature contributions from the round $S^4$.} 
\begin{equation}
\begin{split}\label{can-bdy-cond}
h_{AB} = \d_{AB} + \CO(r^{-2}) \quad \implies \quad U =   \CO(r^{-2}) , \qquad \b = \CO(r^{-4}) , \qquad W^A = \CO(r^{-3})  .
\end{split}
\end{equation}
This implies that $q_{AB} = \d_{AB}$ and $h_{\1AB} = 0$. Comparing this to \eqref{eq:q-form} and \eqref{sec:h1}, these boundary conditions are equivalent to the requirement $C(x)=\Phi(x)=0$ and $\chi^A(x)=x^A$. The data that defines this solution space is therefore
\begin{equation}
\begin{split}\label{S0dot-coord}
\CS_\text{canonical} = \left\{ \mr{h}_{\2AB} (u,x) , \mr{M}(x) , \mr{\CN}_A (x) , \mr{h}_{\{AB\}} |_{\G}  \right\} .
\end{split}
\end{equation}
Note that we have used a superscript $\circ$ to distinguish metrics in $\CS_\text{canonical}$ from those in $\CS_\text{analytic}$. The detailed asymptotic expansion for solutions in $\CS_\text{canonical}$ is given in Appendix \ref{app:mrS0-explicit-details}.

\subsection{Linearised Gravity and Mode Expansions}
\label{sec:lin-grav}

Having discussed the general (non-linear) solution space in Einstein gravity, we now discuss solutions in linearised gravity in the background of Minkowski spacetime. These solutions are described by a mode expansion and live in $\CS_\text{canonical}$. To construct the solutions, we first write the metric as
\begin{equation}
\begin{split}\label{linearised-bondi-coord}
\mr{G}_{\mu\nu}(X) = \mr\eta_{\mu\nu}(X) + \mr\g_{\mu\nu}(X) , \qquad \mr\eta_{\mu\nu} \dt X^\mu \dt X^\nu = - 2 \dt u \dt r + r^2 \d_{AB} \dt x^A \dt x^B.
\end{split}
\end{equation}
$\mr\eta_{\mu\nu}(X)$ is the metric of Minkowski spacetime in Bondi-Sachs coordinates. The linearised perturbation ${\mr\g}$ satisfies the equation
\begin{equation}
\begin{split}\label{linearised-eom}
{\mr\n}_\mu {\mr\n}^\rho {\mr\g}_{\nu\rho} + {\mr\n}_{\nu} {\mr\n}^\rho {\mr\g}_{\mu\rho} - {\mr\n}^2 {\mr\g}_{\mu\nu} - {\mr\n}_\mu {\mr\n}_\nu {\mr\g} = 0 , \qquad {\mr\g} \equiv {\mr\eta}^{\mu\nu} {\mr\g}_{\mu\nu} . 
\end{split}
\end{equation}
This equation is traditionally solved in harmonic gauge (H) where we have
\begin{equation}
\begin{split}
\n^\nu {\mr\g}^\text{(H)}_{\mu\nu}  = \frac{1}{2} \n_\mu {\mr\g}^\text{(H)} \quad \implies \quad  \n^2 {\mr\g}^\text{(H)}_{\mu\nu}  = 0 . 
\end{split}
\end{equation}
This is the standard wave equation. In Cartesian coordinates ${\bar X}^\mu$, the \emph{radiative solution} to the wave equation is given by a mode expansion
\begin{equation}
\begin{split}\label{mode-exp}
{\mr{\bar \g}}^\text{(H)}_{\mu\nu} ({\bar X}) = \int \frac{\dt^5 p}{(2\pi)^5} \frac{1}{2p^0} \ve _{\mu\nu}^{AB} (p) \left[ O_{AB}(p) e^{ i p \cdot\bar  X} + O^\dagger_{AB}(p) e^{ - i p \cdot \bar X} \right] .
\end{split}
\end{equation}
where $p^2 =  p^\mu  \ve _{\mu\nu}^{AB} (p) = \mr{\eta}^{\mu\nu} \ve _{\mu\nu}^{AB} (p)  = 0$. The solution in \eqref{mode-exp} is given in Cartesian coordinates (for the background $\mr\eta$) and in harmonic gauge (for the perturbation ${\mr\g}$). In this paper, however, we are interested working entirely in Bondi-Sachs gauge {\eqref{Bondi-metric}}. To do this, we simply need to perform the appropriate coordinate and gauge transformations. The full solution in Bondi-Sachs gauge takes the form
\begin{equation}
\begin{split}\label{Bondi-lin-sol}
{\mr\g}_{\mu\nu}(X) = {\mr\g}^{\text{(H)}}_{\mu\nu} ( X )  + \CL_\zeta \eta_{\mu\nu}(X)  , \qquad {\mr\g}^{\text{(H)}}_{\mu\nu} ( X ) \equiv \p_\mu {\bar X}^\rho(X) \p_\nu {\bar X}^\s(X) {\mr{\bar \g}}^\text{(H)}_{\rho\s} (\bar X(X))  . 
\end{split}
\end{equation}
Here, ${\bar X}^\mu(X)$ is the coordinate transformation that maps from Cartesian coordinates to Bondi-Sachs coordinates. This is given by
\begin{equation}
\begin{split}
{\bar X}^\mu(X) = \bigg( \frac{r}{2} ( 1 + x^2 ) + u  \, , \,   r x^A   \, , \,  \frac{r}{2} ( 1 - x^2 ) - u   \bigg)  , \qquad x^2 = \d_{AB} x^A x^B .
\end{split}
\end{equation}
The second term is required to map from harmonic gauge to Bondi-Sachs gauge. The vector field $\zeta^\mu$ which does this will be determined by imposing the Bondi-Sachs gauge conditions,
\begin{equation}
\begin{split}\label{chi-eq}
{\mr\g}_{rr} &= 0 \qquad \implies \qquad  \p_r \zeta^u  =  \frac{1}{2} {\mr\g}^\text{(H)}_{rr} ,   \\
{\mr\g}_{rA} &= 0 \qquad \implies \qquad \p_r \zeta_A =  \frac{1}{r^2} ( \p_A \zeta^u  - {\mr\g}^\text{(H)}_{rA} ) , \\
\p_r ( \d^{AB} {\mr\g}_{AB} ) &= 0 \qquad \implies \qquad \p_r ( r \zeta^r )   = - \frac{1}{8} \p_r ( \d^{AB} {\mr\g}^\text{(H)}_{AB} + 2 r^2 \p_A \zeta^A ) . 
\end{split}
\end{equation}
The integration constants in $r$ are fixed by requiring that $\zeta^\mu$ is a \emph{small} diffeomorphism so that
\begin{equation}
\begin{split}\label{chi-bdycond}
\zeta^u = \CO(r^{-1}) , \qquad \zeta^A = \CO(r^{-1}) , \qquad \zeta^r = \CO(1) . 
\end{split}
\end{equation}
To solve \eqref{chi-eq}, we first need to determine ${\mr\g}^\text{(H)}$. We parameterize the integration variable $p$ and polarization tensor $\ve (p)$ as
\begin{equation}
\begin{split}\label{mompar}
p^\mu = \o \left( \frac{1 + y^2}{2}  \,,\,y^A \,,\, \frac{1 - y^2}{2} \right)  , \qquad \ve^{AB}_{\mu\nu}  (p) = \frac{1}{\o^2} \p^A p_{\{\mu}  \p^B p_{\nu\}}  .
\end{split}
\end{equation}
With this choice, it can immediately be determined that ${\mr\g}^\text{(H)}_{u\mu}  = 0$ and
\begin{equation}
\begin{split}
{\mr\g}^\text{(H)}_{rr}  &= - \frac{1}{2(2\pi)^5} \frac{1}{r^2} \int_0^\infty \dt \o \o  \left[ e^{ - i \o u }  \p^A \p^B  \int \dt^4 y \, O_{AB}(\o,x+y) e^{ - \frac{i}{2}  \o r y^2 } + \text{c.c.} \right] , \\
{\mr\g}^\text{(H)}_{rA}  &= \frac{1}{2(2\pi)^5} \int_0^\infty \dt \o  \o^2 \left[ i e^{ - i \o u }  \p^B \int \dt^4 y \,O_{AB}(\o,x+y) e^{ - \frac{i}{2}  \o r y^2 } + \text{c.c.} \right] , \\
{\mr\g}^\text{(H)}_{AB} &= \frac{r^2}{2(2\pi)^5} \int_0^\infty \dt \o  \o^3  \left[ e^{ - i \o u }  \int \dt^4 y\, O_{AB}(\o,x+y) e^{ - \frac{i}{2}  \o r y^2 } + \text{c.c.} \right] .
\end{split}
\end{equation}
With this, we can solve the differential equations \eqref{chi-eq}. To do this, we first perform the integral over $y$ above. This is given by
\begin{equation}
\begin{split}
\int \dt^4 y \, O_{AB}(\o,x+y)  e^{   - \frac{i}{2} \o r y^2  } &= \sum_{n=0}^\infty \frac{1}{r^{n+2}} \frac{(2\pi)^2}{2^n n!}  ( i \o )^{-n-2} ( \p^2 )^n   O_{AB}(\o,x)   .   \\
\end{split}
\end{equation}
To evaluate this, we first do a Taylor expansion of $O_{AB}(\o,x+y)$ around $y=0$ and then use the property
\begin{equation}
\begin{split}
\int \dt^4 y y^{A_1} \cdots y^{A_{2n}} e^{   - a y^2  } =  \frac{1}{a^{n+2}}  \frac{\pi^2\G(2n)}{2^{2n-1} \G(n)}    \d^{(A_1A_2} \cdots \d^{A_{2n-1} A_{2n})} .
\end{split}
\end{equation}
We then find
\begin{equation}
\begin{split}\label{gamma-tilde-exp}
{\mr\g}^\text{(H)}_{rr}  &= \frac{1}{2(2\pi)^3} \sum_{n=0}^\infty \frac{1}{r^{n+4}} \frac{1}{2^n n!}  \int_0^\infty \dt \o \o^{-n-1} \left[ i^n  ( - \p^2 )^n  \p^A \p^B  O_{AB}(\o,x) e^{ - i \o u }     + \text{c.c.} \right] , \\
{\mr\g}^\text{(H)}_{rA}  &=  \frac{1}{2(2\pi)^3} \sum_{n=0}^\infty \frac{1}{r^{n+2}} \frac{1}{2^n n!} \int_0^\infty \dt \o  \o^{-n} \left[ i^{n+1}   ( - \p^2 )^n   \p^B  O_{AB}(\o,x)  e^{ - i \o u }    + \text{c.c.} \right] , \\
{\mr\g}^\text{(H)}_{AB} &= - \frac{1}{2(2\pi)^3} \sum_{n=0}^\infty \frac{1}{r^n} \frac{1}{2^n n!}  \int_0^\infty \dt \o  \o^{-n+1}  \left[  i^n  ( - \p^2 )^n   O_{AB}(\o,x)  e^{ - i \o u }  + \text{c.c.} \right] .
\end{split}
\end{equation}
Using this large $r$ expansion, we can now determine $\zeta^\a$ from \eqref{chi-eq} and \eqref{chi-bdycond},
\begin{equation}
\begin{split}
\zeta^u &= - \frac{1}{4(2\pi)^3} \sum_{n=0}^\infty \frac{1}{r^{n+3}} \frac{(n+3)^{-1}}{2^n n!}    ( - \p^2 )^n  \int_0^\infty \dt \o \o^{-n-1}  [ i^n \p^A \p^B  O_{AB}(\o,x) e^{ - i \o u } + \text{c.c.}  ] , \\
\zeta_A &= - \frac{1}{6(2\pi)^3} \frac{1}{r^3} \int_0^\infty \dt \o  \left[ i  \p^B  O_{AB}(\o,x)  e^{ - i \o u }    + \text{c.c.} \right] - \frac{1}{4(2\pi)^3} \sum_{n=0}^\infty \frac{1}{r^{n+4}} \frac{(n+4)^{-1}}{2^n n!}   \\
&\qquad \qquad \times  ( - \p^2 )^n\int_0^\infty \dt \o  \o^{-n-1} \left[ i^n \left( \frac{ \d_A^B \p^2 }{n+1} - \frac{\p_A  \p^B}{n+3} \right) \p^C  O_{BC}(\o,x)  e^{ - i \o u } + \text{c.c.} \right] , \\
\zeta^r  &= - \frac{1}{4(2\pi)^3} \sum_{n=0}^\infty \frac{1}{r^{n+2}} \frac{n+1}{2^n (n+3)!} ( - \p^2 )^{n} \int_0^\infty \dt \o  \o^{-n} \left[ i^{n-1} \p^A \p^B O_{AB}(\o,x)  e^{ - i \o u } + \text{c.c.} \right]  . 
\end{split}
\end{equation}
Using this, we can determine the large $r$ expansion for $\mr h_{AB}\equiv \frac{1}{r^2} {\mr\g}_{AB} $ in Bondi-Sachs gauge,
\begin{equation}
\begin{split}
\mr h_{AB} &= - \frac{1}{2(2\pi)^3} \frac{1}{r^2} \int_0^\infty \dt \o  \o  O_{AB}(\o,x)  e^{ - i \o u }   + \frac{i}{4(2\pi)^3} \frac{1}{r^3} \int_0^\infty \dt \o  \CD_\ttt{\frac{4}{3}} O_{AB}(\o,x) e^{ - i \o u }   \\
&\qquad \qquad + \frac{1}{8(2\pi)^3} \sum_{n=0}^\infty \frac{i^n }{r^{n+4}} \frac{ ( - \p^2 )^{n}  }{2^{n} (n+2) !} \int_0^\infty \dt \o  \o^{-n-1}  \CD_\ttt{a_1} \CD_\ttt{a_2} O_{AB}(\o,x) e^{ - i \o u }  + \text{c.c.},
\end{split}
\end{equation}
where
\begin{equation}
\begin{split}
\label{SDa-def}
\CD_\ttt{a} O_{AB} \equiv \p^2 O_{AB} - a \p_{\{A} \p^C O_{B\}C} , 
\end{split}
\end{equation}
and $a_1$ and $a_2$ are solutions to the quadratic polynomial
\begin{equation}
\begin{split}
(n+3) (n+4) x^2-4 (n+2) (3 n+5) x+16 (n+1) (n+2) = 0 . 
\end{split}
\end{equation}
With this, we can finally extract the mode expansion formulae for $\mr{h}_\2$, $\mr{h}_\3$ and $\mr{h}_\4$ as
\begin{equation}
\begin{split}\label{h2modeexp}
\mr{h}_{\2AB} &= - \frac{1}{2(2\pi)^3}  \int_0^\infty \dt \o  \o  \left[   O_{AB}(\o,x) e^{ - i \o u }   + \text{c.c.} \right] , \\
\mr{h}_{\3AB} &= \frac{1}{4(2\pi)^3}  \int_0^\infty \dt \o   \left[ i \CD_\ttt{\frac{4}{3}}  O_{AB}(\o,x)  e^{ - i \o u }   + \text{c.c.} \right] , \\
\mr{h}_{\4AB}  &= \frac{1}{16(2\pi)^3}  \int_0^\infty \frac{\dt \o}{\o}  \left[   \CD_\ttt{2} \CD_\ttt{\frac{4}{3}}  O_{AB}(\o,x)   e^{ - i \o u  }  + \text{c.c.}\right] .
\end{split}
\end{equation}
Note that the mode expansion for $\mr{h}_\2$ is valid in \emph{any} gauge but the formulas for the subleading objects depends on our gauge choice. We also note that the equations above are consistent with the linearised versions of \eqref{Bondi-frame-sol}.

\subsection{Large \texorpdfstring{$u$}{u} Fall-Offs}
\label{sec:largeufalloff}

In this subsection, we provide two arguments for determining large $u$ falloff conditions of each coefficient of $h_{AB}$. The first is based on the linearised analysis performed above the second is on the consistency of the solution space under BMS diffeomorphisms.

\subsubsection{From Linearised Gravity}
\label{sec:largeu-lingrav-1}

Having determined $\mr{h}_{AB}$ as a mode expansion in \eqref{h2modeexp}, we can use it to determine the large $u$ fall-offs of these fields. It is clear that at large $u$, the dominant contribution to these integrals arises from the region near $\o=0$. To perform the integral, we expand the operator $O_{AB}$ as
\begin{equation}
\begin{split}\label{soft-exp-1}
O_{AB}(\o,x) = \frac{1}{\o} \CG_{AB}(x)  + \CB_{AB}(x) + \sum_{m=2}^\infty \o^{m-1} O_{AB}^\ttt{m}(x) .
\end{split}
\end{equation}
This structure of the expansion around $\o=0$ follows from our understanding of the soft (low energy) behaviour of graviton scattering amplitudes. In particular, insertions of the first two operators shown above are related to the leading and subleading soft graviton theorems respectively which are known to be universal, i.e. they do not depend on \emph{any} details of the theory \cite{Weinberg:1965nx,Cachazo:2014fwa}. We will review these soft theorems in Section \ref{sec:ward}. It is also possible to see that the operators satisfy \cite{He:2019ywq}
\begin{equation}
\begin{split}\label{soft-op-der-properties}
\CD_\ttt{\frac{4}{3}} \CG_{AB} = 0 , \qquad \CD_\ttt{2} \CD_\ttt{\frac{4}{3}} \CB_{AB} = 0, \qquad \CG_{AB} = \CG^{\dagger}_{AB}, \qquad \CB_{AB} =- \CB^{\dagger}_{AB},
\end{split}
\end{equation}
where $\CD_{(a)}$ are defined in \eqref{SDa-def}. These differential operators are precisely the ones that show up in the mode expansions \eqref{h2modeexp} and \eqref{soft-op-der-properties} play a crucial role in ensuring that the integrals over $\o$ converge. 

Substituting \eqref{soft-exp-1} into the integrands of \eqref{h2modeexp}, we find
\begin{equation}
\begin{split}
\mr{h}_{\2AB} &= \frac{1}{2(2\pi)^3}  \sum_{m=0}^\infty \frac{(m+2)!}{u^{m+3}}   \left[ (-i)^{m+1} O_{AB}^\ttt{m\!+\!2}(x) + i^{m+1} O_{AB}^{\ttt{m\!+\!2}\dagger}(x)  \right]   , \\
\mr{h}_{\3AB} &= \frac{1}{4(2\pi)^3} \sum_{m=0}^\infty  \frac{(m+1)!}{u^{m+2}}  \CD_\ttt{\frac{4}{3}}  \left[  (-i)^{m+1} O_{AB}^\ttt{m\!+\!2}(x) +  i^{m+1} O_{AB}^{\ttt{m\!+\!2}\dagger} (x) \right] , \\
\mr{h}_{\4AB}  &= \frac{1}{16(2\pi)^3}  \sum_{m=0}^\infty  \frac{m!}{u^{m+1}}  \CD_\ttt{2} \CD_\ttt{\frac{4}{3}}  \bigg[ (-i)^{m+1} O_{AB}^\ttt{m\!+\!2}(x)   +  i^{m+1} O_{AB}^{\ttt{m\!+\!2}\dagger}(x)   \bigg]  .
\end{split}
\end{equation}
It follows that
\begin{equation}
\begin{split}\label{mrh2-largeu}
\mr{h}_{\2AB} = \CO(u^{-3}) , \qquad \mr{h}_{\3AB} = \CO(u^{-2}) , \qquad \mr{h}_{\4AB} = \CO(u^{-1}) .
\end{split}
\end{equation}

With this, we have determined the large $u$ fall-offs for \emph{linearised} solutions in $\CS_\text{canonical}${\eqref{S0dot-coord}}. However, since the fields die off at large $u$, it immediately follows that \eqref{mrh2-largeu} hold in the non-linear theory as well (at least to all orders in perturbation theory). Finally, using the fact that solutions in $\CS_\text{analytic}$ {\eqref{S-analytic-data}} are obtained from those in $\CS_\text{canonical}$ via a WBMS diffeomorphism (which we will prove later in Section \ref{sec:finite-BMS}) and that under this map $\mr{h} \to {\wt h}$, we can conclude that \eqref{mrh2-largeu} generalises as-is to $\CS_\text{analytic}$ so we have
\begin{equation}
\begin{split}\label{h2-largeu}
{\wt h}_{\2AB} = \CO(u^{-3}) , \qquad {\wt h}_{\3AB} = \CO(u^{-2}) , \qquad {\wt h}_{\4AB} = \CO(u^{-1}) .
\end{split}
\end{equation}

\subsubsection{From WBMS Consistency}
\label{sec:BMSconst}

Another argument in support of \eqref{h2-largeu} comes from the reasonable demand that the solution space $\CS_\text{analytic}$ {\eqref{S-analytic-data}} is consistent with WBMS diffeomorphisms ({discussed in Section \ref{sec:BMS-transform}}). More precisely, we have to ensure that the boundary condition \eqref{vanishingRiemann} is not violated by the WBMS diffeomorphism. To run the argument, we need to consider the component 
\begin{equation}
\begin{split}
   r^4  R_{rArB} [G] |_{\ci^+ _-} &= 0 ,
      \end{split}
\end{equation}
of the condition \eqref{vanishingRiemann}. The reason for this is that the leading order terms of this component of the Riemann tensor are functions of ${\wt h} _{(2)}$, ${\wt h} _{(3)}$ and ${\wt h} _{(4)}$. Explicitly
\begin{equation}
\begin{split}
  R_{(2)rArB}[G] &= -{\wt h}_{(2)AB} , \\
  R_{(3)rArB}[G] &=   - 3 {\wt h}_{(3)AB}    +  \frac{1}{4} q _{AB} \tr [h_\1 {\wt h} _{(2)}]     - 2 ( h_\1  \Tilde{h} _{(2)}) _{AB} , \\
  R_{(4)rArB}[G] &= -6 {\wt h}_{(4)AB}-\frac{3}{2}(h_\1{\wt h}_{(3)})_{(AB)} - \frac{3}{8}q _{AB} \tr[h_\1{\wt h} _{(3)}] - 2({\wt h}_{(2)} ^2)_{AB}  \\
  &\quad + \frac{1}{4}q _{AB}\tr[{\wt h}_{(2)} ^2] -\frac{1}{4}{\wt h}_{(2)AB} \tr[h_\1 ^2 ] - \frac{1}{8}q _{AB}\tr[h_\1^2{\wt h}_{(2)}] +\frac{1}{8}h_{(1)AB} \tr[h_\1 {\wt h} _{(2)}]  \\
  &\quad - \frac{1}{4}\left( h_{(1)A} {\wt h} _{(2)} h_{(1)B} +  {\wt h} _{A(2)} h_\1 h_{(1)B}  + h_{(1)A} h_\1 {\wt h} _{B(2)} \right).
\end{split}
\end{equation}
With a slight anticipation with respect to Section \ref{sec:BMS-transform}, infinitesimal WBMS diffeomorphisms are generated by the vector fields
\begin{equation}
\begin{split}\
\xi &= T \p_u + \left( Y^A -  \int_r^\infty \dt r  e^{2\b}  G^{AB}  \p_B T \right) \p_A  \\
&\qquad \qquad \qquad \qquad \qquad + r \left( \o + \frac{1}{4} W^A \p_A T + \frac{1}{4}\int_r^\infty \dt r  D_A [ e^{2\b}  G^{AB}  \p_B T ] \right) \p_r  , 
\end{split}
\end{equation}
where crucially the function $T$ is linear in the coordinate $u$. The action of this vector field on the Riemann tensor is given by
\begin{equation}
\begin{split}
\d _{\xi} R_{(2)rArB}[G] &= \left( T \partial _u  + \CL _{Y^C}  \right) R_{(2)rArB}[G], \\
\d _{\xi} R_{(3)rArB}[G] &  = \left( T \partial _u  + \CL _{Y^C}  - \omega  \right) R_{(3)rArB}[G] +  \CL _{D^C T} R_{(2)rArB}[G]  - 2 D^C T R_{(1)CArB}[G] , \\
\d _{\xi } R_{(4)rArB}[G] & = \left( T \partial _u  + \CL _{Y^C}  - 4 \o \right) R_{(4)rArB}[G]  -\left(  \CL _{h_\1 \cdot \p T } + \frac{1}{2} D^2 T  \right) {\wt h}_{(2)AB}  \\
&+ \CL _{D^C T }R_{(3)rArB}[G] - 2 \p _C T \left( G ^{CD} R_{DArB} [G] \right)_{(4)} + 2 \partial _{(A} T R_{(2)rur|B)} [G].
\end{split}
\end{equation}
Notice that because of the linear $u$ dependence of both $T$ and $h_{(1) AB }$, we need the following to be true in order for the WBMS diffeomorphisms to preserve the boundary conditions {\eqref{vanishingRiemann}}
\begin{equation}
\begin{split}\label{h2-largeu-1}
u^2 {\wt h} _{(2)AB}|_{\ci ^+ _-}  = 0, \qquad u {\wt h} _{(3)AB} |_{\ci ^+ _-} = 0 , \qquad  {\wt h} _{(4)AB} |_{\ci ^+ _-} = 0 . 
\end{split}
\end{equation}
These fall-offs are consistent with \eqref{h2-largeu}. Note that the fall-offs derived here are \emph{weaker} than \eqref{h2-largeu}, but in fact all we will need in the subsequent calculations of this paper.

\subsection{Summary and Discussion}
\label{sec:summary1}

Before moving on we take a pause, summarise the results of this section, and contextualize them within a historical framework.

We started in Section \ref{sec:gensol} by noting that the most general asymptotic expansion admissible by Einstein's equations is polyhomogeneous\footnote{While we have closely followed the derivation in \cite{Capone:2021ouo}, the generic dynamical appearance of logarithmic structures at null infinity in higher dimensions has been later confirmed in five dimensions also by the Hamiltonian analysis of \cite{Fuentealba:2022yqt} (paralleling the four-dimensional case \cite{Henneaux:2018cst, Henneaux:2018hdj, Henneaux:2019yax}).} near $\ci^+$. In the context of the asymptotic symmetry analysis, some forms of polyhomogeneous solutions were first discussed in four-dimensional gravity in \cite{Barnich:2011mi,Laddha:2018myi,Laddha:2018vbn,Sahoo:2018lxl,Campiglia:2019wxe} and the log terms (in $u$ and $r$) are now understood to be related to infrared divergences which plague all four-dimensional theories with massless particles. Such infrared issues are absent in higher dimensions so in this case, we must find a new interpretation of the log terms. 

To do this, it is useful to take inspiration from AdS/CFT where log terms in the asymptotic expansion of the bulk solution are associated with anomalies in the dual CFT. For example, the asymptotic expansion in Fefferman-Graham gauge for generic asymptotically AdS solutions in $D=5$ has the form
\begin{equation}
\begin{split}\label{AdS-FG-exp}
\dt s^2 = \frac{\dt r^2}{4 r^2}  + r \left[ g_{\0AB}  + \frac{g_{\1AB}}{r}   + \frac{g_{\2AB} + g_{\ttt{2,1}AB} \ln r}{r^2} + \CO(r^{-3}) \right] \dt x^A \dt x^B  . 
\end{split}
\end{equation}
where $r$ is the radial Poincar\'e coordinate. The presence of the log term presents an obstruction to holographic renormalization and one is forced to introduce \emph{non-covariant} terms in the counterterm action \cite{Skenderis:2002wp}. This non-covariance then generates a Weyl anomaly in the dual CFT. A similar argument will likely apply to the log terms that appear in \eqref{metric-exp}. In the present case, there are three fundamental log terms: $h_{\ttt{2,1}AB}$, $U_\ttt{3,1}$ and $W_{\ttt{5,1}A}$ and these are likely related to the conformal anomaly, supertranslation anomaly and superrotation anomaly respectively. We will explore these interesting issues in future work.\footnote{Interestingly, note that the asymptotic expansion of $h_{AB}$ in \eqref{metric-exp} is strikingly similar to \eqref{AdS-FG-exp} and Einstein's equations actually imply that $h_\ttt{2,1}$ has exactly the same form as $g_\ttt{2,1}$. This strongly suggests that the anomaly structure of asymptotically flat gravity in $D=6$ is similar to that of Euclidean $\ads_5$. As first suggested in \cite{Capone:2021ouo}, the observation concerning the relationship between the leading log term in $h_{AB}$  and the leading log term in the expansion of odd-dimensional AlAdS spacetimes actually carries over to any even dimensional asymptotically flat spacetime, while the absence of such a term in odd-dimensional flat spacetimes also matches with the fact that it is absent in even-dimensional AlAdS. Given the result of \cite{deBoer:2003vf}, such an analogy is not completely surprising, but its deeper consequences remain to be investigated.}

We next constructed analytic solutions in Section \ref{sec:analyticsol} by imposing a ``no-flux condition'' \eqref{vanishingRiemann} at $\ci^+_-$, thereby showing that it is  physically related to the logarithmic terms. We can compare this with some well-known facts and theorems developed in four-dimensional gravity. In four-dimensional gravity, a sufficient condition for smooth initial hyperboloidal data\footnote{An initial hyperboloidal surface is by definition a hypersurface that extends smoothly to $\ci$, intersects it and is everywhere spacelike.} (which then develops to a non-polyhomogeneous $\ci$ according to \cite{Friedrich:1983}) is that the conformal Weyl tensor at the boundary of the hyperboloidal initial data surface vanishes \cite{Andersson:1992}. It is also known that non-trivial solutions of Einstein equations exist that satisfy the peeling property \cite{Chrusciel:2002vb} thanks to the Corvino-Schoen theorem \cite{Corvino:2003sp}. Similar results exist in the setting of the standard four-dimensional Cauchy problem \cite{10.1007/978-3-0348-7953-8_4}.  Less formally, it is known that early-time waves are a source for the polyhomogeneity of $\ci^+$ \cite{Damour:1985cm,2002nmgm.meet...44C} and that such asymptotic structure forms for example when one allows for spacetimes with infalling matter from $i^-$ \cite{Kehrberger:2021uvf, Kehrberger:2021vhp, Kehrberger:2021azo, Gajic:2022pst}. In such cases the ASG remains the BMS group \cite{Chrusciel:1993hx}. These results show that null infinity is generically polyhomogeneous in four dimensions, even when the initial data are smooth, but physically relevant conditions can be imposed to remove the polyhomogeneity. In higher dimensions, although our setting is not exactly the same as the aforementioned mathematical theorems (e.g., we only gave the falloff rate of the radiation field toward  $\ci^+_-$, rather than specifying a stationary portion of  $\ci$), it is likely that the existence of $\CS_\text{analytic}$ in higher dimensions can be proved in similar ways. We also note that our analytic solution space $\CS_\text{analytic}$ is actually compatible with (a higher dimensional generalisation of) the peeling theorem which appears in \cite{Hollands:2003xp,Tanabe:2011es,Godazgar:2012zq}.

Finally, in Section \ref{sec:mrS0-explicit-details}, we discussed the canonical solution space $\CS_\text{canonical}$. Historically, this is the solution space that was most widely studied in higher dimensional gravity essentially due to the fact that it is the simplest and most natural generalization of the linearised solutions which we reviewed in Section \ref{sec:lin-grav}. This consideration motivated previous works to argue in favour of setting $h_{(1)AB}=0$ on all of $\ci^+$ and it is this restriction which led to the conclusion that the asymptotic symmetry group for higher dimensional gravity is the finite-dimensional Poincar\'{e} group.
Another argument in favour of setting $h_\1=0$ is the following. Consider the solution space $\CS_\text{canonical}$ augmented just by the inclusion of $h_\1$.   Here, $h_{(1)AB}$ is time independent and pure gauge (compare with \eqref{sec:h1}). If the system undergoes a stationary-to-stationary transition, then the same supertranslation can be used to set $h_{(1)AB}=0$ on all $\ci$ \cite{Hollands:2016oma}. This argument assumes that the supertranslation involved in the process is a small gauge transformation. However, a  straightforward evaluation of the supertranslation charge in such a phase space gives a divergent result. Without a proper renormalization of the charge, there is no reason to draw the above conclusion. Indeed, the phase space renormalization procedure we develop in this paper suggests that in $\CS_\text{analytic}$ the gauge fixing of $h_\1$ corresponds to a large gauge choice.

To summarise, we  have spelt out the conditions and constraints near spacelike infinity that the gravitational field has to satisfy in order to evolve to the form given in \eqref{app:metric-exp-S0}. We thus have a constructive definition of the solution space at null infinity which serves several purposes: $(i)$ clarify and remove unnecessary assumptions from previous literature, $(ii)$ give a detailed understanding of the action of asymptotic symmetries on this solution space, $(iii)$ provide the basis for analysing what conditions at spacelike infinity may give rise to the given structure at null infinity, $(iv)$ assess holographic structures, such as anomalies in the dual theory as suggested in \cite{Capone:2021ouo} and $(v)$ give the building blocks for further consistent extensions of the phase space.

\section{Weyl-BMS Diffeomorphisms}
\label{sec:BMS-transform}

In this section, we determine the asymptotic symmetry group for the space of solutions described in the previous section. This is the set of all diffeomorphisms that preserve the gauge conditions \eqref{Bondi-metric}, \eqref{Bondi-detcond} and boundary conditions \eqref{afs-falloff} and which act non-trivially on the data of the theory.

\subsection{The Weyl-BMS Algebra}
\label{sec:GMBSalgebra}

Infinitesimal diffeomorphisms are generated by vector field $\xi^\mu$ and act on the metric via the Lie derivative
\begin{equation}
\begin{split}
G_{\mu\nu} \to  G_{\mu\nu} + \CL_{\xi}  G_{\mu\nu} + \CO(\xi^2) . 
\end{split}
\end{equation}
To preserve the gauge conditions, we must have
\begin{equation}
\begin{split}
\label{preserve-bondi-gauge} \CL_\xi  G_{rr} = 0 , \qquad \CL_\xi  G_{rA} = 0 , \qquad \p_r (  G^{AB} \CL_\xi  G_{AB} ) = 0 , \\
\end{split}
\end{equation}
whereas in order to preserve the boundary conditions, we must have
\begin{equation}
\begin{split}
\label{preserve-asymp-cond} \CL_\xi  G_{uu} &= \CO(1) , \qquad \CL_\xi  G_{ur} = \CO(r^{-2}) , \qquad \CL_\xi  G_{uA} = \CO(1) , \qquad \CL_\xi  G_{AB} = \CO(r^2) . 
\end{split}
\end{equation}
To solution to these equations is relatively easy to find and in general are parameterized by functions $f(x)$, $\o(x)$ and $Y^A(x)$,
\begin{equation}
\begin{split}\label{WBMSvector}
\xi &= T \p_u + \left( Y^A -  \int_r^\infty \dt r  e^{2\b}  G^{AB}  \p_B T \right) \p_A  \\
&\qquad \qquad \qquad \qquad \qquad + r \left( \o + \frac{1}{4} W^A \p_A T + \frac{1}{4}\int_r^\infty \dt r  D_A [ e^{2\b}  G^{AB}  \p_B T ] \right) \p_r  , 
\end{split}
\end{equation}
where
\begin{equation}
\begin{split}\label{Tdef}
T \equiv f(x) - u \o(x) , \qquad Y^A \equiv Y^A(x) , \qquad \o \equiv \o(x) . 
\end{split}
\end{equation}
The vector field \eqref{WBMSvector} satisfies the algebra
\begin{equation}
\begin{split}\label{WBMSalgebra}
[ \xi_1 , \xi_2 ]_\star \equiv [ \xi_1 , \xi_2 ]  - \d_{\xi_1} \xi_2 + \d_{\xi_2} \xi_1 = \xi_{12} ,  
\end{split}
\end{equation}
where
\begin{equation}
\begin{split}
\label{xi12}
f_{12} &=  Y_{[1}( f_{2]} ) + \o_{[1} f_{2]} -  \d_{\xi_{[1}} f_{2]}  , \qquad Y_{12} = [ Y_1 , Y_2 ] - \d_{\xi_{[1}} Y_{2]}  , \qquad \o_{12} = Y_{[1} ( \o_{2]} ) - \d_{\xi_{[1}} \o_{2]}  .
\end{split}
\end{equation}
$[\,,\,]_\star$ is a modified Lie bracket which keeps track of the fact that the BMS vector field depends on the metric. Equation \eqref{WBMSalgebra} is an infinite-dimensional algebra known as the Weyl-Bondi-Metzner-Sachs (WBMS) Algebra. The action of the WBMS vector field \eqref{WBMSvector} on the metric functions is given by
\begin{equation}
\begin{split}\label{WBMSaction}
\d_\xi h_{AB} &= ( T \p_u  + \CL_\xi + r^{-2} \xi^r \p_r r^2 )  h_{AB}  - 2 W^C h_{C(A} \p_{B)} T , \\
\d_\xi W^A &= ( T \p_u + \CL_\xi  + \xi^r \p_r - \o - W^B \p_B T ) W^A  - \p_u \xi^A + r^{-2} e^{2\b} {\bar h}^{AB} (  U \p_B T  + \p_B \xi^r ) , \\
\d_\xi e^{2\b} &= ( T \p_u + \CL_\xi  + \xi^r \p_r  - \o + \p_r \xi^r - W^A \p_A T)  e^{2\b}, \\
\d_\xi U &= ( T \p_u + \CL_\xi  + \xi^r \p_r  - \o  - \p_r \xi^r  -  W^A   \p_A T  ) U  + 2 \p_u \xi^r + 2 W^A \p_A \xi^r  .
\end{split}
\end{equation}
In the above, $\CL_\xi$ is the 4D Lie derivative w.r.t. the vector field $\xi^A$. Expanding this in large $r$, we can determine the action of the WBMS vector field on the data of the theory. For instance, on the leading large $r$ coefficients of $h_{AB}$, we find 
\begin{equation}
\begin{split}\label{S-WBMS}
\d_\xi q_{AB} &= ( \CL_Y + 2 \o ) q_{AB} , \\
\d_\xi h_{\1AB} &= ( T \p_u + \CL_Y + \o ) h_{\1AB}  - 2 D_{\{A} D_{B\}} T , \\
\d_\xi {\wt h}_{\2AB} &= ( T \p_u + \CL_Y ) {\wt h}_{\2AB} +  \o  h_{\ttt{2,1} AB}  - 2 \o_{C(AB)} D^C T  . 
\end{split}
\end{equation}
Note that $Y^A$ generates an infinitesimal diffeomorphism on the metric $q_{AB}$ and $\o$ generates an infinitesimal Weyl transformation.

On the analytic solution space $\CS_\text{analytic}$ {\eqref{S-analytic-data}}, these diffeomorphisms act as
\begin{equation}
\begin{split}\label{BMStransform-S0}
\d_\xi {\wt h}_\2 &= ( T \p_u +  \CL_Y ) {\wt h}_\2  , \\
\d_\xi C &= \CL_Y C  + e^{-\Phi} f  , \\
\d_\xi M &= ( \CL_Y - 5 \o ) M , \\
\d_\xi \chi^A &= Y^B \p_B \chi^A , \\
\d_\xi \CN_A &=  ( \CL_Y - 4 \o  ) \CN_A + f \p_A  M + 5 M \p_A f , \\
\d_\xi \Phi &= \CL_Y \Phi + \o . \\
\end{split}
\end{equation}

\subsubsection{Important Subalgebras}
\label{sec:important-subalgebras}

The WBMS algebra has three important subalgebras:
\paragraph{(1) Poincar\'e Algebra:} This is generated by vector fields which preserve $q_{AB}$ and which act covariantly on the $u$-independent part of $h_\1$ {\eqref{sec:h1}}. These satisfy
\begin{equation}
\begin{split}\label{Poincare-restriction}
\o = - \frac{1}{4} D_A Y^A , \qquad D_{\{A} Y_{B\}} = 0  , \qquad ( 2 D_{\{A} D_{B\}} + R_{\{AB\}}  ) f = 0. 
\end{split}
\end{equation}
This generates a finite-dimensional subalgebra which is a generalisation of the Poincar\'e algebra. To reproduce the Poincar\'e algebra specifically, we consider the case where $q_{AB}$ is a conformally flat metric. Then, moving to coordinates where the metric has the form $q_{AB} = e^{2\Phi} \d_{AB}$, we find the solution
\begin{equation}
\begin{split}\label{Poincare-transforms}
Y^A(x) &= \xi^A + \o^{AB} x_B + \l x^A + \zeta^B ( x^2 \d^A_B - 2 x^A x_B ) , \\
f(x) &= e^{\Phi(x)} ( \a + \b_A x^A + \c x^2 ) .
\end{split}
\end{equation}
$f(x)$ generates the translation subgroup $T_6$. $\a$ generates time translations and $(\b_A,\g)$ generates the spatial translations. Likewise, $Y^A(x)$ generates the Lorentz subgroup $SO(1,5)$ and \eqref{Poincare-transforms} makes manifest the isomorphism between the Lorentz group and the conformal group on $\mrr^4$. $\xi^A$ and $\o^{AB}$ together generate $SO(5)$ rotations in the bulk six-dimensional spacetime whereas $\l$ and $\zeta^A$ together generate the Lorentz boosts. On the celestial $S^4$, these parameters generate translations, $SO(4)$ rotations, dilatations and special conformal transformations respectively. The algebra of these generators is precisely the Poincar\'e algebra, $T_6 \rtimes SO(1,5)$.

In older literature  dealing with higher dimensions \cite{Hollands:2016oma}, the Poincar\`e algebra was not obtained as a subalgebra of a larger algebra, but rather as the full algebra of asymptotic symmetries. This arose from constraining oneself to the canonical solution space $\CS_\text{canonical}$ {\eqref{S0dot-coord}} where $h_\1 = 0$. This then automatically forces the equations \eqref{Poincare-restriction} implying that the full asymptotic symmetry group is simply the Poincar\'e group. 

\paragraph{(2) BMS Algebra:} This is generated by vector fields which preserve $q_{AB}$. They satisfy the first two equations in \eqref{Poincare-restriction} but not the last. Consequently, for this subalgebra, $Y^A$ is still a conformal Killing vector w.r.t. $q_{AB}$ (which generates a finite-dimensional group) but $f$ is an arbitrary function. The diffeomorphisms generated by $f$ are known as \emph{BMS supertranslations (ST)}. The corresponding algebra is an infinite-dimensional generalisation of the Poincar\'e algebra and has the structure $\textit{ST}_6 \rtimes SO(1,5)$.

Since the celestial manifold is of dimension greater than two, all conformal Killing vectors are globally defined and hence there is no immediate generalisation of the algebraic notion of Virasoro-superrotations. A way to generalise this notion to higher dimensions has been explored in \cite{Capone:2019aiy}, by mapping codimension-$2$ cosmic branes to Bondi-Sachs gauge. In our six-dimensional setting, the explicit map can be easily read from Appendix \ref{app:finite-BMS-transforms} by specifying the cosmic brane tension parameter (which is the only parameter that differentiates this spacetime from pure Minkowski). These are spacetimes where the regularity of the celestial sphere is broken and a conical deficit is introduced on two-dimensional sections of the celestial manifold. They are intended to generalise to higher dimensions the relationship between cosmic strings and Virasoro-superrotations  \cite{Bicak:1989rb,Strominger:2016wns,Compere:2016jwb}. However, in four dimensions this relationship is supported by Penrose cut-and-paste constructions of impulsive waves. The Virasoro-superrotation is the map relating the spacetime on the two sides of the impulse. For example, the dynamical process creating the shock is the breaking of a cosmic string.  In higher dimensions such constructions are still missing.

\paragraph{(3) Generalised BMS Algebra:} This is generated by vector fields which preserve the volume form on $q_{AB}$, i.e. $\d_\xi \sqrt{q} = 0$. This satisfies only the first equation of \eqref{Poincare-restriction} which fixes $\o$ in terms of $Y^A$ but there is no additional restriction on $Y^A$ (so it is an arbitrary vector field). The corresponding diffeomorphisms generated by $Y^A(x)$ are (also) known as \emph{superrotations (SR)}. This algebra has the structure $\textit{ST}_6 \rtimes  \textit{SR}_4$.

\subsection{Finite WBMS Diffeomorphisms from \texorpdfstring{$\CS_\text{canonical} \to \CS_\text{analytic}$}{S0toS0dot}}
\label{sec:finite-BMS}

Due to the particularly simple nature of \eqref{BMStransform-S0}, we can determine the structure of the finite WBMS transformations:
\begin{equation}
\begin{split}\label{finite-BMS-fields}
\chi^A(x) \to \chi^A(\CX(x)) , \qquad C(x) \to C(\CX(x)) + e^{-\Phi(x)} \CF(x) , \qquad \Phi(x) \to \Phi(\CX(x)) + \CW(x).
\end{split}
\end{equation}
The infinitesimal transformations are recovered by setting $\CX^A(x) = x^A + Y^A$, $\CF(x) = f(x)$ and $\CW(x) = \o(x)$ and then expanding to linear order in $Y$, $f$ and $\o$.

An important corollary of \eqref{finite-BMS-fields} is that it is always possible to perform a WBMS transformation to map a solution in $\CS_\text{canonical}$ {\eqref{S0dot-coord}} (which has $C=\Phi=0$, $\chi^A=x^A$) to a generic solution in $\CS_\text{analytic}$ by choosing $\CX = \chi$, $\CF=C$ and $\CW=\Phi$. These WBMS diffeomorphisms can be extended to the bulk so that for any metric $G \in \CS_\text{analytic}$ {\eqref{S-analytic-data}}, there exists a \emph{unique} metric $\mr G \in \CS_\text{canonical}$ and a diffeomorphism $X^\mu \to \mr{X}^\mu(X)$ such that
\begin{equation}
\begin{split}\label{gen-sol}
G_{\mu\nu}(X) = \p_\mu \mr{X}^\rho (X) \p_\nu \mr{X}^\s(X) \mr{G}_{\rho\s}(\mr{X}(X)).
\end{split}
\end{equation}
We performed a detailed analysis of these equations in Appendix \ref{app:finite-BMS-transforms}. In this section, we only present a couple of results that will be used in the next section.

With this goal in mind, firstly we clarify the notation that we are forced to employ in this section (and also in Section \ref{sec:symppot-rad} and Appendix \ref{app:finite-BMS-transforms}). The confusing aspect of these sections is that we are simultaneously discussing solutions in $\CS_\text{analytic}$ and in $\CS_\text{canonical}$ for which the conformal metric on $\ci^+$ is  $q_{AB}=e^{2\Phi}{\hat q}_{AB}$  and $\d_{AB}$ respectively (with $\hat q$ as defined in \eqref{eq:q-form}). With three different concurrent metrics -- namely $q_{AB}$, ${\hat q}_{AB}$ and $\d_{AB}$ -- we have to be extra careful!

We start by denoting any tensor in the solution space $\CS_\text{canonical}$ with a superscript $\circ$, e.g. ${\mr T}_{A_1\cdots A_n}(u,x)$. The indices of such tensors are raised w.r.t. $\mr{q}_{AB} = \d_{AB}$. Hatted tensors are then defined as
\begin{equation}
\begin{split}\label{That-def}
{\hat T}_{A_1 \cdots A_n} (u,x) \equiv \p_{A_1} \chi^{B_1}(x) \cdots \p_{A_n} \chi^{B_n}(x) {\mr T}_{B_1 \cdots B_n} ( \k ( u , x ) , \chi(x) ) ,   
\end{split}
\end{equation}
where
\begin{equation}
\begin{split}
\k(u,x) \equiv e^{-\Phi(x)} u + C(x) . 
\end{split}
\end{equation}
This is essentially the tensor obtained by performing a diffeomorphism $(u,x) \mapsto (\k(u,x),\chi(x))$. The hatted tensors indices are raised and lowered w.r.t. the flat metric ${\hat q}_{AB}$ {\eqref{eq:q-form}}. Note that by their definition, the hatted tensors have the same large $u$ behaviour as the ringed tensors.

Finally, the unhatted tensors are related to the hatted ones by an extra Weyl factor
\begin{equation}
\begin{split}
T_{A_1 \cdots A_n} (u,x) = e^{\D_T \Phi(x)} {\hat T}_{A_1 \cdots A_n} (u,x) ,
\end{split}
\end{equation}
where $\D_T$ is the scaling dimension of the tensor $T$.\footnote{The BMS group contains the Lorentz group $SO(1,5)$ as a subgroup. This is isomorphic to the conformal group on $S^4$. The scaling dimension is the eigenvalue under dilatations of $\text{Conf}(S^4)$.} These unhatted tensor indices are raised and lowered w.r.t. $q_{AB}$. 

With these definitions, it is easy to derive the following properties
\begin{equation}
\begin{split}
\SD_u {\hat T}_{A_1 \cdots A_n} (u,x) \equiv \p_{A_1} \chi^{B_1}(x) \cdots \p_{A_n} \chi^{B_n}(x) \p_\k {\mr T}_{B_1 \cdots B_n} ( \k(u,x) , \chi(x) ) , \\
\SD_C {\hat T}_{A_1 \cdots A_n} (u,x) \equiv \p_{A_1} \chi^{B_1}(x) \cdots \p_{A_n} \chi^{B_n}(x)  \p_{\chi^C} {\mr T}_{B_1 \cdots B_n}( \k(u,x) , \chi(x) ) ,
\end{split}
\end{equation}
where we have defined the covariant derivatives by
\begin{equation}
\begin{split}
\SD_u \equiv e^\Phi \p_u , \qquad \SD_A \equiv {\hat D}_A - e^\Phi \p_A \k \p_u .
\end{split}
\end{equation}
Here, ${\hat D}_A$ is the covariant derivative w.r.t. ${\hat q}_{AB}$ \eqref{eq:q-form}. Note that the covariant derivatives all commute
\begin{equation}
\begin{split}
 [ \SD_A , \SD_B ] =  [ \SD_A , \SD_u ] = [ \SD_u , \SD_u ] = 0 ,
\end{split}
\end{equation}
and satisfy
\begin{equation}
\begin{split}\label{totalDprop}
e^{-\Phi} \SD_u ( \cdot )  &= \p_u ( \cdot ) , \qquad e^{-\Phi} \SD_A ( \cdot )  = - \p_u [  \p_A \k ( \cdot )  ]   + {\hat D}_A [ e^{-\Phi} ( \cdot ) ] .
\end{split}
\end{equation}
With this notation, we can easily transform \emph{any} equation for the ringed fields to an equation for the hatted fields with the following simple replacements
\begin{equation}
\begin{split}\label{dot-to-hat}
\mr{T} \to \hat{T}  ,  \qquad \d_{AB} \to {\hat q}_{AB} , \qquad \p_u \to \SD_u , \qquad \p_A \to \SD_A  . 
\end{split}
\end{equation}
For instance, the constraint $\p_A \mr{W}_\4^A  = 0$ becomes
\begin{equation}
\begin{split}\label{W4hatconstraint}
\SD_A {\hat W}_\4^A = 0 . 
\end{split}
\end{equation}
We now present our results. The finite WBMS transformations $\mr{X}(X)$ satisfying \eqref{gen-sol} admits a Taylor expansion in $r^{-1}$ of the form
\begin{equation}
\begin{split}\label{finite-BMS-transform}
\mr{u} &= \k + e^{-\Phi} \frac{\mr{u}_\1}{r} + e^{-2\Phi} \frac{\mr{u}_\2}{r^2} + e^{-3\Phi}  \frac{\mr{u}_\3}{r^3} + e^{-4\Phi} \frac{ \mr{u}_\4}{r^4} + \CO(r^{-5}) , \\
\mr{r} &=  r e^{\Phi} + \mr{r}_\0 + e^{-\Phi} \frac{ \mr{r}_\1}{r} + e^{-2\Phi} \frac{ \mr{r}_\2}{r^2} + e^{-3\Phi} \frac{\mr{r}_\3}{r^3} + \CO(r^{-4}) , \\
\mr{x}^A &= \chi^A + e^{-\Phi} \frac{ \mr{x}^A_\1}{r} + e^{-2\Phi} \frac{ \mr{x}^A_\2}{r^2} + e^{-3\Phi}  \frac{\mr{x}^A_\3}{r^3} + e^{-4\Phi} \frac{\mr{x}^A_\4}{r^4} + \CO(r^{-5}) . \\
\end{split}
\end{equation}
In order to find the relationship between the un-hatted tensors which appear in the asymptotic expansion of the metric, we need to determine all the coefficients shown above. This is done in Appendix \ref{app:finite-BMS-transforms}. For the purposes of section \ref{sec:symppot-rad}, we only need the very leading order results and ${\hat x}_\1^A = - {\hat q}^{AB} \p_B  \k$ and the fact
\begin{equation}
\begin{split}\label{h2tU3trel}
{\hat h}_{\2AB}(u,x) = {\wt h}_{\2AB}(u,x)  , \qquad {\hat U}_\3(u,x) = e^{5\Phi(x)} {\wt U}_\3(u,x)  + \text{terms that vanish on $\ci^+_-$.}
\end{split}
\end{equation}
It then follows from \eqref{sec:MNAdef} that
\begin{equation}
\begin{split}\label{h2tU3trel-ipm}
{\hat U}_\3 (u,x) ~ \xrightarrow{u\to-\infty} ~ - e^{5\Phi(x)} M(x) + \CO(u^{-2}) . 
\end{split}
\end{equation}

\section{Gravitational Symplectic Potential}
\label{sec:symppot}

In this section, we will use the covariant phase space formalism \cite{Crnkovic:1986ex,Iyer:1994ys,Iyer:1995kg,Wald:1999wa,Harlow:2019yfa,Shi:2020csw,He:2020ifr} to elevate the solution space $\mss_0$ to a phase space $\G_0$. The dynamics of the theory is described by the Einstein-Hilbert Lagrangian 
\begin{equation}
\begin{split}\label{action}
L = \frac{1}{16\pi G} \e R[G] + \dt L_{\p} . 
\end{split}
\end{equation}
The boundary term is typically required to make the variational principle well-defined, e.g. the Gibbons-Hawking-York boundary term. Varying the action, we find
\begin{equation}
\begin{split}\label{action-var}
\d L &= \frac{1}{16\pi G} \e   \left( R_{\mu\nu} [G] - \frac{1}{2} G_{\mu\nu} R[G] \right) \d G^{\mu\nu} + \dt \left( \frac{1}{16\pi G} \t(\d) + \d L_{\p} \right) ,
\end{split}
\end{equation}
where
\begin{equation}
\begin{split}\label{GR-theta-canonical}
[ \star \t (\d) ]^\mu = G^{\mu\nu}  \d \G^\rho_{\nu\rho}[G]  - G^{\nu\rho} \d \G^\mu_{\nu\rho}[G].
\end{split}
\end{equation}
The first term in \eqref{action-var} gives us the equations of motion \eqref{Einstein-EoM}. The (pre-)symplectic potential on a Cauchy slice $\S$ is obtained by integrating the boundary term in \eqref{action-var} over $\S$,
\begin{equation}
\begin{split}\label{symp-pot}
\T_\S ( \d ) =  \frac{1}{16\pi G} \int_\S \t (\d) + \d \int_\S L_{\p} + \frac{1}{16\pi G}  \int_{\p\S}  Y(\d) .
\end{split}
\end{equation}
We see that there are two types of ambiguities in the pre-symplectic potential.

The first ambiguity is the $\d$-exact term which is related to the boundary term in the action \eqref{action}. Part of $L_{\p}$ is fixed by the variational principle (e.g. Gibbons-Hawking-York boundary term). The rest of $L_{\p}$ (typically referred to as the \emph{counterterm action}) is fixed by requiring that the on-shell action be finite. In asymptotically AdS spacetimes, there is a systematic procedure to fix $L_{\p}$ known as holographic renormalization. However, such an algorithm is lacking for flat spacetimes and we will \emph{not} address this issue in this paper either. Here, we are interested in constructing the classical phase space of general relativity whose local structure depends crucially on the symplectic form $\O$, not the symplectic potential. Any gauge-invariant $\d$-exact term in $\T$ never contributes to $\O$. Consequently, we will simply drop any such term that shows up in our calculation of the symplectic potential, including $L_{\p}$. 

The second ambiguity is denoted $Y$ in \eqref{symp-pot}. Note that this is a boundary term and it arises from the fact that the variation of the action depends directly only on $\dt \t$. Consequently, a shift of the type $\t \to \t + \dt Y$ does not change \eqref{action-var}. This ambiguity will play a crucial role in our renormalization procedure.

To understand this in more detail, let us use \eqref{gen-sol} to simplify $\t(\d)$. We first note that given \eqref{gen-sol}, the variation of the metric has the form
\begin{equation}
\begin{split}\label{metric-var}
\d G_{\mu\nu} = \p_\mu \mr{ X}^\rho \p_\nu \mr{ X}^\s \d \mr{G}_{\rho\s}(\mr{ X}) + \CL_{\ds\mr{ X}} G_{\mu\nu} , \qquad \d \mr{X}^\mu = \ds\mr{ X}^\nu \p_\nu \mr{ X}^\mu .
\end{split}
\end{equation}
Note that since both the LHS and the first term above satisfy the Bondi-Sachs gauge conditions ({\eqref{Bondi-metric} and \eqref{Bondi-detcond}}) and our boundary conditions for asymptotic flatness {\eqref{vanishingRiemann}}, the second term must satisfy the same conditions. However, these are precisely the conditions \eqref{preserve-bondi-gauge}, \eqref{preserve-asymp-cond} that define a WBMS vector field. Consequently, $\ds\mr{ X}$ has the form \eqref{WBMSvector} and is parameterized by functions $f$, $\o$ and $Y^A$. They can be determined from the leading order behaviour of the finite WBMS transformations
\begin{equation}
\begin{split}\label{WBMSvarparameters}
e^{-\Phi}f = ( \d - \CL_{\ds\chi} ) C   , \qquad \o = ( \d - \CL_{\ds\chi} ) \Phi  , \qquad Y^A = \ds \chi^A \equiv  \d \chi^B \p_{\chi^B} x^A, 
\end{split}
\end{equation}
Using \eqref{metric-var}, we can rewrite the $\t(\d)$ in {\eqref{GR-theta-canonical}} as
\begin{equation}
\begin{split}\label{theta-split}
\t (\d) = \t_\text{can} (\d) + \dt Q_{\ds \mr{ X}} , 
\end{split}
\end{equation}
where
\begin{equation}
\begin{split}\label{theta-rad-def}
[ \star \t_\text{can} (\d) ]^\mu = G^{\mu\nu}  \p_\nu \mr{ X}^\rho \left( [ \star \mr{\t} (\d) ]_\rho |_{  X \to \mr{ X} } \right) , \qquad [ \star Q_\xi ]^{\mu\nu} = 2 \n^{[\mu} \xi^{\nu]}  . 
\end{split}
\end{equation}
Here, $\mr{\t}$ is evaluated in the canonical solution space $\CS_\text{canonical}$ detailed in Section \ref{app:mrS0-explicit-details}. Since this solution space only allows for gravitational radiation, $\t_\text{can}$ represents the radiative contribution. Using \eqref{theta-split}, we can rewrite \eqref{symp-pot} as
\begin{equation}
\begin{split}\label{symp-pot-1}
\T^\text{pre}_\S ( \d ) =  \frac{1}{16\pi G} \int_\S \t_\text{can} (\d) +  \frac{1}{16\pi G} \int_{\p\S}  [  Q_{\ds \mr{ X}} + Y(\d)  ]  .
\end{split}
\end{equation}
As we will see, the first term is completely finite on $\ci^+$. All the divergences appear in the second term above and these will be cancelled by a judicious choice of $Y(\d)$.

\subsection{Bulk Contribution (\texorpdfstring{$\ci^+$}{ip})}
\label{sec:symppot-rad}

We wish to evaluate the symplectic potential on the asymptotic null boundary $\ci^+$ whereas the formula \eqref{symp-pot-1} can only be applied to \emph{finite} null boundaries. To fix this, we define a timelike regulated boundary $\ci^+_r$ by $r = $ constant. We then define the symplectic potential on $\ci^+$ by 
\begin{equation}
\begin{split}
\T_{\ci^+}(\d) \equiv \lim_{r \to \infty} \T_{\ci_r^+}(\d) .
\end{split}
\end{equation}
On $\ci^+_r$, the first term in \eqref{symp-pot-1} evaluates to
\begin{equation}
\begin{split}\label{thetaradsimp}
\frac{1}{16\pi G} \int_{\ci_r^+} \t_\text{can} (\d) &= - \frac{1}{16\pi G} \int_{\ci_r^+} \dt u \dt^4 x \sqrt{q} r^4 e^{2\b}  [ \star \t_\text{can} (\d)  ]^r  \\
&= \frac{1}{16\pi G} \int \dt u \dt^4 x \sqrt{q} r^4 \left[ \p_u \mr{ X}^\rho  - U  \p_r \mr{ X}^\rho  +  W^A \p_A \mr{ X}^\rho  \right]  \left( [ \star \mr{\t} (\d) ]_\rho |_{  X \to \mr{ X} } \right)  ,
\end{split}
\end{equation}
where we have used \eqref{theta-rad-def} to simplify. Using the large $r$ fall-offs in Section \ref{sec:mrS0-explicit-details}, we can compute $\mr\t(\d)$ explicitly,
\begin{equation}
\begin{split}
[ \star {\mr\t} (\d) ]_u &= - \frac{1}{r^4} \left[ \d \mr{U}_\3 + 6  \p_u \d \mr{\b}_\4 + \frac{1}{2} \d^{AC} \d^{BD}  \mr{h}_{\2AB} \p_u \d \mr{h}_{\2CD} \right] + \CO(r^{-5}) , \\
[ \star {\mr\t} (\d) ]_r &=  \CO(r^{-6}) , \\
[ \star {\mr\t} (\d) ]_A &=  \frac{1}{r^3} \left[ - \frac{2}{3} \d_{AB} \d \mr{W}_\4^B \right] + \CO(r^{-4})  , \\
\end{split}
\end{equation}
now we have to replace $ X \to \mr X$ everywhere and rewrite everything in terms of the hatted fields defined in \eqref{That-def}. This can be done quite simply using the replacement rules \eqref{dot-to-hat} though in addition to those, we also need a replacement rule for the variation $\d$. This is easily done using the definition \eqref{That-def} from which we find
\begin{equation}
\begin{split}\label{That-def-var}
 \p_{A_1} \chi^{B_1}(x) \cdots \p_{A_n} \chi^{B_n}(x) \d  {\mr T}_{B_1 \cdots B_n} ( \k ( u , x ) , \chi(x) )  = \ds {\hat T}_{A_1 \cdots A_n} (u,x) ,  
\end{split}
\end{equation}
where
\begin{equation}
\begin{split}\label{ds-def}
\ds \equiv  \d - \CL_{\ds \chi} - e^\Phi ( \d - \CL_{\ds \chi} ) C  \p_u   + ( \d - \CL_{\ds \chi} )  \Phi u  \p_u, \qquad \ds \chi^A \equiv  \d \chi^B \p_{\chi^B} x^A .
\end{split}
\end{equation}
It follows from \eqref{That-def-var} that the replacement rule for $\d$ is simply $\d \to \ds$. As before, $\ds$ commutes with the derivatives $\SD_u$ and $\SD_A$ and with the flat metric ${\hat q}_{AB}$ {\eqref{eq:q-form}} so the order of the replacement is irrelevant. We also note that it satisfies the property
\begin{equation}
\begin{split}\label{ds-prop}
\sqrt{{\hat q}} e^{-\Phi} \ds {\hat f} = \d ( \sqrt{{\hat q}}  e^{-\Phi}   {\hat f} ) - \p_A (  \sqrt{{\hat q}} \ds\chi^A  e^{-\Phi} {\hat f} ) - \p_u [  \sqrt{{\hat q}} ( \d \k - \CL_{\ds \chi } \k )   {\hat f} ] . 
\end{split}
\end{equation}
Putting all of this together, we find
\begin{equation}
\begin{split}
\left( [ \star {\mr\t} (\d) ]_u |_{ X \to \mrx} \right) &= - \frac{1}{r^4} e^{-4\Phi} \left[ \ds \hat{U}_\3 + 6  \SD_u \ds \hat{\b}_\4 + \frac{1}{2} \hat{h}_{\2AB} \SD_u \ds \hat{h}_\2^{AB} \right] + \CO(r^{-5}) , \\
\left( [ \star {\mr\t} (\d) ]_r  |_{ X \to \mrx} \right) &=  \CO(r^{-6}) , \\
\left( [ \star {\mr\t} (\d) ]_A  |_{ X \to \mrx} \right) &=  \frac{1}{r^3} e^{-3\Phi} \left[ - \frac{2}{3} \d_{AB} \p_C \chi^B \ds \hat{W}_\4^C \right] + \CO(r^{-4})  , \\
\end{split}
\end{equation}
It follows that 
\begin{equation}
\begin{split} \label{eq: int_theta_can} 
\int_{\ci^+} \t_\text{can} (\d) &= \lim_{r\to\infty} \int_{\ci_r^+} \t_\text{can} (\d)  \\
&= - \int \dt u \dt^4 x  \sqrt{{\hat q}}  e^{-\Phi} \bigg[  \frac{1}{2} \tilde{h}_{\2AB} \SD_u \ds \tilde{h}_\2^{AB}   + \ds \hat{U}_\3 + 6  \SD_u \ds \hat{\b}_\4 + \frac{2}{3}  \ds \hat{W}_\4^A  \SD_A \Phi  \bigg] 
\end{split}
\end{equation}
where we have used $\tilde{h}_\2=\hat{h}_\2$ as shown in \eqref{h2tU3trel}. Let us process each of these terms one-by-one. To do this, we will decompose the variation $\ds$ into two pieces as
\begin{equation}
\begin{split}
\ds = {\bar \d}  + \frac{1}{4}  \d \ln \sqrt{q} u  \p_u , \qquad {\bar \d}   \equiv \left(  \d - \CL_{\ds \chi}   -  \frac{1}{4} D_A \ds \chi^A u  \p_u \right)  - e^\Phi \ds C  \p_u   . 
\end{split}
\end{equation}
The reason for this decomposition will become apparent shortly. Using this, the first term {in \eqref{eq: int_theta_can}} becomes 
\begin{equation}
\begin{split}
 \int \dt u \dt^4 x  \sqrt{{\hat q}}  e^{-\Phi}  \frac{1}{2} \wt{h}_{\2AB} \SD_u \ds \wt{h}_\2^{AB}   &= - \frac{1}{2} \int \dt u \dt^4 x \sqrt{q}  \p_u \wt{h}_{\2AB} {\bar \d}  \wt{h}_\2^{AB} \\
&\qquad \qquad \qquad \qquad - \frac{1}{8} \int \dt^4 x \d \sqrt{q} \int \dt u u \tr [ ( \p_u {\wt h}_\2 )^2 ]  . 
\end{split}
\end{equation}
Next, using \eqref{ds-prop}, we can simplify the second term {in \eqref{eq: int_theta_can}} as
\begin{equation}
\begin{split}
\int \dt u \dt^4 x  \sqrt{{\hat q}}  e^{-\Phi} \ds \hat{U}_\3 &= \d \left( \int \dt u \dt^4 x \sqrt{q}  e^{-5\Phi}   {\hat U}_\3 \right)   - \int  \dt^4 x \sqrt{q} e^\Phi  \ds C M  \\
&\qquad \qquad \qquad \qquad \qquad + \frac{u_0}{4}  \int \dt^4 x \sqrt{q}  \ds \chi^A D_A M + \frac{u_0}{4} \int \dt^4 x \d \sqrt{q}  M .
\end{split}
\end{equation}
where we have used \eqref{h2tU3trel-ipm}. The first term is $\d$-exact and will not contribute to the symplectic form so we shall henceforth drop it. The last two terms are linearly divergent at large $u$. To deal with this, we have regulated the boundary $\ci^+_-$ by setting it to $u=u_0$ instead of at $u=-\infty$. We are eventually interested in taking $u_0 \to -\infty$. Such large $u$ divergences will be discussed in Section \ref{sec:large-u-divergences-symp-pot}.

The third term in \eqref{eq: int_theta_can}  is a total $u$-derivative which then vanishes due to equations of motion \eqref{Bondi-frame-sol} and the large $u$ fall-offs \eqref{h2-largeu}. The last term can be manipulated as
\begin{equation}
\begin{split}
\int \dt u \dt^4 x  \sqrt{{\hat q}}   e^{-\Phi}  \SD_A \Phi   \ds \hat{W}_\4^A  &= \int \dt u \dt^4 x  \sqrt{{\hat q}}   e^{-\Phi}  \SD_A \left(  \Phi   \ds \hat{W}_\4^A \right)  - \int \dt u \dt^4 x  \sqrt{{\hat q}} e^{-\Phi}   \Phi  \ds \left( \SD_A \hat{W}_\4^A \right) .
\end{split}
\end{equation}
The last term {in \eqref{eq: int_theta_can}} vanishes due to the constraint \eqref{W4hatconstraint}. The first term simplifies to a total derivative due to \eqref{totalDprop} as
\begin{equation}
\begin{split}
\int \dt u \dt^4 x  \sqrt{{\hat q}}   e^{-\Phi}  \SD_A \Phi   \ds \hat{W}_\4^A  &= - \int \dt u \dt^4 x  \sqrt{{\hat q}}  \p_u \left( \Phi   \ds \hat{W}_\4^A \p_A \k \right) .
\end{split}
\end{equation}
This then vanishes due to the equations of motion \eqref{Bondi-frame-sol} and the large $u$ fall-offs \eqref{h2-largeu}. Putting all of this together and taking the $r \to \infty$ limit, we find
\begin{equation}
\begin{split}
 \int_{\ci^+} \ut_\text{can} (\d) &=  \frac{1}{2} \int \dt u \dt^4 x \sqrt{q} e^{-4\Phi} \p_u \tilde{h}_{\2AB} {\bar \d}  \tilde{h}_\2^{AB} + \int  \dt^4 x \sqrt{q} e^\Phi  \ds C M  \\
&\qquad - \frac{u_0}{4}  \int \dt^4 x \sqrt{q}  \ds \chi^A D_A M  + \int \dt^4 x \d \sqrt{q} \left( - \frac{u_0}{4}  M  +  \frac{1}{8} \int \dt u \, u \tr [ ( \p_u {\wt h}_\2 )^2 ] \right) .
\end{split}
\end{equation}

\subsection{Boundary Contribution (\texorpdfstring{$\ci^+_-$}{ipm})}
\label{sec:symppot-bdy}

Having discussed the canonical contribution in the previous section, we now turn to the second term in 
\eqref{symp-pot-1} which we recall here
\begin{equation}
\begin{split}\label{Qxcompute}
\frac{1}{16\pi G} \int_{\p\ci^+_r}  [  Q_{\ds \mr{ X}} + Y(\d)  ]   , \qquad  [ \star Q_\xi ]^{\mu\nu} = 2 \n^{[\mu} \xi^{\nu]} . 
\end{split}
\end{equation}
The boundary $\p \ci^+_r$ is located at $u=u_0$ so the first term above simplifies to
\begin{equation}
\begin{split}
Q_{\ds \mr{ X}} |_{\p \ci^+_r } = - \dt^4 x r^4 \sqrt{q} e^{2\b} [ \star Q_{\ds \mr{ X}} ]^{ur}  . 
\end{split}
\end{equation}
For a WBMS vector field $\xi$ \eqref{WBMSvector}, we have
\begin{equation}
\begin{split}\label{Qxi-exp}
\int_{\p\ci^+_r}  Q_{\xi} &= -  \int  \dt^4 x r^4 \sqrt{q} e^{-2\b} [ \p_r \xi_u - \p_u \xi_r + W^A ( \p_r \xi_A - \p_A \xi_r ) ]  \\
&= - \int  \dt^4 x  \sqrt{q} \, \bigg[  r^4  [  - 2 \o ]  + r^3 [ 2 Y^A W_{\2A} ]  + r^2 \bigg[  4 \o \b_\2 + Y^A ( 3 W_\3 +   2  h_\1 W_\2 )_A  \\
&\qquad + T \left(  - 2 \p_u \b_\2 - \frac{1}{8} D \cdot D \cdot h_\1  \right)  \bigg]  + r \bigg[ 6 \o \b_\3 + Y^A ( 4 W_\4 - 4 \b_\2 W_\2 \\
&\qquad  + 2 h_\2 W_\2 + 3 h_\1 W_\3 )_A  +  T  \bigg(  D^2 \b_\2  + D \cdot ( h_\1 W_\2 ) + \frac{3}{2} D \cdot W_\3  \\
&\qquad +  2 U_\2  + 2 \p_u \b_\3  + 4 \b_\2 U_\0 - 2 W_\2^2 \bigg) \bigg] +   8 \o \b_\4 + Y^A [ 5  W_\5 + 4 h_\1 W_\4 \\
&\qquad + 3 h_\2 W_\3  + 2 h_\3 W_\2 - 6 \b_\2 W_\3 - 4 \b_\2 h_\1 W_\2 - 4 \b_\3 W_\2 ]_A \\
&\qquad    + T \bigg( - \frac{1}{2} D \cdot D \cdot ( \b_\2 h_\1 ) + \frac{3}{2} D^2 \b_\3 + \frac{9}{4} D \cdot W_\4 + \frac{3}{2} D \cdot  ( h_\1 W_\3 )    \\
&\qquad + \frac{1}{3}   D \cdot  [ ( 4 h_\2 -  h_\1^2 - 20 \b_\2  ) W_\2 ]  + 3 U_\3 + 4  \b_\2  U_\1  +  6 \b_\3 U_\0     \\
&\qquad  +  2 \p_u \b_\4  +  2 W_\2 D \b_\2  - 2 W_\2 h_\1 W_\2 -  5 W_\2 W_\3 \bigg)  \bigg] + \CO(r^{-1}) . 
\end{split}
\end{equation}
We can now apply this result to evaluate the first term in \eqref{Qxcompute}. To do this, we simply recall from \eqref{WBMSvarparameters} that $\ds\mr X^\mu$ is a WBMS vector field where the parameters are given by $e^{-\Phi}f = \ds C$, $\o = \ds \Phi$ and $Y^A = \ds\chi^A$. Substituting these into \eqref{Qxi-exp} we find\footnote{Due to the constraints \eqref{hh-div-constraints}, the fields and variations satisfy
$$
\int_{\ci^+_-} \dt^4 x \sqrt{q} \left[ 16 \d \b_\4 -  {\bar h}_\4^{AB} \d q _{AB}  - \frac{1}{2} {\bar h}_\3^{AB} \d h_{\1AB} + \frac{1}{2} {\bar h}_\1^{AB}  \d h_{\3AB} +  q^{AB} \d h_{\4AB} \right]  = 0 . 
$$}
\begin{equation}
\begin{split}\label{msQformula}
&\int_{\p\ci^+_r}  Q_{\ds\mrx} = - \int \dt^4 x \sqrt{q} \bigg( r^4 \left[ - \frac{1}{4} q^{AB} \d q_{AB} \right] + r^3 \left[  \frac{1}{6} h_\1^{AB} \d q_{AB} \right] \\
& + r^2 \left[ \left( - \frac{3}{16}  (h_\1^2)^{AB}  + \frac{1}{32} \tr [ h_\1^2 ]  q^{AB} \right)  \d q_{AB}  + \frac{1}{16}  h_\1^{AB} \d h_{\1AB}  \right] \\
&  + r \left[ \left( \frac{3}{16}  (h_\1^3)^{AB}  - \frac{1}{32} \tr [  h_\1^3 ]   q^{AB} \right) \d q_{AB} + \left( - \frac{1}{16}  ( h_\1^2 )^{AB} - \frac{3}{64} \tr [ h_\1^2 ] q^{AB}   \right) \d h_{\1AB}  \right] \\
& + \left( - \frac{1}{64} \tr[h_\1^2]  (h_\1^2)^{AB} - \frac{1}{48} \tr [ h_\1^3 ] h_\1^{AB} - \frac{3}{256} \tr [ h_\1^4 ] q^{AB} + \frac{1}{128} \tr[h_\1^2]^2 q^{AB}   \right) \d q_{AB} \\
& + \left( \frac{3}{64} (h_\1^3)^{AB} - \frac{1}{64} \tr[h_\1^2]  h_\1^{AB} \right) \d h_{\1AB} - \frac{1}{4} u_0 \ds \chi^A  D_A M - \ds \chi^A {\bar \CN}_A - 3 e^\Phi \ds C M \\
& + \frac{3}{4} u_0  \d \ln \sqrt{q} M  \bigg) + \CO(r^{-1}) ,
\end{split}
\end{equation}
where
\begin{equation}
\begin{split}\label{Nbardef}
{\bar \CN} &\equiv \CN -  \frac{1}{24} \left( h_\1^3  - \frac{1}{2} \tr[h_\1^2] h_\1 - \frac{1}{3} q \tr[h_\1^3] \right) D \cdot h_\1 - \frac{1}{256} D \left( 2 \tr[h_\1^4] - \tr[h_\1^2]^2 \right) .  
\end{split}
\end{equation}

To recast \eqref{Qxi-exp} into the form \eqref{msQformula}, we have used the equations listed in Appendix \ref{sec:solution-nolog} quite extensively. Note that the term on the RHS is evaluated at $\ci^+_-$ (regulated at $u=u_0$) so we can also use the boundary conditions \eqref{sec:htilde-largeu} to simplify our results. In particular, this implies that only terms that depend on $q_{AB}$ and $h_{\1AB}$ appear in the boundary symplectic form as can be seen in \eqref{msQformula}.

\subsection{Large \texorpdfstring{$r$}{r} Divergences and Phase Space Renormalization} \label{sec:counterterm}

Unlike the radiative/bulk contribution we discussed in the previous section, the boundary contribution to the symplectic potential is divergent at large $r$. In this section, we show that via a judicious choice of $Y(\d)$, the large $r$ divergences (and \emph{only} the divergences) can be cancelled. Such a renormalization process was discussed in four dimensions in \cite{Compere:2018ylh,Chandrasekaran:2021vyu}, where it was suggested that since the divergences in the less involved four-dimensional analogue of \eqref{msQformula} are all \emph{boundary terms}, we can simply cancel them all by choosing
\begin{equation}
\begin{split}\label{counterterm-prescription}
Y(\d) = - \text{divergences} . 
\end{split}
\end{equation}
However, there are two major issues with the prescription \eqref{counterterm-prescription} which we will address in this section.
\begin{itemize}
\item It is crucial that $Y(\d)$ is covariant which means that it must be constructed out of the induced metric on $\p\S$ and the extrinsic curvature (which keeps track of the embedding of $\p\S \hookrightarrow M$). It implies that the symplectic potential {\eqref{symp-pot-1}} will not depend on our choice of regulator for the asymptotic surface $\p \S$. For example, in this paper, we choose to regulate $\ci^+$ by a $r=$ constant surface. Of course, this is only one out of many choices of regulators. For instance, we can define the regulated surface in the same way but in Newman-Unti coordinates \cite{Newman:1962cia} or as a $v=$ constant surface where $v$ is a null coordinate, etc. As long as $Y(\d)$ is constructed in a covariant way, the final symplectic potential does not depend on this choice. Secondly, in order to preserve bulk locality, it is also crucial that $Y(\d)$ be constructed locally out of the above-mentioned quantities. This is the first major problem with \eqref{counterterm-prescription}. It is not at all obvious that there even exists a local and covariant choice of $Y(\d)$ that cancels all the divergences.

\item A second major issue with \eqref{counterterm-prescription} is regarding the finite terms in \eqref{msQformula}. If we were indeed allowed to cancel all the divergences through such a simple prescription, \emph{what is stopping us from using the same prescription to cancel the finite terms as well?} We would then end up in a phase space with no BMS transformations! It would then seem that the existence of BMS symmetries in the gravitational phase space is entirely a matter of choice.
\end{itemize}

In this section, we will address both the issues above and show that by enforcing the locality and covariance of $Y(\d)$ \emph{and} by using the boundary conditions \eqref{vanishingRiemann} which defines the solution space $\mss_0$ we are indeed able to cancel all the divergences, \emph{but not the finite part}!

\subsection*{General Structure}

We start by making the implications of locality and covariance on $Y(\d)$ more precise. First, locality of $Y(\d)$ implies that it has the structure
\begin{equation}
\begin{split}
\int_{\p\S} Y(\d) = \int_{\p\S} \dt^4 \t \sqrt{{\cal g}} \CY(\d) . 
\end{split}
\end{equation}
where $\CY(\d)$ is a scalar function of its arguments and $\dt s^2|_{\p\S} = {\cal g}_{AB}\dt\t^A \dt\t^B$ is the induced metric on $\p\S$. Further, covariance implies that $\CY(\d)$ can only depend on the induced metric, the extrinsic curvatures and their variations. In particular, a Euclidean codimension-2 surface admits two independent null normals $n$ and $\ell$ (with $n \cdot \ell = -1$) and we can define two extrinsic curvatures as
\begin{equation}
\begin{split}
{\cal k}_{AB} \equiv e_A^\mu e_B^\nu \n_\mu n_\nu , \qquad {\cal l}_{AB} \equiv e_A^\mu e_B^\nu \n_\mu \ell_\nu .
\end{split}
\end{equation}
where $e^\mu_A$ is the projection tensor from $M \to \p\S$. On-shell, one of the extrinsic curvatures is related to the other. In particular, if $\ell$ is tangent to the null surface whose boundary is $\p\S$, then ${\cal l}$ is fixed in terms of ${\cal k}$ and ${\cal g}$. Consequently, we can simply work with the latter two quantities. Indeed, in the case at hand ${\cal l}_{AB} \sim \p_u g_{AB}$ which is fixed by the equations $R_{AB}[G] = 0$. On the other hand ${\cal k}_{AB} \sim \p_r g_{AB}$ is \emph{not} fixed by any equation.

Locality and covariance together imply that $\CY(\d)$ is a polynomial function of the following arguments
\begin{equation}
\begin{split}
\CY(\d) \equiv \CY^{AB}_1 ( {\cal g} , \CD^n {\cal k} ) \d {\cal g}_{AB} + \CY^{AB}_2 ( {\cal g} , \CD^n {\cal k} ) \d {\cal k}_{AB} . 
\end{split}
\end{equation}
where $\CD$ is the covariant derivative w.r.t. the induced metric ${\cal g}_{AB}$.
The goal of this section will be to construct a $\CY$ of this form so that all the divergences are cancelled from \eqref{msQformula}. Let us now apply the discussion of the previous section to the case at hand. The induced metric on $\p\ci^+_r$ is ${\cal g}_{AB} =  r^2 h_{AB}$. The two null normals are given by 
\begin{equation}
\begin{split} \label{eq:curlyk-def}
n = r \p_r , \qquad \ell = \frac{1}{r} e^{-2\b} \left( \p_u  - \frac{1}{2} U \p_r + W^A \p_A \right) \quad \implies \quad {\cal k}_{AB} =  \frac{1}{2} r \p_r ( r^2 h_{AB} )  . 
\end{split}
\end{equation}
We must now construct $\CY(\d)$ which cancels the divergences in \eqref{msQformula}. We first note that at large $r$, the metric and extrinsic curvature have the form
\begin{align}
\begin{split}\label{gk-larger}
{\cal g}_{AB} & = r^2 q_{AB} + r h_{\1AB} + \CO(1) ,  \\ {\cal k}_{AB} & = r^2 q_{AB} + r \frac{1}{2} h_{\1AB} + \CO(1)  
\end{split}
\end{align}
and we define the covariant tensor\footnote{To make contact with the notation in \cite{Capone:2021ouo}, notice that ${\cal k}_{AB}$ here is $r k_{AB}$ there and that ${\cal h}_{AB} $ here is $2r\tilde{K}_{AB}$  there in equations (4.1)-(4.2). Einstein equations can be written in terms of such tensors and $l_{AB}=\partial_u {\cal g}_{AB}/2$, see equations (4.6)-(4.10) there.}
\begin{equation}
\begin{split}\label{h1-def}
{\cal h}_{AB} \equiv 2 ( {\cal g} - {\cal k} )_{\{AB\}} = r h_{\1AB} + \CO(1) . 
\end{split}
\end{equation}
In this section, to preserve covariance, we raise and lower indices using the induced metric ${\cal g}$.

From \eqref{gk-larger} and \eqref{h1-def}, we see that the non-covariant tensors $q_{AB}$ and $h_{\1AB}$ are the leading order large $r$ coefficients of the covariant tensors ${\cal g}_{AB}$ and ${\cal h}_{AB}$. This strongly suggests that any instances of $q_{AB}$ and $h_{\1AB}$ in a divergent quantity can be covariantized by replacing
\begin{equation}
\begin{split}
q_{AB} ~\to~ {\cal g}_{AB} , \qquad h_{\1AB} ~\to~ {\cal h}_{AB} . 
\end{split}
\end{equation}
Of course, it is important to note that $q$ and $h_\1$ are only the \emph{leading} order coefficients of ${\cal g}$ and ${\cal h}$ so making such a replacement modifies the subleading divergent terms in the large $r$ expansion. Consequently, our analysis must be done order-by-order in large $r$. Following this process, it can be shown that the appropriate counterterm is given by
\begin{equation} \label{Y}
\begin{split}
\CY(\d) &= - \left( \frac{1}{4} {\cal g}^{AB}   - \frac{1}{6} {\cal h}^{AB}   - \frac{1}{16}  ({\cal h}^2)^{AB} - \frac{1}{96} \tr [ {\cal h}^2 ] {\cal g}^{AB} - \frac{5}{96} ({\cal h}^3)^{AB}  + \frac{1}{96}  \tr [  {\cal h}^3 ] {\cal g}^{AB} \right. \\
&\left. \qquad - \frac{1}{128} \tr[ {\cal h}^2]  ({\cal h}^2)^{AB} - \frac{1}{64} \tr [ {\cal h}^3 ] {\cal h}^{AB} - \frac{3}{256} \tr [ {\cal h}^4 ] {\cal g}^{AB} + \frac{3}{512} \tr[{\cal h}^2]^2 {\cal g}^{AB}   \right) \d {\cal g}_{AB} \\
&\qquad - \left(  \frac{5}{48}  {\cal h}^{AB}  + \frac{1}{96}  ( {\cal h}^2 )^{AB} + \frac{1}{48} \tr [ {\cal h}^2 ] {\cal g}^{AB} + \frac{3}{64} ({\cal h}^3)^{AB} - \frac{1}{64} \tr[{\cal h}^2]  {\cal h}^{AB} \right) \d {\cal h}_{AB} \\
&\qquad - \CY_\text{finite}(\d) . 
\end{split}
\end{equation}
The counterterm above cancels all the terms except for the last four in \eqref{msQformula}. Above, we have also included a counterterm $\CY_\text{finite}$ which changes the symplectic potential by a finite amount. The most general form for such a counterterm is
\begin{equation}
\begin{split} \label{eq:Y-fintite}
 \sqrt{{\cal g}} \CY_\text{finite} (\d) &=  \sqrt{{\cal g}} \bigg( a_1 \tr[ {\cal h}^2]  ({\cal h}^2)^{\{AB\}} + a_2 \tr [ {\cal h}^3 ] {\cal h}^{AB}  \bigg) \d {\cal g}_{AB}  \\
&\qquad + \d \sqrt{{\cal g}} \bigg( a_3 \tr [ {\cal h}^4 ] + a_4 \tr[{\cal h}^2]^2  \bigg)  + \d \bigg( a_5 \sqrt{{\cal g}}  \tr [ {\cal h}^4 ]  + a_6 \sqrt{{\cal g}}  \tr[{\cal h}^2]^2 \bigg) .
\end{split}
\end{equation}
Here, we have classified the ambiguities into three types of terms. The last two are $\d$-exact terms so they do not contribute to the symplectic form and we will henceforth drop them. The middle two terms also do not contribute to the symplectic potential for reasons we will discuss in the next section.  To see the effect of the first two terms, we put everything together and find the renormalized symplectic potential
\begin{equation}
\begin{split}\label{Theta-final}
\T_{\ci^+}(\d) &= \frac{1}{32\pi G} \int \dt u \dt^4 x \sqrt{q} \p_u \wt{h}_\2^{AB} {\bar \d}  \wt{h}_{\2AB} + \frac{1}{16\pi G}\int \dt^4 x \sqrt{q} \left( 4 e^\Phi \ds C M + \ds \chi^A N_A  \right) \\
&\qquad - \frac{1}{16\pi G} \int \dt^4 x \d \sqrt{q} \left(  a_3 \tr [ h_\1^4 ] + a_4 \tr[h_\1^2]^2 + u_0 M   -   \frac{1}{8} \int \dt u \, u \tr [ ( \p_u {\wt h}_\2 )^2 ] \right) , 
\end{split}
\end{equation}
where $N_A(x) \equiv \CN_A(x) - P_A(x)$ and
\begin{equation}
\begin{split}\label{Ndef}
P &\equiv \frac{1}{24} \left( h_\1^3  - \frac{1}{2} \tr[h_\1^2] h_\1 - \frac{1}{3} q \tr[h_\1^3] \right) D \cdot h_\1  + \frac{1}{256} D \left( 2 \tr[h_\1^4] - \tr[h_\1^2]^2 \right) \\
&\qquad \qquad \qquad - 2 D \cdot \bigg(   a_1 \tr[ h_\1^2]  h_\1^2  +   a_2 \tr [ h_\1^3 ] h_\1  \bigg) ,  
\end{split}
\end{equation}
where $h_\1$ above is evaluated at $u=u_0$. We see that the ambiguities parameterized by $a_1$ and $a_2$ appear exclusively in the tensor $P_A(x)$. It follows that the finite ambiguities in the symplectic potential can be absorbed into a redefinition of the phase space variable $\CN_A \to N_A$.

\paragraph{Alternative way} 

For a quicker construction of the counterterm, we can take advantage of a simple computational trick. Knowing that the divergences are a boundary term it is easy to show that \cite{Compere:2018ylh, Chandrasekaran:2021vyu}
\begin{flalign}
    \int _{u=\text{const}} \left( \star \theta  \right)^{u \ \text{div}}\dt r  =  \int _{r=\text{const}}\left( \star \theta \right) ^{r \ \text{div}} \dt u.
\end{flalign}
In this paper we work with metrics that have an analytic expansion in terms of $r$. This implies that the entire $ \int ( \star \theta ) ^u \dt r$ is a  boundary term. Therefore,  $ \int ( \star \theta ) ^u \dt r$  has the same tensor structure as the counterterm we need to add to $\Theta$ to make it finite.

The question now is how to express $ \int (\star \theta ) ^u \dt r$ covariantly in terms of ${\cal g}_{AB}$ and ${\cal k}_{AB}$ {\eqref{gk-larger}} and their variations. This is achieved in two simple steps. The first is the following,

\begin{flalign}\label{eq:theta-u-alt}
\begin{split}
&-\int \sqrt{\cal g } \, ( \star \theta ) ^u \, \dt r = \int r^{-1} \sqrt{{\cal g}}   \left( {\cal g}^{AB} \d {\cal k}_{AB}  - 8 \d \beta + \frac{1}{4} \d ({\cal k} _A ^B {\cal k} _B^A ) \right)   \dt r   \\
&\qquad =  \frac{1}{6} \sqrt{\cal g } ( 3+ 2r\partial_r  + r \p _r ( r \p _r )) \left( {\cal g}^{AB} \d {\cal k} _{AB}  - 8 \d \beta  + \frac{1}{4} \d ({\cal k} _A ^B {\cal k} _B^A  )  + \CO (r^{-5}) \right)  - \frac{1}{2}\d \sqrt{\cal g}  \\
&\qquad =  \frac{1}{6} \sqrt{\cal g } ( 3+ 2 \CL_n + \CL_n^2) \left( {\cal g}^{AB} \d {\cal k} _{AB}  - 8 \d \beta  + \frac{1}{4} \d ({\cal k} _A ^B {\cal k} _B^A  )  + \CO (r^{-5}) \right)  - \frac{1}{2}\d \sqrt{\cal g}.
\end{split}
\end{flalign}
Here we note two things: firstly, the awkward $\d \sqrt{\cal g}$ will be fixed to $0$ in the subsequent analysis. Even if it was not, it could be fixed by adding an additional term $ \CL_n^3$ and re-adjusting the coefficients of the $\CL ^n$, $n \in \{ 0, \ldots, 3 \}$. These would make the calculations in this section more tedious and since it is ultimately unnecessary it is not done here.
Secondly, there will be no finite order contributions from this counterterm. The reason for this is that in the first line of {\eqref{eq:theta-u-alt}}, $r^{-1}$ integrates to $\ln r$ and the coefficient of this term necessarily has to vanish due to the no-log constraints {\eqref{hh-div-constraints}}.
 
The second step is expressing $r$ derivatives of ${\cal g} _{AB}$ and ${\cal k} _{AB}$  in terms of themselves. This is easily done by using the equations of motion {\eqref{Einstein-EoM}}, boundary conditions {\eqref{vanishingRiemann}} (specifically all we need is the fact that $r^4 R_{rArB} [G] = 0 | _{\ci^ + _-}$) and the definition of ${\cal k} _{AB}$ {\eqref{eq:curlyk-def}}
\begin{equation}\label{eq:alt-ren-needs}
\begin{split}
r \partial _r {\cal g}_{AB} &=  2 {\cal k} _{AB} , \qquad 8r \partial _r \beta  = {\cal k} ^A _B {\cal k} ^B _A - 4 , \\
r \partial _r {\cal k} _{AB} &=   \frac{1}{4}  {\cal k} _{CD} {\cal k} ^{CD} {\cal k} _{AB} + {\cal k} _{AC} {\cal k} ^C_B  +  \CO (r^{-3}) ~ \text{as} ~ {u \rightarrow -\infty}.
\end{split}
\end{equation}
Notice that even though the third equation does not hold in general, it is valid to the orders relevant to this computation. This is not an accident, and one can prove that a relationship of this type holds in any dimension.

Combining all this, the counterterm can be written as
\begin{equation}
\begin{split}
\sqrt{\cal g} \CY '(\d) = - \frac{\sqrt{\cal g }}{6}& \left( 3 {\cal k} ^{AB}  - 2 ({\cal k} ^2)^{AB}  +  \frac{1}{2} {\cal k} ^{AB}   \tr [ {\cal k} ^2 ]+ 2 ({\cal k} ^3) ^{AB} \right.\\
 &\left. -  \frac{1}{4}({\cal k} ^2) ^{AB}  \tr[{\cal k} ^2]  + \frac{3}{4} {\cal k} ^{AB} \left(  \frac{3}{4} \tr [{\cal k} ^2 ]^2  -   2\tr [{\cal k} ^3] \right) \right) \d {\cal g} _{AB}  + \d (\, \cdots ) .
\end{split}
\end{equation}
This counterterm is equivalent to \eqref{Y} up to the finite order and terms that are total variations, which will not contribute to the symplectic form. Counterterms at finite order have to be added separately. In order to recover \eqref{Y} we need to add
\begin{equation}
\begin{split}
\sqrt{\cal g}  \mathcal{W} (\d) &=- \sqrt{\cal g}  \bigg( \frac{1}{128} \tr[ {\cal h}^2]  ({\cal h}^2)^{\{AB\}} + \frac{1}{192} \tr [ {\cal h}^3 ] {\cal h}^{AB}  \bigg) \d {\cal g}_{AB} \\
&\qquad \qquad \qquad \qquad \qquad \qquad   -  \d \left( \frac{3}{256}  \sqrt{\cal g} \tr [{\cal h} ^4] + \frac{1}{256} \sqrt{\cal g} \tr[{\cal h}^2]^2 \right). 
\end{split}
\end{equation}

\subsection{Large \texorpdfstring{$u$}{u} Divergences}
\label{sec:large-u-divergences-symp-pot}

In the previous section, we have discussed the large $r$ divergences and their renormalization extensively. We now turn to the large $u$ divergences in the symplectic potential {\eqref{Theta-final}}.

There are two types of $u$ dependences in $\T_{\ci^+}(\d)$. First, an explicit factor of $u_0$ appears in the second line of \eqref{Theta-final} and second, there is an implicit $u_0$ dependence in $h_\1$. The implicit dependence cannot be renormalized and in fact, plays a physical role in the charge algebra. The divergence arising from the explicit $u$ dependence can be removed by restricting to a smaller phase space where
\begin{equation}
\begin{split}\label{Phi-constraint}
\d \sqrt{q} = 0 \qquad \implies \qquad \ds \Phi = - \frac{1}{4} D_A \ds \chi^A . 
\end{split}
\end{equation}
While the reasoning provided here is sufficient to impose such a constraint, there is a second reason for expecting this. From \eqref{Theta-final} we can easily extract the conjugate modes for all the fields on our phase space. In particular, the symplectic conjugate of $C$ is $M$ and that of $\chi^A$ is $N_A$ (roughly). However, the Weyl mode $\Phi$ (which appears in $\d \sqrt{q}$) does not have any independent symplectic conjugate so the symplectic form constructed out of \eqref{Theta-final} will be degenerate. In order, then, to have a well-defined phase space, we must remove $\Phi$ as a dynamical variable and consequently fix its variation in terms of other variations, i.e. we must require
\begin{equation}
\begin{split}
\ds \Phi = F ( \d {\wt h}_\2 , \d C , \d \chi^A ) . 
\end{split}
\end{equation}
The precise choice of the function $F$ is required to preserve the identification \eqref{chi-phi-identification} and \eqref{Phi-constraint} does the job! In this case, we can simply solve for $\Phi(x)$ as
\begin{equation}
\begin{split}\label{Phi-final-form}
\Phi(x) = \Phi_0(x) - \frac{1}{4} \ln \det ( \p_A \chi^B ) , \qquad \d \Phi_0(x) = 0 .
\end{split}
\end{equation}
$\Phi_0(x)$ is a non-dynamical background Weyl factor that is constant over the entire phase space.

On such a phase space, the symplectic potential is
\begin{equation}
\begin{split}\label{Theta-final-1}
\T_{\ci^+}(\d) &=   \frac{1}{32\pi G} \int \dt u \dt^4 x \sqrt{q} \p_u \wt{h}_\2^{AB} {\bar \d}  \wt{h}_{\2AB}  + \frac{1}{16\pi G}\int \dt^4 x \sqrt{q}  ( 4 e^\Phi \ds C M + \ds \chi^A N_A   ) .
\end{split}
\end{equation}

\section{Symplectic Form and Poisson Brackets}
\label{sec:sympform}

The symplectic form is constructed from the symplectic potential as
\begin{equation}
\begin{split}
\O(\d,\d') \equiv \d \T_{\ci^+}(\d')  - \d' \T_{\ci^+}(\d) - \T_{\ci^+}([\d,\d']). 
\end{split}
\end{equation}
Note that $\O$ does not depend on the choice of Cauchy slice, even though $\T$ does so we have dropped the subscript $\ci^+$ from $\O$. Using this definition, the symplectic form can easily be constructed from \eqref{Theta-final-1}. Our goal is to eventually invert the symplectic form in order to construct the Poisson brackets on the phase space. In order to do this, it will be useful to work in a particularly nice basis where the inversion becomes trivial. To do this, we define a primed field $h'_\2$ by
\begin{equation}
\begin{split}\label{hprime-def}
{\wt h}_{\2AB} = e^{2\Phi} \p_A \chi^C \p_B \chi^D h'_{\2CD} . 
\end{split}
\end{equation}
The primed fields do not transform nicely under BMS transformations at all and do not have any simple physical interpretation. The sole purpose of introducing them in this section is to move to a basis where the symplectic form can be easily inverted to obtain the Poisson bracket. Once we have done this, we will move back to the original basis of fields. An explicit calculation reveals
\begin{equation}
\begin{split}\label{Theta-final-2}
{\bar \d} {\wt h}_{\2AB} &= e^{2\Phi}  \p_A \chi^C \p_ B \chi^D  \left( \d - \ds\chi \cdot \p - \frac{1}{4} D \cdot \ds \chi ( 2 + u \p_u )  - e^\Phi \ds C \p_u \right) h'_{\2CD}  .
\end{split}
\end{equation}
Using this, the symplectic potential {\eqref{Theta-final-1}} can be written as
\begin{equation}
\begin{split}
\!\!\!\! \T_{\ci^+}(\d) &=   \frac{1}{32\pi G} \int \dt u \dt^4 x \sqrt{q}  \p_u h'^{AB}_\2 \d h'_{\2AB} + \frac{1}{16\pi G}\int \dt^4 x \sqrt{q} [ e^\Phi M' \d C + ( \p_{\chi^A} x^B N'_B ) \d \chi^A ] .
\end{split}
\end{equation}
where
\begin{equation}
\begin{split}\label{MNprime-def}
M' &= 4 M - \frac{1}{2} \int \dt u  \p_u h'^{AB}_\2 \p_u  h'_{\2AB} , \\
N'_A &= N_A - e^\Phi  M' \p_A C  - \frac{1}{2} \int \dt u  \p_u h'^{BC}_\2 \p_A h'_{\2BC} + \frac{1}{8} \p_A \int \dt u  u  \p_u h'^{BC}_\2  \p_u h'_{\2BC} . 
\end{split}
\end{equation}
Using \eqref{Theta-final-2}, we find that the symplectic form is given by 
\begin{equation}
\begin{split}\label{Omega-final}
\O(\d,\d') &=  \frac{1}{32\pi G} \int \dt u \dt^4 x \sqrt{q} \d^{AC} \d^{BD} \p_u  \d h'_{\2AB} \curlywedge \d h'_{\2CD} \\
&\qquad \qquad \qquad \qquad \qquad \qquad + \frac{1}{16\pi G}\int \dt^4 x \sqrt{q} [ \d ( e^\Phi M' )  \curlywedge \d C +  \d ( \p_{\chi^A} x^B N'_B ) \curlywedge \d \chi^A   ] ,
\end{split}
\end{equation}
where $\curlywedge$ is the wedge product in field space, $\d f \curlywedge \d g \equiv \d f \d' g - \d' f \d g$. From \eqref{Omega-final}, we can see the usefulness of defining the primed fields. In terms of these, the symplectic form attains a simple block diagonal form! Inverting this becomes incredibly simple and the Poisson brackets are given by
\begin{equation}
\begin{split}
\{ h'_{\2AB}(u,x) , \p_{u'} h'_{\2CD} (u',x')  \} &=  16\pi G \d_{A\{C} \d_{D\}B} \d(u-u') \d^4(x,x') , \\
\{ \chi^A(x) , \p_{\chi^B} x^C N'_C(x')  \} &= 16\pi G \d^A_B \d^4(x,x') , \\
\{ C(x) , e^{\Phi(x')} M'(x')  \} &= 16\pi G \d^4(x,x') , \\
\text{all others} &= 0 .
\end{split}
\end{equation}
Note that these brackets are consistent with the identification \eqref{chi-phi-identification}.

We can now go back to the original unprimed fields and determine the full set of Poisson brackets. Brackets involving ${\wt h}_{\2AB}(u,x)$ are
\begin{equation}
\begin{split}
\label{h2-brackets}
\{ {\wt h}_{\2AB}(u,x) , \p_{u'} {\wt h}_{\2CD}(u',x') \} &=  16\pi G  q_{A\{C} q_{D\}B} \d(u-u') \d^4(x,x')  , \\
\{ {\wt h}_{\2AB}(u,x) , C(x') \} &= 0 , \\
\{ {\wt h}_{\2AB}(u,x) , M(x') \} &= 4 \pi G \p_u {\wt h}_{\2AB}(u,x)  \d^4(x,x') , \\
\{ {\wt h}_{\2AB}(u,x) , \chi^C(x') \} &= 0 , \\
\{ {\wt h}_{\2AB}(u,x) ,  N_C(x')  \} &= 16 \pi G \left( \p_C {\wt h}_{\2AB}(u,x) + {\wt h}_{\2AC}(u,x) \p_B \right. \\
&\left. \qquad \qquad + {\wt h}_{\2CB}(u,x) \p_A - \frac{1}{4} u \p_u {\wt h}_{\2AB}(u,x) \p'_C \right) \d^4(x,x') . \\
\end{split}
\end{equation}
Brackets involving $C(x)$ are
\begin{equation}
\begin{split}
\label{C-brackets}
\{ C(x) , {\wt h}_{\2AB}(u',x') \} &= 0 , \\
\{ C(x) , C(x') \} &= 0 , \\
\{ C(x) , M(x') \} &= 4\pi G e^{-\Phi(x)} \d^4(x,x')   , \\
\{ C(x) , \chi^A(x') \} &= 0 , \\
\{ C(x) , N_A(x') \} &= 16\pi G  \p_A C(x) \d^4(x,x') .
\end{split}
\end{equation}
Brackets involving $M(x)$ are
\begin{equation}
\begin{split}
\label{M-brackets}
\{ M(x) , {\wt h}_{\2AB}(u',x') \} &= - 4 \pi G \p_u {\wt h}_{\2AB}(u,x)  \d^4(x,x')  , \\
\{ M(x) , C(x') \} &= -  4\pi G e^{-\Phi(x)} \d^4(x,x')   , \\
\{ M(x) , M(x') \} &= 0 , \\
\{ M(x) , \chi^A(x') \} &= 0 , \\
\{ M(x) , N_A(x') \} &= 16 \pi G \left( \p_A  M(x) - \frac{5}{4} M(x) \p_A'  \right) \d^4(x,x') . \\
\end{split}
\end{equation}
Brackets involving $\chi^A(x)$ are
\begin{equation}
\begin{split}
\label{chi-brackets}
\{ \chi^A(x) , {\wt h}_{\2BC}(u',x') \} &= 0 , \\
\{ \chi^A(x) , C(x') \} &= 0 , \\
\{ \chi^A(x) , M(x') \} &= 0 , \\
\{ \chi^A(x) , \chi^B(x') \} &= 0 , \\
\end{split}
\end{equation}
Finally, brackets involving $N_A(x)$ are
\begin{equation}
\begin{split}
\label{N-brackets}
\{ N_A(x) , {\wt h}_{\2BC}(u',x') \} &= - 16 \pi G \left( \p'_A {\wt h}_{\2BC}(u',x') + {\wt h}_{\2AC}(u',x') \p'_B \right. \\
&\left. \qquad \qquad + {\wt h}_{\2AB}(u',x') \p'_C - \frac{1}{4} u' \p_{u'} {\wt h}_{\2BC}(u',x') \p_A \right) \d^4(x,x') ,  \\
\{ N_A(x) , C(x') \} &= - 16\pi G  \p_A C(x) \d^4(x,x') , \\
\{ N_A(x) , M(x') \} &= - 16 \pi G \left( \p_A  M(x) - \frac{5}{4} M(x') \p_A \right) \d^4(x,x')  , \\
\{ N_A(x) , \chi^B(x') \} &= - 16 \pi G \p_A \chi^B (x) \d^4(x,x') , \\
\{ N_A(x) , N_B(x') \} &=   - 16\pi G  [  N_A(x') \p_B' - N_B(x) \p_A   ] \d^4(x,x') . \\
\end{split}
\end{equation}

\section{GBMS Charge Algebra}\label{sec:BMS-charge-algebra}

Now that we have constructed the symplectic form for six-dimensional {asymptotically flat} gravity, we turn to a study of canonical transformations on this phase space. We recall that a transformation $\d_c$ which acts on the fields is canonical iff. there exists a Hamiltonian charge which satisfies 
\begin{equation}
\begin{split}
\d H_c = \Omega(\d,\d_c) . 
\end{split}
\end{equation}
Particularly important transformations that we would like to consider are the WBMS diffeomorphisms discussed in Section \ref{sec:BMS-transform}. Under infinitesimal WBMS diffeomorphisms ({generated by \eqref{WBMSvector}}), the phase space fields transform as in \eqref{BMStransform-S0} which we recall here
\begin{equation}
\begin{split}\label{BMStransform-S0-1}
\d_\xi {\wt h}_\2 &= ( T \p_u +  \CL_Y ) {\wt h}_\2  , \qquad T =  f - u \o , \\
\d_\xi C &= \CL_Y C  + e^{-\Phi} f  , \\
\d_\xi M &= ( \CL_Y - 5 \o ) M , \\
\d_\xi \chi^A &= Y^B \p_B \chi^A , \\
\d_\xi \CN_A &=  ( \CL_Y - 4 \o  ) \CN_A + f \p_A  M + 5 M \p_A f , \\
\d_\xi \Phi &= \CL_Y \Phi + \o  . 
\end{split}
\end{equation}
In general, WBMS diffeomorphisms are generated by the parameters $f(x)$, $Y^A(x)$ and $\o(x)$. However, the phase space under consideration is constrained by \eqref{Phi-constraint} and generic WBMS diffeomorphisms do not preserve this constraint. However, the GBMS subgroup discussed in Section \ref{sec:important-subalgebras} does preserve the constraint! This subgroup is defined by restricting $\o$ as
\begin{equation}\label{eq:omega}
\begin{split}
\o = - \frac{1}{4} D_A Y^A .
\end{split}
\end{equation}

In this section, we study whether GBMS diffeomorphisms are indeed canonical and construct -- if possible -- the corresponding GBMS charge $H_\xi$. If it exists, $H_\xi$ satisfies 
\begin{equation}
\begin{split}
\d H_\xi \stackrel{?}{=} \O ( \d , \d_\xi ) . 
\end{split}
\end{equation}
To check whether an equation of this form could possibly hold, we process the RHS of this equation. Using \eqref{Omega-final}, we find
\begin{equation}\
\begin{split}\label{Hxi-var}
&\d H_\xi \stackrel{?}{=} \frac{1}{16\pi G} \int \dt u \dt^4 x \sqrt{q} \d^{AC} \d^{BD}   \p_u  \d h'_{\2AB} \d_\xi h'_{\2CD}    \\
& + \frac{1}{16\pi G}\int \dt^4 x \sqrt{q} [ \d ( e^\Phi M' ) \d_\xi C  - \d_\xi  ( e^\Phi M' ) \d C  +  \d ( \p_{\chi^A} x^B N'_B ) \d_\xi \chi^A  - \d \chi^A \d_\xi ( \p_{\chi^A} x^B N'_B )   ] .
\end{split}
\end{equation}
To process this equation, we need to evaluate the GBMS diffeomorphism of the primed fields. We start by recalling all the relevant definitions here (from \eqref{hprime-def} and \eqref{MNprime-def})
\begin{equation}
\begin{split}
h'_{\2AB}  &= e^{-2\Phi} \p_{\chi^A} x^C \p_{\chi^B} x^D {\wt h}_{\2CD}, \\
M' &= 4 M - \frac{1}{2} \int \dt u  \p_u h'^{AB}_\2 \p_u  h'_{\2AB} , \\
N'_A &= N_A - e^\Phi  M' \p_A C - \frac{1}{2} \int \dt u  \p_u h'^{BC}_\2 \p_A h'_{\2BC} + \frac{1}{8} \p_A \int \dt u  u  \p_u h'^{BC}_\2  \p_u h'_{\2BC} , \\
\end{split}
\end{equation}
where $N_A$ is related to $\CN_A$ via \eqref{Ndef}
\begin{equation}
\begin{split} \label{N_AP_A}
N_A &= \CN_A - P_A , \\
P_A &= \frac{1}{24} \left( (h_\1^3)_{AB}  - \frac{1}{2} \tr[h_\1^2] h_{\1AB} - \frac{1}{3} q_{AB} \tr[h_\1^3] \right) D_C h_\1^{BC}   \\
&\qquad + \frac{1}{256} \p_A \left( 2 \tr[h_\1^4] - \tr[h_\1^2]^2 \right)  - 2 D^B \left( a_1 \tr[ h_\1^2] ( h_\1^2)_{AB}  +   a_2 \tr [ h_\1^3 ] h_{\1AB} \right) .
\end{split}
\end{equation}
Using these definitions, we easily find
\begin{equation}
\begin{split}
\d_\xi h'_{\2AB}  &=  ( T \p_u + Y^C \p_C  - 2  \o ) h'_{\2AB}  , \\
\d_\xi ( e^\Phi M' )  &= ( \CL_Y - 4 \o ) ( e^\Phi  M' ) , \\
\d_\xi N'_A &=  ( \CL_Y - 4 \o  )  N'_A  + f M' \p_A \Phi + \frac{1}{4} \p_A  ( f M' )  - ( \d_\xi  - \CL_Y + 4 \o  )  P_A .
\end{split}
\end{equation}
Substituting these into \eqref{Hxi-var}, we find
\begin{equation}
\begin{split}\label{Hxi-var-1}
\d H_\xi &\stackrel{?}{=} \frac{1}{4\pi G} \int \dt^4 x \sqrt{q} f \d M + \frac{1}{16\pi G} \int \dt^4 x \sqrt{q}\, Y^A  \d N_A \\
&\qquad \qquad \qquad \qquad \qquad \qquad + \frac{1}{16\pi G} \int \dt^4 x \sqrt{q}  \ds \chi^A  ( \d_\xi  - \CL_Y + 4 \o  )  P_A  . 
\end{split}
\end{equation}
The LHS is $\d$-exact so the Hamiltonian charge $H_\xi$ exists iff. the RHS is $\d$-exact. We see that the first two terms above are $\d$-exact \emph{if} we assume that the GBMS parameters $f$ and $Y^A$ are constants on the phase space so that $\d f = \d Y^A = 0$. On the other hand, the last term presents a serious obstruction to exactness. It follows from this that GBMS diffeomorphisms are in fact, \emph{not} canonical transformations on the phase space and the charge $H_\xi$ cannot be defined globally on the phase space! In other words, the charges are \emph{non-integrable}.

Let us take a pause and briefly discuss the non-integrability of the GBMS charges. We first note that the non-integrability arises entirely from superrotations on the phase space. On a phase space \emph{without} any superrotations, we can impose an additional constraint $\ds \chi^A = 0$ which removes the last non-integrable term in {\eqref{Hxi-var-1}} and makes the Hamiltonian charges for supertranslations and Lorentz transformations integrable! This is precisely what happens in four dimensions as well (e.g. the non-integrable term in equation (3.3) of \cite{Barnich:2011mi} vanishes near $\ci^+_-$). The reader can find a discussion of this point in the concluding remarks in Section \ref{sec.conclusions}. One possible way to make the GBMS charges integrable is to take the parameters $f$ and $Y^A$ to be field-dependent so as to cancel the last term in \eqref{Hxi-var-1}. A complete study of the structure of the non-integrable term in \eqref{Hxi-var} is certainly of interest but is beyond the scope of this paper and is left for future work.

Despite the fact that GBMS diffeomorphisms are not canonical, we can still proceed to construct \emph{some} charges which have a GBMS-like structure. Inspired by the form of equation \eqref{Hxi-var-1}, we define the following charges
\begin{equation}
\begin{split} \label{GBMScharges}
T_f \equiv \frac{1}{4\pi G} \int \dt^4 x \sqrt{q} f M , \qquad J_Y \equiv \frac{1}{16\pi G} \int \dt^4 x \sqrt{q}\, Y^A N_A . 
\end{split}
\end{equation}
Using the commutators \eqref{h2-brackets}--\eqref{N-brackets}, we can immediately check that these charges satisfy the BMS algebra \eqref{WBMSalgebra}
\begin{equation}
\begin{split}
\{ T_f , T_{f'} \} &= 0 , \qquad \{ J_Y , T_f \}  = T_{( Y \cdot \p  - \frac{1}{4} D \cdot Y ) f  } , \qquad \{ J_Y , J_{Y'} \} = J_{[Y,Y']}  . 
\end{split}
\end{equation}
The action of these charges on the phase space variables are
\begin{equation}
\begin{split}\label{TJ-action}
\{ T_f + J_Y , {\wt h}_{\2AB} \} &= -  \left( f \p_u + \CL_Y + \frac{1}{4} D \cdot  Y u \p_u \right) {\wt h}_{\2AB} , \\ 
\{ T_f + J_Y , C \} &= - \CL_Y C -  e^{-\Phi} f ,  \\
\{ T_f + J_Y , M \} &= - \left( \CL_Y  + \frac{5}{4} D \cdot  Y \right) M   , \\
\{ T_f + J_Y , \chi^A \} &= - Y^B  \p_B \chi^A , \\
\{ T_f + J_Y , N_A \} &=  - ( \CL_Y + D \cdot Y ) N_A - ( f \p_A  + 5 \p_A  f ) M  , \\ 
\end{split}
\end{equation}
We shall refer to the transformations that these charges generate as ``GBMS transformations''. These must be distinguished from the GBMS diffeomorphisms which act as \eqref{BMStransform-S0-1} with \eqref{eq:omega}. GBMS transformations differ from GBMS diffeomorphisms by their action on the $N_A$! Indeed, the ``difference'' between the transformation generated by the charges and GBMS diffeomorphisms is
\begin{equation}
    \begin{split}
    \{ T_f + J_Y , N_A \} &= - \d_\xi N_A - ( \d_\xi - \CL_Y + 4 \o )  P_A  . 
    \end{split}
\end{equation}
As expected, the ``extra'' term in the GBMS transformation of $N_A$ (as compared to GBMS diffeomorphisms) is precisely the obstruction term in \eqref{Hxi-var-1}.

\section{GBMS Ward Identities and Soft Theorems}\label{sec:ward}

In this section, we construct the semi-classical Hilbert space by quantizing the classical phase space constructed in Section \ref{sec:sympform}. This involves elevating all functions on the phase space to operators, Poisson brackets to quantum commutators, $\{ ~,~ \} \to \frac{1}{i} [ ~ , ~ ]$ and complex conjugation to the adjoint. The Hilbert space furnishes an irreducible representation of the Poisson bracket algebra \eqref{h2-brackets}--\eqref{N-brackets}. The procedure implemented here is similar in structure to the one described in Section 4 of \cite{He:2020ifr}.

\subsection{Quantum Hilbert Space}
\label{sec:HilbertSpace}

We start by discussing the vacuum sector. This is spanned by the operators $C(x)$, $M(x)$, $\chi^A(x)$ and $N_A(x)$. The operators $C$ and $\chi^A$ commute {\eqref{C-brackets}} so we can construct a basis which simultaneously diagonalizes these operators,
\begin{equation}
\begin{split}
C(x) \ket{ {\bar C} , {\bar \chi} } = {\bar C}(x) \ket{ {\bar C} , {\bar \chi} }  , \qquad \chi^A(x) \ket{ {\bar C} , {\bar \chi} } = {\bar \chi}^A(x) \ket{ {\bar C} , {\bar \chi} } . 
\end{split}
\end{equation}
The brackets \eqref{C-brackets}--\eqref{N-brackets} can then be used to deduce the action of $M$ and $N_A$ on these states
\begin{equation}
\begin{split}
M(x) \ket{ {\bar C} , {\bar \chi} } &= 4\pi G i e^{-{\bar \Phi}(x)} \frac{\d}{\d{\bar C}(x)} \ket{ {\bar C} , {\bar \chi} } , \qquad N_A(x) \ket{{\bar C},{\bar \chi}} = 16\pi G i \p_A {\bar \chi}^B(x) \frac{\d}{\d{\bar \chi}^B(x)} \ket{{\bar C},{\bar \chi}}  . 
\end{split}
\end{equation}
Generic vacuum states are \emph{not} invariant under the Poincar\'e subalgebra of the extended BMS algebra {(discussed in Section \ref{sec:important-subalgebras})}. However, there is a unique (up to the identification \eqref{chi-phi-identification}) vacuum state -- namely the one with ${\bar C} = 0$ and ${\bar \chi}^A = x^A$ which is invariant! We shall refer to this particular vacuum state as the QFT vacuum,
\begin{equation}
\begin{split}\label{QFT-vac}
\ket{\O_\text{QFT}} = \ket{{\bar C}(x) = 0 , {\bar \chi}^A(x) = x^A }.
\end{split}
\end{equation}
We next turn to the radiative sector which is spanned by ${\wt h}_{\2AB}$. Define the operator
\begin{equation}
\begin{split}\label{OAB-def}
O_{AB}(\o,x) \equiv - \frac{8\pi^2}{\o \k } q^{CD} \d_{C(A} \int \dt u e^{i \o u } {\wt h}_{\2B)D} (u,x)  ,  \qquad \k = \sqrt{32\pi G}  . 
\end{split}
\end{equation}
Before proceeding, let us quickly comment on the funny tensor structure in front of the integral above. We will see shortly that $O_{AB}$ is a graviton annihilation operator (which we will conclude from its commutator with $O^\dagger_{AB}$). Basic representation theory of the six-dimensional Poincar\'e group implies that this transforms in the symmetric traceless representation of the little group $SO(4)$. On the other hand, ${\wt h}_\2$ transforms as a 2-tensor of $GL(4)$. Consequently, in order to determine the relationship between $O_{AB}$ and ${\wt h}_{\2AB}$, we need to understand how one can embed representations of $SO(4)$ into those of $GL(4)$. This is trivial to do for any representation that doesn't explicitly rely on the invariant tensor of $SO(4)$, namely $\d_{AB}$. For example, the vector rep. of $SO(4)$ trivially embeds into the vector rep. of $GL(4)$. In the case at hand, the ``traceless'' symmetric rep. \emph{does} depend on $\d$ as the ``tracelessness'' property is defined w.r.t. $\d$. Consequently, any attempt to embed this into a tensor of $GL(4)$ while preserving the symmetry structure is futile.  The way to get around this is to consider instead the operator $O^A{}_B \equiv \d^{AC} O_{CB}$ which transforms in the (1,1) representation. This operator is also traceless but w.r.t. the metric $\d^A_B$ which is also the invariant tensor for $GL(4)$! It is then clear that we can embed $O^A{}_B \sim \int \dt u e^{i\o u} {\wt h}^A_{\2B}$ and preserve the traceless property on both sides of the equation. Lowering the index on the LHS with $\d$ and on the RHS with $q$ as usual, we reproduce \eqref{OAB-def}.

From \eqref{h2-brackets}, we can determine the algebra of these operators (for $\o,\o'>0$)
\begin{equation}
\begin{split}\label{O-comm}
\left[ O_{AB}(\o,x) , O_{CD}^\dagger(\o',x') \right] &= M_{AB,CD} ( 2  \o^{-3} )  (2\pi)^5  \d^4(x,x') \d ( \o - \o' ) .
\end{split}
\end{equation}
where $M_{AB,CD} = \frac{1}{2} q^{PQ} \d_{P\{A} q_{B\}\{C} \d_{D\}Q} + \frac{1}{2} \d_{A\{C} \d_{D\}B}$. The unusual tensor structure appearing in the brackets is due to the tensor structure in front of the integral in our definition \eqref{OAB-def}. We will see that this tensor structure will be necessary to reproduce the subleading-soft theorem correctly.

Though not obvious, this is precisely the algebra of momentum space creation and annihilation operators when evaluated in the QFT vacuum. To see this, we parameterize a massless momentum as we did in \eqref{mompar}
\begin{equation}
\begin{split}\label{mompar-1}
p^\mu(\o,x) = \o {\hat p}^\mu(x) , \qquad {\hat p}^\mu(x)  = \left( \frac{1+x^2}{2} , x^A , \frac{1-x^2}{2} \right) .
\end{split}
\end{equation}
It is then easy to see that \eqref{O-comm} is equivalent to
\begin{equation}
\begin{split}
[ O_{AB}(p) , O_{CD}^\dagger (p') ] |_\text{QFT} &=   \d_{A\{C} \d_{D\}B} (2\pi)^5  ( 2p^0) \d^5(\vec{p} - \vec{p}\,' ) . 
\end{split}
\end{equation}
Here, the subscript QFT indicates that the commutators have been evaluated and then simplified by evaluating them in the QFT vacuum state \eqref{QFT-vac}. 

From the result above, we can immediately conclude that $O_{AB}(p)$ is an annihilation operator for a graviton with momentum $p^\mu$ and polarization tensor $\ve_{AB}$ given by \eqref{mompar},
\begin{equation}
\begin{split}
\ve_{AB}^{\mu\nu}(p) = \frac{1}{\o^2} \p_{\{A} p^\mu \p_{B\}} p^\nu  . 
\end{split}
\end{equation}
The annihilation operators kill all the vacuum states
\begin{equation}
\begin{split}\label{vac-def}
O_{AB}(p) \ket{{\bar C},{\bar \chi}} = 0 . 
\end{split}
\end{equation}
Excited states are constructed by acting on the vacuum state with a creation operator, e.g. a one-graviton state with polarization tensor $\ve_{AB}^{\mu\nu}(p)$ is described by
\begin{equation}
\begin{split}
\ket{ p , \ve \, ; {\bar C} , {\bar \chi} } \equiv O^\dagger_{AB}(p) \ket{ {\bar C}, {\bar \chi}} . 
\end{split}
\end{equation}

\subsection{Matching Conditions and Ward Identity}
\label{sec:matchingcond}

So far, we have restricted the discussion to the future null infinity $\ci^+$. In order to discuss the scattering problem in general relativity, we also need to consider the structure of the theory on $\ci^-$. This is almost entirely identical to the analysis performed in the bulk of this paper so we will not discuss this in detail. For our purposes, it will be sufficient to note that the Hilbert space on $\ci^-$ is spanned by an analogous set of operators ${\wt h}_{\2AB}^-(v,x)$, $C^-(x)$, $M^-(x)$, $\chi^{-A}(x)$ and $N^-_A(x)$ where $v$ is the null coordinate on $\ci^-$. The last four boundary fields are defined on the boundary of $\ci^-$, namely $\ci^-_+$.

We conjecture that the boundary fields on $\ci^+_-$ and $\ci^-_+$ satisfy the following matching condition
\begin{equation}
\begin{split}
\label{eq:matchingcond}
C &= \Upsilon^\star C^- , \qquad M = \Upsilon^\star M^- , \qquad \chi^A = \Upsilon^\star \chi^{-A} , \qquad N_A = - \Upsilon^\star N_A ^-, 
\end{split}
\end{equation}
where $\Upsilon^\star$ indicates the antipodal map on the celestial $S^4$.\footnote{The antipodal point is defined by embedding the $S^4$ in $\mrr^5$, $x \hookrightarrow \vec{X}(x)$ and we then have $\vec{X}(P(x)) = - \vec{X}(x)$. For the round $S^4$ and stereographic embedding $\vec{X}(x) = \left( \frac{2x^A}{1+x^2} , \frac{1-x^2}{1+x^2} \right)$, the antipodal map is defined by $P^A(x)=-\frac{x^A}{x^2}$.} A proof of these matching conditions rests on the analysis of the asymptotic structure at spacelike infinity and this has only been accomplished in four-dimensional spacetimes \cite{Henneaux:2018cst, Henneaux:2018hdj,Prabhu:2019fsp,Henneaux:2019yax,Prabhu:2021cgk,Capone:2022gme,Compere:2023qoa}. The only discussion of this in higher dimensions (specifically $D=5$) that we are aware of is the Hamiltonian analysis of \cite{Fuentealba:2022yqt} where the existence of BMS at spacelike infinity was studied (which is the first step in the analysis of the matching conditions).  These analyses suggest that a similar structure likely extends to six (and all higher dimensions) in which the matching conditions \eqref{eq:matchingcond} are realised.\footnote{In four dimensions, to give the matching when the mass is non-vanishing, logarithmic supertranslations have to be used \cite{Troessaert:2017jcm, Compere:2023qoa}. This is a non-trivial point whose extension to higher dimensions is impossible to assess without explicit analysis.}

From \eqref{eq:matchingcond}, we find equality for the GBMS charges {\eqref{GBMScharges}}
\begin{equation}
\begin{split}\label{charge-equal}
T_f = T^-_{\Upsilon^\star f} , \qquad J_Y = J^-_{-\Upsilon^\star Y} . 
\end{split}
\end{equation}
In the semi-classical theory, this implies a very simple Ward identity 
\begin{equation}
\begin{split}\label{Ward-Identity}
\bra{\text{out}} T_f \CS - \CS T^-_{\Upsilon^\star f}\ket{\text{in}} = 0  , \qquad \bra{\text{out}}  J_Y \CS - \CS J^-_{- \Upsilon^\star Y} \ket{\text{in}} = 0  . 
\end{split}
\end{equation}
Here, $\CS$ is the $S$-matrix operator that maps from the Hilbert space on $\ci^-$ to that on $\ci^+$ and $\bra{\text{out}}$, $\ket{\text{in}}$ are the respective ``non-interacting'' multi-particle states that we are scattering. 

In the rest of this section, we process \eqref{Ward-Identity} into a ``soft theorem'' which we can then compare to the usual soft theorem derived in standard QFT via Feynman diagrams \cite{Weinberg:1965nx,Cachazo:2014fwa}. To do this, we rewrite the charges $T_f$ and $J_Y$  {\eqref{GBMScharges}} as integrals over all of $\ci^+$ using the boundary conditions \eqref{sec:MNAdef} for ${\wt U}_\3$ and ${\wt W}_\5$,
\begin{equation} \label{GBMScharges-uint}
\begin{split}
T_f &= \frac{1}{4\pi G} \int \dt^4 x \sqrt{q} f M(x)  = \frac{1}{4\pi G} \int_{\ci^+} \dt u  \dt^4 x \sqrt{q} f \p_u {\wt U}_\3 , \\
J_Y &= \frac{1}{16\pi G} \int \dt^4 x \sqrt{q} Y^A N_A (x) \\
&= - \frac{5}{16\pi G} \int_{\ci^+} \dt u \dt^4 x \sqrt{q} Y^A u \p_u^2 {\wt W}_{\5A} - \frac{1}{16\pi G} \int \dt^4 x \sqrt{q} Y^A P_A (x)  , \\
\end{split}
\end{equation}
and similarly for the charges on $\ci^-$.

The next step is to use the $u$-evolution equations \eqref{app:nolog-U3du} and \eqref{app:nolog-W5du} to ``simplify'' the charges. Since we have all the relevant equations explicitly worked out, it would be possible to do this and derive the most general form of the soft theorem. However, our main goal is to derive a soft theorem that we can match to the one derived in QFT. Consequently, we shall simplify our work and consider only the special case where the $in$ and $out$ vacuum states are the QFT vacuum \eqref{QFT-vac}. We shall also assume that $\Phi_0(x) = 0$ for simplicity.\footnote{$\Phi_0(x)$ can be incorporated in QFT by choosing the parameterization of null momenta as $p^\mu = \o e^{\Phi_0(x)} {\hat p}^\mu(x)$. We set $\Phi_0(x)=0$ for simplicity so that we can match our results to the soft theorems determined in \cite{Kapec:2017gsg,Kapec:2021eug}.}

In this state, $q_{AB} = \d_{AB}$ and $h_{\1AB}=0$ so the evolution equations for ${\wt U}_\3$ and ${\wt W}_\5$ simplify to \eqref{U3-evolution-eq} which we recall here
\begin{equation}
\begin{split}
\p_u {\wt U}_\3 &= \frac{1}{24} \p^2 \p^A \p^B {\wt h}_{\2AB} + \frac{1}{8} \p_u {\wt h}_{\2AB} \p_u  {\wt h}_\2^{AB}  , \\
5 \p_u {\wt W}_{\5} &=  \p {\wt U}_\3  + \frac{3}{4} \p^2 \p \cdot {\wt h}_{\3}  + \frac{1}{2} \tr [ \p_u {\wt h}_\2 \p  {\wt h}_\2 ]  + \frac{1}{2} \p \cdot (   {\wt h}_\2 \overleftrightarrow{\p_u} {\wt h}_\2 )  \\
&\qquad \qquad \qquad \qquad \qquad \qquad \qquad  + \p_u \left( {\wt h}_\2 \p \cdot {\wt h}_\2  -  \frac{1}{16} \p \tr[ ( {\wt h}_\2^{BC} )^2 ] \right) . 
\end{split}
\end{equation}
Using this, we can rewrite the charge in the form
\begin{equation}\label{GBMScharges_sh}
\begin{split}
T_f = T_f^\text{S} + T_f^\text{H} , \qquad J_Y = J_Y^\text{S} + J_Y^\text{H} , 
\end{split}
\end{equation}
where
\begin{equation}
\begin{split}
T_f^\text{S} &= \frac{1}{3\k^2} \int \dt^4 x   f(x) \p^2 \p^A \p^B \int \dt u  {\wt h}_{\2AB}(u,x) , \\
T_f^\text{H} &=  \frac{1}{\k^2} \int  \dt^4 x  f \int \dt u \tr [ ( \p_u {\wt h}_\2 )^2 ] ,  \\
J_Y^\text{S} &= \frac{16\pi^2}{\k^2} \int \dt^4 x Y^A(x)  \mdd^B  \int \dt u u {\wt h}_{\2AB} (u,x)  , \\
J_Y^\text{H} &= \frac{1}{\k^2} \int \dt^4 x Y^A \int \dt u \left( \tr [ \p_u {\wt h}_\2 \p_A {\wt h}_\2 ]  +  [ \p \cdot( {\wt h}_\2 \overleftrightarrow{\p_u} {\wt h}_\2 )  ]_A  - \frac{1}{4} u \p_A \tr [ ( \p_u {\wt h}_\2 )^2 ]  \right) . 
\end{split}
\end{equation}
where the derivative operator $\mdd$ was defined in equation (4.27) of \cite{Kapec:2017gsg}, 
\begin{equation}
\begin{split}
{\mathbb D}^B O_{AB} &\equiv  \frac{1}{64\pi^2} \left[  (-\p^2 )^2 \p^B O_{AB}  + \frac{4}{3} (-\p^2) \p_A \p^B \p^C O_{BC} \right] ~. 
\end{split}
\end{equation}
We can similarly simplify the charges on $\ci^-$ as well. We can then rewrite the Ward identities \eqref{Ward-Identity} as
\begin{equation}
\begin{split}\label{Ward-id-1}
\bra{\text{out}} T^\text{S}_f \CS - \CS T^{-\text{S}}_{\Upsilon^\star f}\ket{\text{in}} &= - \bra{\text{out}} T^\text{H}_f \CS - \CS T^{-\text{H}}_{\Upsilon^\star f}\ket{\text{in}}  , \\
\bra{\text{out}}  J_Y^\text{S} \CS - \CS J^{-\text{S}}_{- \Upsilon^\star Y} \ket{\text{in}} &= - \bra{\text{out}}  J_Y^\text{H} \CS - \CS J^{-\text{H}}_{- \Upsilon^\star Y} \ket{\text{in}}  .
\end{split}
\end{equation}
Using the definitions \eqref{OAB-def} and \eqref{soft-exp-1}, the soft charges can be written as
\begin{equation}
\begin{split}
T_f^\text{S} &= - \frac{1}{24\pi^2\k} \int \dt^4 x f \p^2 \p^A \p^B \CG_{AB} , \qquad J_Y^\text{S} = \frac{2i}{\k} \int \dt^4 x Y^A\mdd^B \CB_{AB}  .
\end{split}
\end{equation}
We immediately see that the soft charges $T^\text{S}_f$ and $J^\text{S}_Y$ insert a single \emph{soft}-graviton into the $S$-matrix so the LHS of \eqref{Ward-id-1} is the soft limit of a graviton amplitude. To simplify the RHS, we recall that -- by construction -- the charges $T_f$ and $J_Y$ {\eqref{GBMScharges}} generate GBMS transformations on the fields as in \eqref{TJ-action}. From this, we find
\begin{equation}\label{OAB-trans}
\begin{split}
\left[ T_f , O_{AB} \right] &= - \o f(x) O_{AB} , \\
\left[ J_Y , O_{AB} \right] &= - i \left[ Y^C \p_C  - \frac{1}{4} \p \cdot  Y   \o \p_\o  + \frac{i}{2}  \p^C Y^D(x_i) \CS_{i\,CD}  \right]  O_{AB} . 
\end{split}
\end{equation}
where $\CS_{AB}$ is the spin-operator for the graviton. Explicitly, 
\begin{equation}
\begin{split}
\label{graviton-spin}
( \CS_{AB} )_{CD}{}^{EF} &= - \frac{i}{2} \left[ ( \d_B^E \d_{AC} - \d_A^E \d_{BC}  ) \d_D^F  + \d^E_C ( \d_B^F \d_{AD} - \d_A^F \d_{BD} ) \right. \\
&\left. \qquad \qquad \qquad \qquad + ( \d_B^E \d_{AD} - \d_A^E \d_{BD} ) \d_C^F  + \d^E_D ( \d_B^F \d_{AC}  - \d_A^F \d_{BC} ) \right]  . 
\end{split}
\end{equation}
The precise set of terms that appear in the superrotation transformation of $O_{AB}$ in {\eqref{OAB-trans}} are due to the specific tensor structure we have in front of the integral in \eqref{OAB-def}. 

Putting all of this together, we can recast the Ward identities \eqref{Ward-id-1} into the form
\begin{equation}
\begin{split}\label{WardID-leading}
\int \dt^4 x f(x)  \p^2 \p^A \p^B \bra{\text{out}} \CG_{AB}(x) \CS \ket{\text{in}} &=  24\pi^2\k \sum_i \o_i f(x_i) \bra{\text{out}} \CS \ket{\text{in}}  , \\
\end{split}
\end{equation}
and
\begin{equation}
\begin{split}
\label{WardID-subleading}
&\int \dt^4 x Y^A(x) \mdd^B  \bra{\text{out}} \CB_{AB} (x) \CS \ket{\text{in}} \\
&\qquad \qquad = - \frac{\k}{2} \sum_i \bigg( Y^A(x_i) \p_{x_i^A}  - \frac{1}{4} \p_A Y^A(x_i) \o_i \p_{\o_i}  + \frac{i}{2}  \p^C Y^D(x_i) \CS_{i\,CD} \bigg) \bra{\text{out}}  \CS\ket{\text{in}}  .
\end{split}
\end{equation}

\subsection{Soft Graviton Theorems}
\label{sec:soft-theorems}

The universal factorization of a scattering amplitude with external graviton states was first studied by Weinberg in 1965 \cite{Weinberg:1965nx} who showed that
\begin{equation}
\begin{split}\label{soft-exp}
\bra{\text{out}} O_{AB}(p) \CS \ket{\text{in}} ~ \xrightarrow{p \to 0}~ \frac{\k}{2} \sum_i \frac{ p_{i\mu} p_{i\nu} \ve^{\mu\nu}_{AB}(p) }{ p_i \cdot p} \bra{\text{out}} \CS \ket{\text{in}} + \CO(1) . 
\end{split}
\end{equation}
In particular, this so-called \emph{leading soft-graviton theorem} showcases a universal factorization of the scattering amplitude in the soft ($p\to0$) limit. The ``soft-factor'' does not depend on any of the details of the theory (such as the explicit interaction terms or field content) -- rather it only depends on the quantum numbers of the external states in the scattering amplitude. This universal structure plays a crucial role in resolving the infrared divergences of four-dimensional gravitational theories.

Surprisingly, it wasn't until 2014 \cite{Cachazo:2014fwa} that it was finally understood that the $\CO(1)$ term in \eqref{soft-exp} is, in fact, also universal in the same way! Including this universal piece, the complete soft limit takes the form
\begin{equation}
\begin{split}\label{soft-exp-2}
\bra{\text{out}} O_{AB}(p) \CS \ket{\text{in}} ~ \xrightarrow{p \to 0}~ \frac{\k}{2} \sum_i \left[ \frac{ p_{i\mu} p_{i\nu} \ve^{\mu\nu}_{AB}(p) }{ p_i \cdot p}  - i \frac{\ve_{AB}^{\mu\nu}(p) p_{i\mu} p_\rho}{ p_i \cdot p} \CJ_i^{\rho\nu} \right] \bra{\text{out}} \CS \ket{\text{in}} + \CO(p).
\end{split}
\end{equation}
where $\CJ_i$ is the angular momentum operator acting on the $i$th one-particle state.

In this section, we show that \eqref{soft-exp-2} is equivalent to the BMS Ward identities \eqref{WardID-leading} and \eqref{WardID-subleading}. This equivalence has already been studied in previous works \cite{Kapec:2015vwa,Kapec:2017gsg,Kapec:2021eug,Chowdhury:2022gib,Chowdhury:2022nus} but in each case, certain restrictive assumptions were made. \cite{Kapec:2015vwa} proved the equivalence of the leading soft theorem to the supertranslation Ward identity for linearised gravity. \cite{Chowdhury:2022nus} generalised the result to non-linear GR but on a smaller phase space with no superrotations. Starting from the subleading soft theorem and the linear analysis of \cite{Kapec:2015vwa}, \cite{Colferai:2020rte} proposed the form of one bracket among asymptotic fields to reproduce the soft theorem.  Finally, \cite{Chowdhury:2022gib} working in non-linear gravity (but linearising around a Bondi frame) showed the equivalence for the subleading soft theorem but for the special case where all external states except the graviton are scalars. With all the pieces at hand, we are now able to prove the equivalence between soft theorems and BMS Ward identities in full generality. 

The first step is to recast the soft-theorem {\eqref{soft-exp-2}} into a more palatable form using the momentum parameterization \eqref{mompar-1}. Written in terms of the operators $\CG_{AB}(x)$ and $\CB_{AB}(x)$ {(defined in \eqref{soft-exp-1}}, the soft theorem takes the form (see \cite{Kapec:2021eug} and \cite{Kapec:2017gsg} respectively)
\begin{equation}
\begin{split}\label{leadingST}
\bra{\text{out}} \CG_{AB}(x) \CS \ket{\text{in}} = - \frac{\k}{2}  \left( \p_A \p_B - \frac{1}{4} \d_{AB} \p^2 \right) \sum_i \o_i [ - {\hat p}_i \cdot {\hat p}(x) ] \ln [ - {\hat p}_i \cdot {\hat p}(x) ] \bra{\text{out}} \CS \ket{\text{in}} ,
\end{split}
\end{equation}
and
\begin{equation}
\begin{split}\label{subleadingST}
\bra{\text{out}} \CB_{AB}(x) \CS \ket{\text{in}} &= \frac{\k}{2}  \sum_i \left[ \CP^C{}_{AB}(x-x_i) \p_{x_i^C} + \frac{1}{4} \p_C \CP^C{}_{AB}(x-x_i) \o_i \p_{\o_i} \right. \\
&\left. \qquad \qquad \qquad \qquad \qquad \qquad - \frac{i}{2} \p^{[C} \CP^{D]}{}_{AB}(x-x_i) \CS_{i\,CD} \right]\bra{\text{out}} \CS \ket{\text{in}} ,
\end{split}
\end{equation}
where
\begin{equation}
\begin{split}
\CP^C{}_{AB}(x) \equiv \frac{1}{2} \left[ x_A \d^C_B + x_B \d^C_A + \frac{1}{2} \d_{AB} x^C - \frac{4}{x^2} x_A x_B x^C \right] . 
\end{split}
\end{equation}
$\CS_{i\,AB}$ is the representation of $SO(d)$ under which the $i$th particle transforms.

To prove the equivalence, we substitute \eqref{leadingST} and \eqref{subleadingST} into the LHS of \eqref{WardID-leading} and \eqref{WardID-subleading} respectively. We start with the leading soft theorem. We first find
\begin{equation}
\begin{split}
\p^A \p^B \bra{\text{out}} \CG_{AB}(x) \CS \ket{\text{in}} = - 6 \k  \sum_i \frac{\o_i}{(x-x_i)^2} \bra{\text{out}} \CS \ket{\text{in}}.
\end{split}
\end{equation}
Then, using the property
\begin{equation}
\begin{split}
\p^2 \left( \frac{1}{x^2} \right) = - 4\pi^2 \d^4(x)  ,
\end{split}
\end{equation}
the first equation of \eqref{WardID-leading} immediately follows. The equivalence for the subleading soft theorem similarly follows if we use the property (derived in \cite{Kapec:2017gsg})
\begin{equation}
\begin{split}
\mdd^B \CP^C{}_{AB}(x) = - \d^C_A \d^4(x).
\end{split}
\end{equation}

\section{Conclusions and Outlook}
\label{sec.conclusions}

In this paper, we have systematically analysed supertranslations, superrotations and asymptotic Weyl rescalings in six-dimensional
asymptotically flat spacetimes. All the steps of our analysis are generalizable to \emph{any} number of even dimensions greater than four with only
minor changes, resulting at most in different numerical factors.  For example, the flatness condition simply reads in general
\begin{equation}
r^d e^\mu_{\hat \mu} e^\nu_{\hat \nu} e^\rho_{\hat \rho} e^\s_{\hat \s} R_{\mu\nu\rho\s}[G] |_{\ci^+_-} = 0.
\end{equation}
On the other hand, the odd-dimensional case is more delicate and we can only expect that some of the conditions and ideas presented here carry over. This is essentially due to the fact that Huygens' principle applies in even dimensions, but not in odd dimensions. More precisely, in even dimensions, Green's functions for massless particles are localized on the light-cone implying a simple and local asymptotic structure. In odd dimensions, Green's functions are supported on the interior of the light-cone which leads to various non-localities on $\ci^\pm$. This was explored for gauge theories in \cite{He:2019pll,He:2019jjk,He:2019ywq} but the analogous structure for gravity is yet to be understood. Given the results of \cite{Kapec:2017gsg,Kapec:2021eug}, it is reasonable to conjecture that the local structure in even dimensions and the non-local structure in odd dimensions can be treated in a unified manner by utilizing the shadow transform. We leave an exploration of this for future work.

It would also be of utmost interest to analyse the consequences of the phase space prescriptions put forward in this paper in four dimensions. In order to comment on this while concluding, let us summarise our results. 

We started from the general solution space derived in \cite{Capone:2021ouo} and we have
restricted it by requiring flatness at the past boundary of future null infinity. Such conditions generate 
analytic expansions in powers of $r^{-1}$. We have studied both large $r$ and large $u$  divergences appearing in the pre-symplectic potential $\theta$ and symplectic form $\Omega$. The large $r$ divergences were renormalised by adding local and covariant counterterms in $\theta$ and the explicit large $u$ were removed by requiring that $\Omega$ be non-degenerate. The non-degeneracy requirement removes the asymptotic Weyl rescalings and reduces the ASG from WBMS transformations to the GBMS transformations, which consist only of supertranslations and 
superrotations. We then constructed the Poisson brackets from the symplectic form so obtained.

We also found from this that the symplectic form for GBMS transformations is non-integrable in field space, namely $\Omega(\delta,\CL_\xi) \neq \delta H_\xi$ and hence there is no unambiguous notion of finite charges. In other words, GBMS diffeomorphisms are  not canonical. This result is expected from previous works on four-dimensional gravity \cite{Wald:1999wa, Barnich:2011mi, Compere:2018ylh}. It is, however, possible to extract the \emph{integrable part} of the symplectic form and define charges which faithfully represent the GBMS algebra. Such charges act consistently on the solution space but their action on the fields is \emph{not} the Lie derivative $\mathcal{L}_\xi$.

The renormalization of the large $r$ divergences with local and covariant counterterms and our treatment of large $u$ divergences in higher dimensions appear to be novel. It also seems that our analysis will imply something new for four-dimensional gravity as well. Previous discussions of renormalization do not discuss covariantisation of the counterterms and remain vague about
the large $u$ behaviour and invertibility \cite{Compere:2018ylh, Chandrasekaran:2021vyu,Freidel:2021fxf} (see however also \cite{Flanagan:2019vbl}) . It is important to check that the analysis performed in this paper applies in a straightforward way to four dimensions as well and we leave this exploration for future work. Let us highlight some points in this regard:
\begin{itemize}
\item The addition of local and covariant counterterms at diverging orders also
 modifies the finite part of the symplectic form and hence affects the parts we identify as integrable 
and non-integrable. The resulting split of $\delta H_\xi$ into integrable $[\delta H_\xi]^{\text{int}}$
 and non-integrable $[\delta H_\xi]^{\text{non-int}}$ parts appears to be less
ambiguous because it is obtained from this sort of first-principle prescription, although 
ambiguities remain if we insist that we can shift $\delta H_\xi$ as
\begin{equation}\label{WZ}
\delta H_\xi = \Omega(\delta,\CL_\xi)+\int_{\text{corner}} i_\xi F(\delta).
\end{equation}
This equation is equivalent to saying that we can split $\Omega(\delta,\CL_\xi)$ into integrable and non-integrable parts. 
In previous literature \cite{Wald:1999wa} $F$ -- and hence the integrable part of the symplectic form -- was selected by imposing that it is zero in the absence of news \cite{Wald:1999wa, Compere:2018ylh, Chandrasekaran:2021vyu, Donnay:2022hkf}. In the case in which only supertranslations are allowed, $F$ is related to the passage of gravitational waves through null infinity and the condition acquires
a clear physical meaning.

\item 
The physics of the non-integrable term becomes slightly confusing in four bulk dimensions when superrotations are included because the GBMS Goldstone modes and the radiative modes appear at the same order in the large $r$ expansion. The non-integrability of GBMS charges \eqref{Hxi-var-1} at $\ci^+_-$ (not at generic cuts of $\ci$) is \emph{always} associated to superrotations, \emph{not} to physical radiation (which is excluded on $\ci^+_-$ by the conditions defining the phase space). This is true in all (even) dimensions. In particular, in four dimensions, the reader can check that equation (5.40) of \cite{Compere:2018ylh} does \emph{not} necessarily vanish at $\ci^+_-$ in the absence of radiation because in a generic superrotation frame the proper news tensor gets corrected by the trace-free part of the Geroch tensor, denoted there by $N_{AB}^\text{vac}$. It is important to note that this issue is \emph{not} resolved by shifting the integrable part of the Hamiltonian as done in equation (5.41) of \cite{Compere:2018ylh}. Indeed, one can check that implementing this shift with equation (5.48) does not cause the last two terms of (5.41) to vanish in the absence of radiation.

\item We obtained the Poisson structure by inverting the gravitational symplectic form obtained with covariant phase space methods. Given that the finite charges are obtained only based on the integrable part of
$\Omega$, $\mathcal{L}_\xi$ is not the Hamiltonian vector field associated to such charges. Future analysis could study the
algebraic structures stemming from the brackets proposed by Barnich and Troessaert \cite{Barnich:2011mi},
which in the case of \eqref{WZ} read as
\begin{equation}
\{H_\xi,H_{\xi'} \} \equiv - \Omega(\CL_{\xi},\CL_{\xi'})+ \int_{\text{corner}} (i_\xi F(\CL_{\xi'})-i_{\xi'} F(\CL_\xi) ).
\end{equation}
Such brackets are not naively related to a symplectic form \cite{Wieland:2021eth}, but we stress that many of our conclusions - for example, the removal of the Weyl freedom - are highly motivated by the systematic analysis of the symplectic form output by the covariant phase space formalism.
\end{itemize}
In reflecting on these matters, it would be interesting to include also the logarithmic terms identified in \cite{Capone:2021ouo} (briefly discussed here in Section \ref{sec:summary1}) both in four and higher even dimensions to have a more detailed and general understanding of asymptotically flat gravitational phase spaces and their holographic properties.

\section*{Acknowledgements} 
We would like to thank Anupam A.H., Goncalo Araujo-Regado, Chandramouli Chowdhury, Adrien Fiorucci, Temple He, Daniel Kapec, Rifath Khan, Filipe S. Miguel, Enrico Parisini, Romain Ruzziconi, Kostas Skenderis, Marika Taylor and Zihan Yan  for many useful conversations. F.C. kindly thanks DAMTP, Cambridge University for their hospitality. The research of F.C. is funded by the Deutsche Forschungsgemeinschaft (DFG) under Grant No. $406116891 $ within the Research Training Group RTG $2522/1$, he also acknowledges support from the EPSRC Doctoral Prize Award EP/T517859/1 in the first part of this project. P.M. gratefully acknowledges support from the STFC consolidated grants ST/P000681/1 and ST/T000694/1 and thanks the STAG Research Center at the University of Southampton and TIFR, Mumbai for their hospitality. A.P. is supported by the National Research Foundation of Korea under the grants, NRF-2022R1A2B5B02002247, NRF-2020R1A2C1008497, he also acknowledges support from an Engineering and Physical Sciences Research Council (EPSRC) Mathematical Sciences Fellowship at the University of Southampton in the first part of this project. B.T. acknowledges support from the Cambridge Trust, King's College, Cambridge and STFC consolidated grants ST/P000673/1 and ST/T00049X/1 and thanks  the STAG Research Center at the University of Southampton for their hospitality.

\appendix

\section{Ricci Tensor Components in Bondi-Sachs Gauge}

In this appendix, we work out the Riemann tensor in Bondi-Sachs gauge in general $D=d+2$ dimensions. The six-dimensional formulas are obtained by setting $d=4$. 

\subsection*{Metric and Inverse}

The line element takes the form
\begin{equation}
\begin{split}\label{app:Bondigaugemetric}
G_{\mu\nu} \dt X^\mu \dt X^\nu = -  e^{2\b} U \dt u^2 - 2 e^{2\b} \dt u \dt r + r^2 h_{AB} ( \dt x^A - W^A \dt u ) ( \dt x^B - W^B \dt u )  ,
\end{split}
\end{equation}
with
\begin{equation}
\begin{split}\label{app:detcond}
\det(h_{AB}) = \det(q_{AB}) \equiv q . 
\end{split}
\end{equation}
The Bondi-Sachs gauge condition implies $G_{rr} = G_{rA} = G^{uu} = G^{uA} = 0$. The metric components and its inverse are given by
\begin{equation}
\begin{split}\label{app:metriccoeff}
G_{uu} &= - e^{2\b} U + r^2 h_{AB} W^A W^B , \quad G_{ur} = - e^{2\b} , \quad  G_{uA} = - r^2 h_{AB} W^B , \quad G_{AB} = r^2 h_{AB},
\end{split}
\end{equation}
and
\begin{equation}
\begin{split}\label{app:inversemetriccoeff}
G^{ur} = - e^{-2\b} , \qquad G^{rr} = e^{-2\b} U , \qquad G^{rA} = - e^{-2\b} W^A , \qquad G^{AB} = \frac{1}{r^2} {\bar h}^{AB} . 
\end{split}
\end{equation}
where ${\bar h}$ is the matrix inverse of $h$. The root determinant of the metric is 
\begin{equation}
\begin{split}\label{app:metricdet}
\sqrt{ - \det(G_{\mu\nu}) } =  r^d e^{2\b} \sqrt{q} . 
\end{split}
\end{equation}

\subsection*{Christoffel Symbols}

The Christoffel symbol is defined by
\begin{equation}
\begin{split}\label{app:Christoffeldef}
\G^\l_{\mu\nu}[G] \equiv \frac{1}{2} g^{\l\rho} [ \p_\mu G_{\nu\rho} + \p_\nu G_{\mu\rho} - \p_\rho G_{\mu\nu} ] . 
\end{split}
\end{equation}
The components of the Christoffel symbol then work out to be
\begin{equation}
\begin{split}\label{app:Christoffelu}
\G^u_{uu}[G]  &=  2 \p_u \b + \frac{1}{2} e^{-2\b} \p_r G_{uu} ,  \\
\G^u_{ur}[G] &= 0 ,   \\
\G^u_{uA}[G] &= \p_A \b + \frac{1}{2} e^{-2\b}  \p_r G_{uA} ,   \\
\G^u_{rr}[G] &= 0 ,   \\
\G^u_{rA}[G] &= 0 ,   \\
\G^u_{AB}[G] &= \frac{1}{2} e^{-2\b} \p_r ( r^2 h_{AB} ) ,   
\end{split}
\end{equation}
and
\begin{equation}
\begin{split}\label{app:Christoffelr}
\G^r_{uu}[G] &=  \frac{1}{2} e^{2\b} \p_u ( e^{-2\b} U ) + \frac{1}{2} r^2 e^{-2\b} \p_u h_{AB}  W^A W^B -  \frac{1}{2} e^{-2\b}  ( U \p_r - \CL_W ) G_{uu} ,  \\
\G^r_{u r}[G] &= \frac{1}{2} \p_r  U  - \frac{1}{2} r^2 e^{-2\b} h_{AB} W^A \p_r W^B +  ( U \p_r - \CL_W ) \b  ,  \\
\G^r_{uA}[G] &=  \frac{1}{2}  \p_A U - \frac{1}{2} r^2 e^{-2\b} \p_u h_{AB} W^B -  \frac{1}{2} e^{-2\b} ( U \p_r - \CL_W ) G_{uA} ,  \\
\G^r_{rr}[G] &= 2 \p_r \b  ,  \\
\G^r_{rA}[G] &= \p_A \b + \frac{1}{2} r^2 e^{-2\b} h_{AB}  \p_r W^B ,  \\
\G^r_{AB}[G] &= \frac{1}{2} r^2 e^{-2\b} \p_u h_{AB} - \frac{1}{2} e^{-2\b} ( U \p_r - \CL_W ) (r^2 h_{AB} )  ,   
\end{split}
\end{equation}
and
\begin{equation}
\begin{split}\label{app:ChristoffelA}
\G^A_{uu}[G] &= - \frac{1}{2} e^{-2\b} W^A [ 2 \p_u G_{ur} - \p_r G_{uu} ]  + \frac{1}{2r^2} {\bar h}^{AB} [ 2 \p_u G_{uB} - \p_B G_{uu} ] ,  \\
\G^A_{ur}[G] &= \frac{1}{2r^2} {\bar h}^{AB} [ \p_r G_{uB} - \p_B G_{ur} ] ,   \\
\G^A_{uB}[G] &=  \frac{1}{2} e^{-2\b} W^A [  \p_r G_{uB} -  \p_B G_{ur} ] + \frac{1}{2r^2} {\bar h}^{AC} [ r^2 \p_u h_{BC} + 2 \p_{[B} G_{C]u}  ] ,   \\
\G^A_{rr}[G] &= 0 ,   \\
\G^A_{rB}[G] &=  \frac{1}{2r^2} {\bar h}^{AC}  \p_r (r^2 h _{CB} ) ,   \\
\G^C_{AB}[G] &=  \G^C_{AB}[q] + \frac{1}{2} e^{-2\b} W^C \p_r (r^2 h_{AB} ) + H^C{}_{AB}  .  
\end{split}
\end{equation}
where
\begin{equation}
\begin{split}\label{app:Hcabdef}
H^C{}_{AB} \equiv \frac{1}{2} {\bar h}^{CD} [ D_A h_{BD} + D_B h_{AD} - D_D h_{AB} ]  . 
\end{split}
\end{equation}

\subsection*{Ricci Tensor}

The Ricci tensor is given by
\begin{equation}
\begin{split}\label{app:Riccidef}
R_{\mu\nu}[G] &\equiv \frac{1}{\sqrt{-G}} \p_\l \left( \sqrt{-G} \G^\l_{\mu\nu}[G] \right) - \p_\mu \p_\nu \ln \sqrt{-G}   - \G^\rho_{\mu\l}[G] \G^\l_{\nu\rho}[G]  . 
\end{split}
\end{equation}
In Bondi-Sachs gauge, the $R_{r\mu}$ components of the Ricci tensor are
\begin{align}
\label{app:Rrreq} R_{rr}[G]  &= \frac{2d}{r}  \p_r \b - \frac{1}{4} \tr [ {\bar h}  \p^2_r h]  ,  \\
\label{app:RrAeq} R_{rA}[G]  &= \big( \frac{1}{2r^d} \p_r ( r^{d+2} e^{-2\b} h \p_r W  )  - r^d \p_r ( r^{-d} D  \b ) +  \frac{1}{2} D \cdot ( {\bar h} \p_r h ) + \frac{1}{4} \tr [ \p_r {\bar h} D h ] \big)_A ,  \\
\nonumber R_{ru}[G] &=   \frac{1}{2r^d}  \p_r [ r^d e^{-2\b} \p_r ( e^{2\b} U ) ] - 2 \p_u \p_r \b  - \frac{1}{r^d}  \p_r ( r^d W D \b ) + \frac{1}{4} \tr [ \p_r {\bar h} \p_u h ] \\
\label{app:Rrueq} &\qquad \qquad \qquad  - \frac{1}{2r^d} \p_r [ r^{d+2} e^{-2\b}  W h  \p_r W  ] - \frac{1}{2r^2} D \cdot ( {\bar h}  [ \p_r ( r^2 h W  )  - 2 e^{2\b}   \p  \b ] ) . 
\end{align}
As in the rest of the paper, we have also employed matrix notation in the formulas above. Repeated indices are contracted w.r.t the metric $q_{AB}$ and $D_A$ is the covariant derivative w.r.t. $q$. 

Next, we have
\begin{equation}
\begin{split}\label{app:RABeq}
e^{2\b} R_{AB}[G] &= r^{2-\frac{d}{2}}  \p_r [ r^{\frac{d}{2}} \p_u  h_{AB} ]  + e^{2\b} R_{AB} - r^{2-d} \p_r ( r^{d-1} U ) h_{AB}  \\
&\qquad \qquad - r^2 {\bar h}^{CD}  \p_r h_{C(A}  \p_u  h_{B)D}  - 2 e^{2\b}  ( D_A D_B \b  + \p_A \b \p_B \b )  \\
&\qquad \qquad -  \frac{1}{2} r^{2-d} \p_r (    r^d U \p_r  h_{AB} ) + \frac{r^2}{2} U  {\bar h}^{CD} \p_r h_{AC}  \p_r h_{BD} \\
&\qquad \qquad + \frac{1}{2} [  r^{2-d}   \p_r ( r^d \CL_W h_{AB}  ) + 2 r D_C W^C h_{AB} + r^2 D_C  ( W^C \p_r   h_{AB} )   ]  \\
&\qquad \qquad - \frac{r^4}{2}  e^{-2\b}  h_{AC} h_{BD} \p_r W^C    \p_r W^D  +  D_C ( e^{2\b}  H^C{}_{AB} )  - e^{2\b} H^C{}_{AD} H^D{}_{BC} \\
&\qquad \qquad - r^2 {\bar h}^{CD} \p_r h_{C(A} [ W^E D_E h_{B)D} + h_{B)E} D_D W^E ]  , 
\end{split}
\end{equation}
Here, $R_{AB}$ is the Ricci tensor w.r.t. $q_{AB}$. From this, we can deduce that
\begin{equation}
\begin{split}\label{app:RABtreq}
e^{2\b} {\bar h}^{AB} R_{AB}[G]  &= - d r^{2-d} \p_r ( r^{d-1} U) + e^{2\b} {\bar h}^{AB} R_{AB} - 2 e^{2\b}  {\bar h}^{AB}  ( D_A D_B \b  + \p_A \b \p_B \b )  \\
&\qquad \qquad +  r^{2-2d} \p_r ( r^{2d} D_C W^C )   - \frac{1}{2}  r^4 e^{-2\b} h_{AB} \p_r W^A \p_r W^B \\
&\qquad \qquad - D_A ( e^{2\b} D_B {\bar h}^{AB} )   -  \frac{1}{2} e^{2\b} D_C {\bar h}^{AB}  H^C{}_{AB}. 
\end{split}
\end{equation}
Finally, we have
\begin{equation}
\begin{split}\label{app:Ruueq}
R_{uu}[G] &= \frac{d}{2r} \p_u U  - \frac{d}{r} U \p_u \b - 2 U \p_u \p_r \b  - \frac{1}{4} e^{-4\b} ( \p_r G_{uu} )^2   \\
&\qquad - \frac{1}{2}  D_A  [ e^{-2\b}  ( 2 \p_u G_{ur} - \p_r G_{uu} ) W^A  ]    + \frac{1}{2r^2}   D_A  [  {\bar h}^{AB} ( 2 \p_u G_{uB} - \p_B G_{uu} )  ]  \\
&\qquad  +  \p_u  [  e^{-2\b}  \p_r (  r^2 h_{AB} W^A W^B ) ] + \frac{d}{2r} e^{-2\b} \p_u  [r^2 h_{AB}  W^A W^B ]  \\
&\qquad -  r^{-d} e^{-2\b} \p_r [ r^{d+2} h_{AB} \p_u  W^A W^B  ]  -  \frac{1}{2} r^{-d} e^{-2\b} \p_r  [ r^d   ( U \p_r - \CL_W ) G_{uu}  ]  \\
&\qquad + \frac{1}{2} e^{-4\b} W^A   \p_r G_{uA}   [ 2 \p_u G_{ur} - \p_r G_{uu} ] - \frac{1}{2r^2} e^{-2\b}  {\bar h}^{AB}  \p_r G_{uA}   [ 2 \p_u G_{uB} - \p_B G_{uu} ]   \\
&\qquad - \left( \frac{1}{2} \p_r  U  - \frac{1}{2} r^2 e^{-2\b} h_{AB} W^A \p_r W^B +  ( U \p_r - \CL_W ) \b \right)^2 \\
&\qquad - \frac{1}{2r^2} {\bar h}^{AB} [  \p_r G_{uA} - \p_A G_{ur}  ] \p_B U  +   \frac{1}{2} e^{-2\b} W^B {\bar h}^{AC} \p_u h_{AB} [ \p_r G_{uC} - \p_C G_{ur} ]  \\
&\qquad + \frac{1}{2r^2} e^{-2\b}   {\bar h}^{AB} [ \p_r G_{uA} - \p_A G_{ur}  ]  ( U \p_r - \CL_W ) G_{uB} \\
&\qquad - \frac{1}{4} e^{-4\b} \left( W^A  \p_A G_{ur} - W^A  \p_r G_{uA} \right)^2 \\
&\qquad - \frac{1}{2r^2} e^{-2\b} W^A \left( r^2 \p_u h_{AB} + 2 \p_{[A} G_{B]u}  \right)  {\bar h}^{BC} \left( \p_r G_{uC} - \p_C G_{ur} \right) \\
&\qquad + \frac{1}{4} \p_u h_{AB}  \p_u {\bar h}^{AB} - \frac{1}{r^4} {\bar h}^{AB} \p_{[B} G_{C]u} {\bar h}^{CD} \p_{[D} G_{A]u} , 
\end{split}
\end{equation}
and
\begin{equation}
\begin{split}\label{app:RuAeq}
R_{uA}[G] &=  \frac{1}{2} r^{-d} e^{-2\b}  \p_r (  r^{d+2} h_{AB} \p_u W^B ) -  \p_u \p_A \b  - e^{-2\b} \p_u \b \p_r G_{uA}   -  \frac{1}{2} e^{-2\b} \p_r G_{uu} \p_A \b \\
& + \frac{1}{2} r^{-d} e^{-2\b}  \p_r (  r^d e^{2\b}  \p_A U )  + \frac{d}{2r} e^{-2\b}  \p_u G_{uA} + e^{-2\b} \p_r \p_u G_{uA}   - \frac{1}{2} \p_r  U \p_A \b  \\
& - \frac{1}{2} r^{-d} e^{-2\b}  \p_r [ r^d ( U \p_r - \CL_W ) G_{uA} )  ] +   \frac{1}{2} e^{-2\b} D_B [ W^B (  \p_r G_{uA} -  \p_A G_{ur} ) ] \\
&   + \frac{1}{2r^2} e^{-2\b} D_B [ e^{2\b}  {\bar h}^{BC} ( r^2 \p_u h_{CA} - 2 \p_{[C} G_{A]u} ) ]  - \frac{1}{2} e^{-2\b} W^B \p_B \b [  \p_r G_{uA} -  \p_A G_{ur} ]  \\
& - \frac{1}{4} e^{-4\b} W^B \p_r G_{uB} [  \p_r G_{uA} -  \p_A G_{ur} ]  - \frac{1}{2r^2} [ r^2 \p_u h_{AB} + 2 \p_{[A} G_{B]u}  ] {\bar h}^{BC} \p_C \b  \\
& - \frac{1}{4r^2} e^{-2\b} [ r^2 \p_u h_{AB} + 2 \p_{[A} G_{B]u} ] {\bar h}^{BC}  \p_r G_{uC}   - \frac{1}{4} r^2  e^{-2\b} \p_r  U h_{AB}  \p_r W^B  \\
& + \frac{1}{2} r^2 e^{-2\b} h_{CD} W^C \p_r W^D \p_A \b + \frac{1}{4} r^4 e^{-4\b} h_{CD} W^C \p_r W^D h_{AB}  \p_r W^B   \\
&  - \frac{1}{4} e^{-4\b}  \p_r G_{uu} \p_r G_{uA}  - \p_A \b ( U \p_r - \CL_W ) \b  - \frac{1}{2} r^2 e^{-2\b} h_{AB}  \p_r W^B ( U \p_r - \CL_W ) \b \\
&  + \frac{1}{4}  e^{-2\b}  \p_r (r^2 h _{AB} ) {\bar h}^{BC}   \p_u h_{CD} W^D + \frac{1}{4r^2} e^{-2\b}  \p_r (r^2 h _{AB} ) {\bar h}^{BC}  ( U \p_r - \CL_W ) G_{uC} \\
& + \frac{1}{4} e^{-4\b} \p_r ( r^2 h_{AB} ) W^B [ 2 \p_u G_{ur} - \p_r G_{uu} ]  - \frac{1}{4}  e^{-2\b} \p_u h_{AB} {\bar h}^{BC} [ \p_r G_{uC} - \p_C G_{ur} ]    \\
& - \frac{1}{4r^2} e^{-2\b} \p_r ( r^2 h_{AB} ) {\bar h}^{BC} [ 2 \p_u G_{uC} - \p_C G_{uu} ]  - \frac{1}{4r^2}   \p_r (r^2 h _{AB} )  {\bar h}^{BC}\p_C U \\
& +  \frac{1}{4r^2} e^{-2\b} ( U \p_r - \CL_W ) (r^2 h_{AB} )  {\bar h}^{BC} [ \p_r G_{uC} - \p_C G_{ur} ] \\
& -  \frac{1}{2} e^{-2\b} H^C{}_{AB} [  \p_r G_{uC} -  \p_C G_{ur} ] W^B    - \frac{1}{2r^2}  H^D{}_{AB}{\bar h}^{BC} [ r^2 \p_u h_{CD} - 2 \p_{[C} G_{D]u}  ]  \\
& - \frac{1}{4} e^{-4\b}  \p_r (r^2 h_{AB} )  W^B [  W^C \p_r G_{uC} - W^C \p_C G_{ur} ] \\
& - \frac{1}{4r^2}  e^{-2\b} \p_r (r^2 h_{AB} )   {\bar h}^{BC} [ r^2 \p_u h_{CD} - 2 \p_{[C} G_{D]u}  ] W^D . 
\end{split}
\end{equation}

\section{Asymptotic Expansions}
\label{app:solutionspace}

In this Appendix, we describe the large expansions of solutions in $\CS_\text{general}$, $\CS_\text{analytic}$ and $\CS_\text{canonical}$.

\subsection{Polyhomegenous Solutions}
\label{app:genlargeexp}

Metrics in $\CS_\text{general}$ admit the following general large $r$ expansion
\begin{equation}
\begin{split}\label{app:metric-exp}
\b &= \frac{\b_\2}{r^2} + \frac{\b_\3 + \b_\ttt{3,1} \ln r }{r^3} + \frac{\b_\4 + \b_\ttt{4,1} \ln r + \b_\ttt{4,2} \ln^2 r }{r^4}  + \CO(r^{-5}) ,  \\
U &= U_\0 + \frac{U_\1}{r} + \frac{U_\2 + U_\ttt{2,1} \ln r }{r^2} + \frac{U_\3 + U_\ttt{3,1} \ln r + U_\ttt{3,2} \ln^2 r  }{r^3} + \CO(r^{-4}),\\
W^A &= \frac{W^A_\2}{r^2} + \frac{W^A_\3 + W^A_\ttt{3,1} \ln r }{r^3} + \frac{W^A_\4 + W^A_\ttt{4,1} \ln r   }{r^4} \\
&\qquad \qquad \qquad \qquad \qquad + \frac{W^A_\5 + W^A_\ttt{5,1} \ln r + W^A_\ttt{5,2} \ln^2 r + W^A_\ttt{5,3} \ln^3 r }{r^5}  + \CO(r^{-6}) ,   \\
h_{AB} &= q_{AB} + \frac{h_{\1AB}}{r} + \frac{h_{\2AB} + h_{\ttt{2,1}AB} \ln r}{r^2} + \frac{h_{\3AB} + h_{\ttt{3,1}AB} \ln r }{r^3} \\
&\qquad \qquad \qquad \qquad \qquad \qquad \qquad + \frac{h_{\4AB} + h_{\ttt{4,1}AB} \ln r + h_{\ttt{4,2}AB} \ln^2 r }{r^4}  + \CO(r^{-5}) .
\end{split}
\end{equation}
The determinant condition \eqref{app:detcond} implies
\begin{equation}
\begin{split}\label{htr}
\tr [h_\1] &= 0  ,   \\
\tr [h_\2] &= \tr [ \frac{1}{2}  h_\1^2 ]  ,  \\
\tr [h_\ttt{2,1}] &= 0 ,   \\
\tr [h_\3] &= \tr [ h_\1 h_\2  - \frac{1}{3}   h_\1^3 ]    ,  \\
\tr[ h_\ttt{3,1} ] &= \tr [ h_\1 h_\ttt{2,1} ] ,    \\
\tr [h_\4 ] &=  \tr [ \frac{1}{2} h_\2^2 + h_\1 h_\3 - h_\1^2 h_\2 + \frac{1}{4}  h_\1^4 ]  ,    \\
\tr [h_\ttt{4,1} ] &= \tr [ h_\2 h_\ttt{2,1} + h_\1 h_\ttt{3,1} - h_\1^2 h_\ttt{2,1} ]  ,  \\
\tr [ h_\ttt{4,2} ] &= \tr [  \frac{1}{2} h_\ttt{2,1}^2 ]  ,
\end{split}
\end{equation}
The matrix inverse of $h$ is denoted by ${\bar h}$. This admits a large $r$ expansion of the form
\begin{equation}
\begin{split}
{\bar h}^{AB} &= q^{AB} + \frac{{\bar h}^{AB}_\1}{r} + \frac{{\bar h}^{AB}_\2+{\bar h}^{AB}_\ttt{2,1} \ln  r}{r^2} + \frac{{\bar h}^{AB}_\3 + {\bar h}^{AB}_\ttt{3,1} \ln  r  }{r^3}   \\
&\qquad \qquad \qquad \qquad \qquad \qquad \qquad \qquad \qquad + \frac{{\bar h}^{AB}_\4+{\bar h}^{AB}_\ttt{4,1} \ln  r +{\bar h}^{AB}_\ttt{4,2}\ln^2  r  }{r^4}  + \CO(r^{-5}) . 
\end{split}
\end{equation}
with
\begin{equation}
\begin{split}\label{hbAB} 
{\bar h}_\1 &= - h_\1 ,  \\
{\bar h}_\2 &=  - h_\2 + h_\1^2   ,   \\
{\bar h}_\ttt{2,1} &= - h_\ttt{2,1} ,   \\
{\bar h}_\3 &=  - h_\3 +  2 [ h_\1 h_\2 ]_S  - h_\1^3   ,   \\
{\bar h}_\ttt{3,1} &= - h_\ttt{3,1} + 2 [ h_\1 h_\ttt{2,1} ]_S   ,   \\
{\bar h}_\4 &=    - h_\4  + h_\2^2  + 2 [ h_\1 h_\3 ]_S   - 3 [ h_\1^2 h_\2 ]_S + h_\1^4   ,   \\
{\bar h}_\ttt{4,1} &=  - h_\ttt{4,1}  + 2 [ h_\2 h_\ttt{2,1} ]_S + 2 [ h_\1 h_\ttt{3,1} ]_S  - 3 [ h_\1^2 h_\ttt{2,1} ]_S  ,   \\
{\bar h}_\ttt{4,2} &= - h_\ttt{4,2}  + h_\ttt{2,1}^2 ,  
\end{split}
\end{equation}
where $[~]_S$ denotes the weighted symmetric product, $[A_1\cdots A_n]_S = \frac{1}{n!} \sum_{\s \in S_n}  A_{\s(1)} \cdots A_{\s(n)}$.

We now determine the constraints from equations of motion.

\paragraph{I. $R_{rr}[G]=0$:} This implies
\begin{equation}
\begin{split}
\p_r \b = \frac{r}{32} \tr [ {\bar h}  \p^2_r h]  . 
\end{split}
\end{equation}
This equation fixes the large $r$ components of $\b$,
\begin{equation}
\begin{split}
\label{app:b}
\b_\2 &= - \frac{1}{64} \tr [  h^2_\1 ]  ,  \\
\b_\3 &= - \frac{1}{144} \tr [ 18 h_\1 h_\2  -  12 h_\1 h_\2  - 3 h_\1^3 -  h_\1 h_\ttt{2,1}   ] ,  \\
\b_\ttt{3,1} &=   -  \frac{1}{24} \tr [ h_\1 h_\ttt{2,1}  ]  ,  \\
\b_\4 &= - \frac{1}{256} \tr [ 8 h_\2^2 + 12 h_\1 h_\3 - 20 h_\1^2 h_\2 + 6 h_\1^4  \\
&\qquad \qquad \qquad \qquad  - 4 h_\2 h_\ttt{2,1}  +  3 h_\1^2 h_\ttt{2,1} - h_\1  h_\ttt{3,1} + h^2_\ttt{2,1} ]  ,  \\
\b_\ttt{4,1} &=  - \frac{1}{64} \tr [ 4  h_\2 h_\ttt{2,1} + 3 h_\1 h_\ttt{3,1} - 5 h_\1^2 h_\ttt{2,1}     - h^2_\ttt{2,1}   ]  ,  \\
\b_\ttt{4,2} &= - \frac{1}{32} \tr [   h^2_\ttt{2,1}  ]   , 
\end{split}
\end{equation}
\paragraph{II: $R_{rA}[G]=0$:} This implies
\begin{equation}
\begin{split}\label{app:Rraeq-1}
- \p_r ( r^6 e^{-2\b} h \p_r W  ) = - 2 r^8 \p_r ( r^{-4} D  \b ) + r^4 D \cdot ( {\bar h} \p_r h ) + \frac{1}{2} r^4 \tr [ \p_r {\bar h} D h ]  .
\end{split}
\end{equation}
The RHS of this equation has the form
\begin{equation}
\begin{split}
\text{RHS} &= r^2 \msW_\2 + r [ \msW_\3 + \msW_\ttt{3,1} \ln  r   ] + [ \msW_\4 + \msW_\ttt{4,1} \ln  r  ]  \\
&\qquad \qquad \qquad \qquad \qquad \qquad \qquad \qquad + \frac{\msW_\ttt{5,1} + \msW_\ttt{5,2} \ln  r  + \msW_\ttt{5,3} \ln^2  r }{r} + \CO(r^{-2}) .
\end{split}
\end{equation}
Using the explicit forms of the large $r$ expansions, we find
\begin{equation}
\begin{split}
\label{app:WW}
\msW_\2 &= - D \cdot h_\1 ,  \\
\msW_\3 &= 12  D \b_\2 + D \cdot ( h^2_\1 - 2  h_\2 + h_\ttt{2,1} ) +  \frac{1}{2} \tr [ h_\1 D h_\1 ]  ,  \\
\msW_\ttt{3,1} &= - 2  D \cdot h_\ttt{2,1} ,  \\
\msW_\4 &= 14 D \b_\3 - 2 D \b_\ttt{3,1} + D \cdot [ -3h_{\3} + h_\ttt{3,1} + 2 h_\1 h_{\2} - h_\1 h_\ttt{2,1} - {\bar h}_\2 h_\1 ]   \\
&\qquad \qquad +  \frac{1}{2} \tr [ h_\1 D  h_\2  - 2  {\bar h}_\2 D h_\1 - h_\ttt{2,1} D h_\1 ]  ,  \\
\msW_\ttt{4,1} &=  14 D \b_\ttt{3,1}  + D \cdot [ - 3 h_\ttt{3,1}  + 2 h_\1 h_\ttt{2,1}  + h_\ttt{2,1} h_\1 ]  \\
&\qquad \qquad  + \frac{1}{2}  \tr [ 2 h_\ttt{2,1} D h_{\1} + h_\1 D  h_\ttt{2,1}  ]  ,  \\
\msW_\ttt{5,1} &=  16 D \b_\4 - 2 D \b_\ttt{4,1} + D \cdot [ - 4 h_{\4} + h_\ttt{4,1} + 3 h_\1 h_{\3} - h_\1 h_\ttt{3,1} - 2 {\bar h}_\2 h_\2  \\
&\qquad \qquad  + {\bar h}_\2h_\ttt{2,1} - {\bar h}_\3  h_\1 ]  +  \tr [ - {\bar h}_\2  D  h_\2  - \frac{1}{2}  h_\ttt{2,1}  D  h_\2 + \frac{1}{2} h_\1 D  h_\3  \\
&\qquad \qquad   -  \frac{3}{2}  {\bar h}_\3 D h_{\1} + \frac{1}{2}  {\bar h}_\ttt{3,1} D h_{\1}  ]  ,  \\
\msW_\ttt{5,2} &= 16 D \b_\ttt{4,1} - 4 D \b_\ttt{4,2}  +  D \cdot [ - 4 h_\ttt{4,1}  + 2 h_\ttt{4,2}  + 3 h_\1 h_\ttt{3,1}  + 2 h_\ttt{2,1} h_\2  \\
&\qquad \qquad  - 2 {\bar h}_\2  h_\ttt{2,1}  - h^2_\ttt{2,1} -  {\bar h}_\ttt{3,1} h_\1 ]  +  \tr [ - {\bar h}_\2 D h_\ttt{2,1}   - \frac{1}{2}  h_\ttt{2,1}D h_\ttt{2,1} \\
&\qquad \qquad + h_\ttt{2,1}  D  h_\2 + \frac{1}{2}  h_\1 D h_\ttt{3,1}  ] ,  \\
\msW_\ttt{5,3} &= 16 D \b_\ttt{4,2} + D \cdot [ - 4 h_\ttt{4,2}  + 2 h^2_\ttt{2,1}   ]  + \tr [ h_\ttt{2,1} D  h_\ttt{2,1}   ] , 
\end{split}
\end{equation}
We can solve for the large $r$ coefficients of $W$ as
\begin{equation}
\begin{split}
\label{app:W}
W_\2 &= \frac{1}{6}  \msW_\2 ,   \\
W_\3 &= \frac{1}{6} \msW_\3  - \frac{1}{9} h_\1  \msW_\2 - \frac{1}{36}  \msW_\ttt{3,1}  ,   \\
W_\ttt{3,1} &= \frac{1}{6} \msW_\ttt{3,1}  ,  \\
W_\4 &= \frac{1}{4} \msW_\4 - \frac{3}{16}  \msW_\ttt{4,1} + \frac{1}{32} h_\1 [ \msW_\ttt{3,1}  -  4 \msW_\3 ]  + \frac{1}{48}  [ 8 \b_\2 + 4 {\bar h}_\2 -  h_\ttt{2,1} ]  \msW_\2 ,  \\
W_\ttt{4,1} &= \frac{1}{4}  \msW_\ttt{4,1}  - \frac{1}{12} h_\ttt{2,1}  \msW_\2 - \frac{1}{8}  h_\1 \msW_\ttt{3,1}  ,  \\
W_\ttt{5,1} &= \frac{1}{5}  \msW_\ttt{5,1} + \frac{1}{25}  \msW_\ttt{5,2} + \frac{2}{125} \msW_\ttt{5,3} + \frac{1}{15} [ 2 \b_\ttt{3,1}  + {\bar h}_\ttt{3,1} ] \msW_\2  - \frac{1}{10} h_\ttt{2,1} \msW_\3 \\
&\qquad \qquad \qquad  + \frac{1}{100} [ 20  \b_\2  + 10 {\bar h}_\2 + h_\ttt{2,1} ] \msW_\ttt{3,1}  - \frac{1}{5}  h_\1 \msW_\ttt{4,1},  \\
W_\ttt{5,2} &=  \frac{1}{10} \msW_\ttt{5,2}   + \frac{1}{25} \msW_\ttt{5,3}   - \frac{1}{10} h_\ttt{2,1} \msW_\ttt{3,1}  ,   \\
W_\ttt{5,3} &=  \frac{1}{15} \msW_\ttt{5,3} .  
\end{split}
\end{equation}
Note that $W_\5$ is not fixed by this equation. 

\paragraph{III: ${\bar h}^{AB} R_{AB}[G] = 0$:} This implies
\begin{equation}
\begin{split}
 \p_r ( r^3 U) &= \frac{1}{4} r^2 e^{2\b} \tr [ {\bar h} R ]  - \frac{1}{2} r^2  e^{2\b} \tr [  {\bar h} ( D D \b  + D \b D \b ) ] + \frac{1}{4} r^{-4} \p_r ( r^8 D \cdot W )   \\
&\qquad \qquad \qquad  - \frac{1}{8}  r^6 e^{-2\b}  \p_r W h \p_r W - \frac{1}{4} r^2 D \cdot ( e^{2\b} D \cdot {\bar h} ) -  \frac{1}{8} r^2 e^{2\b} D_C {\bar h}^{AB}  H^C{}_{AB} . 
\end{split}
\end{equation}
Using this, we can fix the large $r$ coefficients for $U$,
\begin{equation}
\begin{split}
\label{app:U}
U_\0 &= \frac{1}{12} R ,  \\
U_\1 &=   - \frac{1}{8} \tr [ h_\1 R ]  +  \frac{3}{4}  D \cdot  W_\2 + \frac{1}{8} D \cdot  D \cdot h_\1 ,   \\
U_\2 &= \frac{1}{4}  \tr [ {\bar h}_\2 R ]  +  \frac{1}{2} \b_\2 R  - \frac{1}{2} D^2 \b_\2  + \frac{5}{4}  D \cdot W_\3 - D \cdot  W_\ttt{3,1}   - \frac{1}{2} W^2_\2  \\
&\qquad  - \frac{1}{4} D \cdot D \cdot  {\bar h}_\2 + \frac{1}{8} D^C h_\1^{AB}    D_A h_{\1BC} - \frac{1}{16} D^C h_\1^{AB}  D_C h_{\1AB}  \\
&\qquad  + \frac{1}{4}  \tr [ h_\ttt{2,1} R ] - \frac{1}{4} D \cdot D \cdot h_\ttt{2,1} ,   \\
U_\ttt{2,1} &= - \frac{1}{4}  \tr [ h_\ttt{2,1} R ]  +  \frac{5}{4}  D \cdot  W_\ttt{3,1} + \frac{1}{4} D \cdot D \cdot h_\ttt{2,1} ,   \\
U_\ttt{3,1} &= \frac{1}{4} \tr [ {\bar h}_\3 R ] - \frac{1}{2}  \b_\2 \tr [ h_\1 R ]  + \frac{1}{2} \b_\3 R + \frac{1}{2} \tr [ h_\1 D D \b_\2 ]  - \frac{1}{2} D^2 \b_\3  + D \cdot  W_\4  \\
&\qquad + \frac{1}{4}  D \cdot W_\ttt{4,1} - \frac{3}{2} W_\2 W_\3 - \frac{1}{2} W_\2 h_\1 W_\2+ \frac{1}{2}  W_\2 W_\ttt{3,1} - \frac{1}{4} D \cdot D \cdot  {\bar h}_\3   \\
&\qquad + \frac{1}{2} D \cdot [ \b_\2 D \cdot h_\1 ]  + \frac{1}{8}  D^C h_\1^{AB}  D_A h_{\2BC} - \frac{1}{16}  D^C h_\1^{AB}  D_C h_{\2AB}  \\ 
&\qquad  - \frac{1}{8} h_\1^{CD}   D_C h_\1^{AB}  D_A h_{\1BD} + \frac{1}{16}   h_\1^{CD} D_C h_\1^{AB}  D_D h_{\1AB}  \\
&\qquad - \frac{1}{8} D^C  {\bar h}^{AB}_\2  D_A h_{\1BC} + \frac{1}{16} D^C  {\bar h}^{AB}_\2 D_C h_{\1AB} ,  \\
U_\ttt{3,2} &=  \frac{1}{8} \tr [ {\bar h}_\ttt{3,1} R ] + \frac{1}{4} \b_\ttt{3,1} R - \frac{1}{4} D^2 \b_\ttt{3,1} + \frac{1}{2} D \cdot W_\ttt{4,1}   - \frac{3}{4} W_\2  W_\ttt{3,1} \\
&\qquad   - \frac{1}{8} D \cdot D \cdot  {\bar h}_\ttt{3,1}  + \frac{1}{8} D^C h_\ttt{2,1}^{AB}  D_A h_{\1BC} - \frac{1}{16} D^C h_\ttt{2,1}^{AB} D_C h_{\1AB} .   
\end{split}
\end{equation}
Note that $U_\3$ is not fixed by this equation.

\paragraph{IV: $R_{AB}[G]=0$:} This implies
\begin{equation}
\begin{split}\label{RAB-eom}
- \p_r ( r^2 \p_u  h_{AB} ) &=  e^{2\b} R_{AB} + \bigg( - \frac{1}{r^2} \p_r ( r^3 U ) h - r^2 \p_r h {\bar h} \p_u  h -  2 e^{2\b}  ( D D \b  + D \b D \b )   \\
& -  \frac{1}{2r^2} \p_r (  r^4 U \p_r  h  ) + \frac{r^2}{2} U  \p_r h {\bar h}  \p_r h  + \frac{1}{2r^2} \p_r ( r^4 \CL_W h   ) + r D_C W^C h \bigg)_{(AB)}\\
& +  \frac{1}{2}   r^2 D_C  ( W^C \p_r   h_{AB} )  - \frac{r^4}{2}  e^{-2\b}  h_{AC} h_{BD} \p_r W^C    \p_r W^D  +  D_C ( e^{2\b}  H^C{}_{AB} )  \\ 
& - e^{2\b} H^C{}_{AD} H^D{}_{BC} - r^2 ( {\bar h} \p_r h )^C{}_{(A}  W^D D_D h_{B)C}  - r^2 ( {\bar h} \p_r h )^C{}_{(A}   h_{B)D} D_C W^D  .
\end{split}
\end{equation}
We start by using the large $r$ expansions to determine the RHS of the equation above. This has the form
\begin{equation}
\begin{split}
\text{RHS} = \msH_{\0} + \frac{1}{r} \msH_{\1} + \frac{\msH_{\2} + \msH_{\ttt{2,1}} \ln  r }{r^2} + \frac{\msH_{\3} + \msH_{\ttt{3,1}}  \ln  r + \msH_{\ttt{3,2}}  \ln^2  r }{r^3} + \CO(r^{-4}) . 
\end{split}
\end{equation}
where
\begin{equation}
\begin{split}
\msH_{\0AB} &= R_{AB}  -  3   U_\0 q_{AB} ,  \\
\msH_{\1AB} &=  [ - 2 U_\0    h_\1 - 2   U_\1 q  +  \frac{1}{2} \p_u  h_\1^2  + \CL_{W_\2} q  + D_C W_\2^C q ]_{AB} +  D_C  H^C_{\1AB}   ,   \\
 \msH_{\2AB} &= 2 \b_\2  R_{AB} - 2 U_\0 h_{\2AB}  + \frac{1}{2} U_\0  h_{\ttt{2,1}AB}  - \frac{3}{2}  U_\1   h_{\1AB}  - [ U_\2 + U_\ttt{2,1} ] q_{AB}   \\
 &\qquad + [ 2 h_\2   \p_u  h_\1  + h_\1  \p_u h_{\2}  - h_\ttt{2,1}   \p_u  h_\1   - h^2_\1  \p_u  h_\1 ]_{(AB)} - 2 D_A D_B \b_\2 \\
 &\qquad + \frac{1}{2} U_\0 ( h_\1^2 )_{AB}  +  \frac{1}{2} \CL_{W_\2} h_{\1AB}    + \frac{1}{2} \CL_{W_\3} q_{AB}  + \frac{1}{2} \CL_{W_\ttt{3,1}} q_{AB} \\
 &\qquad +  D_C W_\2^C h_{\1AB}  + D_C W_\3^C q_{AB} - \frac{1}{2} D_C [ W_\2^C h_{\1AB} ] - 2 W_{\2A}  W_{\2B}  \\
\label{H2} &\qquad + D_C H^C_{\2AB}  -   H^C_{\1AD} H^D_{\1BC} + h_{\1}^C{}_{(A} D_C W_{\2B )} ,  \\
 \msH_{\ttt{2,1}AB} &= - 2 U_\0 h_{\ttt{2,1}AB}  - U_\ttt{2,1}q_{AB}  + [ 2 h_\ttt{2,1}   \p_u  h_\1  + h_\1  \p_u  h_{\ttt{2,1}} ]_{(AB)}  \\
&\qquad + \frac{1}{2}  \CL_{W_\ttt{3,1}} q_{AB}  + D_C W_\ttt{3,1}^C q_{AB} + D_C H^C_{\ttt{2,1}AB}  ,  \\
 \msH_{\3AB} &= 2 \b_\3  R_{AB} - 3 U_\0  h_{\3AB} - 2 U_\1 h_{\2AB}  - U_\2 h_{\1AB}   - \frac{1}{2} U_\ttt{2,1}  h_{\1AB} \\
 &\qquad - U_\ttt{3,1}q_{AB}  + [ 3 h_\3 \p_u  h_\1 - h_\ttt{3,1} \p_u  h_\1  - 2 h_\2 h_\1\p_u  h_\1   + h_\ttt{2,1} h_\1 \p_u  h_\1  \\
 &\qquad + \p_u  h^2_\2  - h_\ttt{2,1} \p_u  h_\2 + h_\1 {\bar h}_\2 \p_u h_\1 - h_\1^2 \p_u h_\2  + h_\1  \p_u  h_{\3} ]_{(AB)}   \\
 &\qquad - 2 D_A D_B \b_\3 + \frac{3}{2} U_\0  h_{\ttt{3,1}AB} - U_\0  h_{\ttt{3,2}AB}  + U_\1 h_{\ttt{2,1}AB} \\
 &\qquad + \frac{1}{2} U_\1 ( h_\1^2 )_{AB}  - \frac{1}{2} U_\0  ( h^3_\1)_{AB}   +  U_\0   [ 2 h_{\2}  h_\1 -   h_\ttt{2,1}  h_\1  ]_{(AB)}  \\
 &\qquad + \frac{1}{2} \CL_{W_\2} h_{\ttt{2,1}AB}  + \frac{1}{2} \CL_{W_\ttt{3,1}} h_{\1AB}  + \frac{1}{2}  \CL_{W_\ttt{4,1}} q_{AB}  +  D_C W_\4^C q_{AB} \\
 &\qquad  + D_C W_\3^C h_{\1AB}  + D_C W_\2^C h_{\2AB}  - \frac{1}{2} D_C [ W_\3^C h_{\1AB} ]  -  D_C  [ W_\2^C  h_{\2AB} ] \\
 &\qquad + \frac{1}{2} D_C  [ W_\2^C  h_{\ttt{2,1}AB} ]  - 6 W_{\3(A}W_{\2B)} + 2 W_{\ttt{3,1}(A}W_{\2B)}  \\
 &\qquad - 4 W_{\2(A} [ h_\1 W_\2 ]_{B)}  + D_C  [ H^C_{\3AB}  +   2 \b_\2 H^C_{\1AB} ] - 2 H^C_{\2D(A}H_\1^D{}_{B)C}  \\
 &\qquad+ W_\2^D D_D h_{\1C(A}   h_{\1B)}{}^C   + h_\1^C{}_{(A} D_C W_{\3B)} + h_\1^C{}_{(A}   h_{\1B)D} D_C W_\2^D  \\
&\qquad  - ( h^2_\1 )^C{}_{(A} D_C W_{\2B)} + [ 2 h_{\2} - h_{\ttt{2,1}}  ]^C{}_{(A} D_C W_{\2B)}  ,  \\
 \msH_{\ttt{3,1}AB} &= 2 \b_\ttt{3,1}  R_{AB}  - 3 U_\0  h_{\ttt{3,1}AB}  - 2 U_\1 h_{\ttt{2,1}AB} -  U_\ttt{2,1} h_{\1AB}  - 2 U_\ttt{3,2} q_{AB}  \\
 &\qquad + [ 3 h_\ttt{3,1} \p_u  h_\1  - 2 h_\ttt{2,1} h_\1 \p_u  h_\1  + 2 h_\2 \p_u h_\ttt{2,1}  + 2 h_\ttt{2,1} \p_u  h_\2   \\
 &\qquad - h_\ttt{2,1} \p_u h_\ttt{2,1}    - h_\1 h_\ttt{2,1} \p_u  h_\1  - h^2_\1 \p_u  h_\ttt{2,1}  +  h_\1  \p_u h_{\ttt{3,1}} \\
 &\qquad + 2 U_\0 h_\ttt{2,1} h_\1 ]_{(AB)}    - 2 D_A D_B \b_\ttt{3,1}  + D_C W_\ttt{4,1}^C q_{AB}  + D_C W_\ttt{3,1}^C h_{\1AB} \\
 &\qquad  + D_C W_\2^C h_{\ttt{2,1} AB}  - \frac{1}{2} D_C [ W_\ttt{3,1}^C h_{\1AB} ]    - D_C ( W_\2^C h_{\ttt{2,1}AB} )  \\
& \qquad  - 6 W_{\ttt{3,1}(A}W_{\2B)} + D_C H^C_{\ttt{3,1}AB}  - 2 H^C_{\ttt{2,1}D(A} H_\1^D{}_{B)C}  \\
&\qquad + h_\1^C{}_{(A} D_C W_{\ttt{3,1}B)}  + 2 h_{\ttt{2,1}}^C{}_{(A} D_C W_{\2B)} ,  \\
\msH_{\ttt{3,2}AB} &=  \p_u ( h^2_\ttt{2,1} )_{(AB)} . 
\end{split}
\end{equation}
We can determine the following evolution equations for the large $r$ coefficients of $h$
\begin{equation}
\begin{split}
\label{hdu}
\p_u h_{\1} &= - \msH_{\0} , \qquad \p_u h_{\ttt{2,1} } = - \msH_{\1} , \qquad \p_u h_{\3 } = \msH_{\2} + \msH_{\ttt{2,1}}  , \qquad \p_u h_{\ttt{3,1} } = \msH_{\ttt{2,1}}  , \\
\p_u h_{\4 } &= \frac{1}{2} \msH_{\3} + \frac{1}{4}  \msH_{\ttt{3,1}} + \frac{1}{4}   \msH_{\ttt{3,2}} , \quad \p_u h_{\ttt{4,1} } =  \frac{1}{2} \msH_{\ttt{3,1}} + \frac{1}{2} \msH_{\ttt{3,2}}  , \quad \p_u h_{\ttt{4,2} } =  \frac{1}{2} \msH_{\ttt{3,2}} . 
\end{split}
\end{equation}
Note that there is no evolution equation for $h_{\2AB}$. 

\paragraph{Step V: $R_{uu}[G]=0$:} This implies an evolution equation for $U_\3$,
\begin{equation}
\begin{split}
\label{U3u} \p_u U_\3 &=  - U_\0 \p_u \b_\3 + U_\0 \p_u \b_\ttt{3,1}  + \frac{1}{2} U_\1^2  - D \cdot  [  ( \p_u \b_\2 - \frac{1}{4} U_\1  ) W_\2  ] \\
&\qquad + D U_\0 \cdot [ \frac{1}{2}  D \b_\2 + \frac{3}{4}  W_\3 - \frac{1}{4} W_\ttt{3,1} + \frac{1}{2} h_\1 W_\2  ]  \\
&\qquad  + \frac{1}{2}   \p_u D \cdot [ W_\4 + h_\1 W_\3 + h_\2 W_\2 ] - \frac{1}{4} D^2 ( U_\2 + 2 \b_\2 U_\0 - W_\2^2  ) \\
&\qquad  - \frac{1}{4}  D \cdot  [  h_\1 ( 2 \p_u W_\3 + 2 \p_u ( h_\1 W_\2 ) - D U_\1 ) ]  \\
&\qquad  + \frac{1}{4} D \cdot  [  {\bar h}_\2 ( 2 \p_u W_\2 - D U_\0 )  ]  + \frac{1}{4} U_\0  [ 2 U_\2 + 4 \b_\2 U_\0 - 2 W_\2^2  + U_\ttt{2,1}  ]  \\
&\qquad + \frac{1}{4}  \CL_{W_\2} U_\1 - \frac{1}{2} D_{[A} W_{\2B]} D^{[A} W_\2^{B]} \\
&\qquad + \p_u  \bigg( W_\3 W_\2 + \frac{1}{2} W_\2 h_\1 W_\2 \bigg) + \frac{1}{2} W_\2 \p_u [ W_\3 - W_\ttt{3,1}  ] \\
&\qquad - \frac{1}{8} \tr [ \p_u h_\1 \p_u {\bar h}_\3 +  \p_u h_\2  \p_u {\bar h}_\2 - \p_u h_\3  \p_u h_\1 ] . 
\end{split}
\end{equation}

\paragraph{Step VI: $R_{uA}=0$:} This implies an evolution equation for $W_\5$ but we will not need this equation here.

\subsection{Analytic Solutions}
\label{sec:solution-nolog}

In this section, we describe the large $r$ expansion for analytic solutions. These are obtained from $\CS_\text{general}$ by imposing an additional constraint 
\begin{equation}
\begin{split}\label{app:vanishingRiemann}
r^4 e^\mu_{\hat \mu} e^\nu_{\hat \nu} e^\rho_{\hat \rho} e^\s_{\hat \s} R_{\mu\nu\rho\s}[G] |_{\ci^+_-} = 0 . 
\end{split}
\end{equation}
We start by proving that this constraint does, in fact, remove all the log terms in the large $r$ expansion. To process the constraint, we need to first introduce a relevant set of vielbein. We define\footnote{The corresponding one-forms are $n = - r e^{2\b} \dt u$, $\ell = - \frac{1}{r} ( \dt r + \frac{1}{2} U \dt u)$ and $e^a =  r E^a_A ( \dt x^A - W^A \dt u )$.}
\begin{equation}
\begin{split}
n = r \p_r , \qquad \ell = \frac{1}{r} e^{-2\b} \left( \p_u  - \frac{1}{2} U \p_r + W^A \p_A \right) , \qquad e_a = \frac{1}{r} E^A_a \p_A .
\end{split}
\end{equation}
Here, $E^a_A$ is the normalized vielbein for $h_{AB}$ so that $h_{AB} = \d_{ab} E^a_A E^b_B$ and $E^A_a = h^{AB} \d_{ab} E^b_B$. With this choice, $g^{\mu\nu} = - 2 n^{(\mu} \ell^{\nu)} + \d^{ab} e_a ^\mu e_b ^\nu$.

Using this, we can translate the \eqref{vanishingRiemann} to

\begin{equation}
\begin{split}\label{app:Riemann-test}
r^4 R_{urur} [G]|_{\ci^+_-} &= 0 , \qquad r^4 R_{ArBr}  [G]|_{\ci^+_-} = 0 , \qquad r^4 R_{Arur} [G]|_{\ci^+_-} = 0 , \\
r^2 R_{ABCr} [G]|_{\ci^+_-} &= 0 , \qquad r^2 R_{ArBu} [G]|_{\ci^+_-} = 0 , \qquad r^2 R_{uruA} [G]|_{\ci^+_-}  = 0 , \\
R_{ABCD} [G]|_{\ci^+_-} &= 0 , \qquad R_{ABCu} [G]|_{\ci^+_-}  = 0 , \qquad R_{AuBu} [G]|_{\ci^+_-}  = 0  . 
\end{split}
\end{equation}
analysing these constraints requires moving back and forth between each of these equations at different orders in large $r$. We describe the step-by-step process below.
\begin{enumerate}[leftmargin=0.75cm]
\item \textbf{$R_{ABCD} [G]|_{\ci^+_-} =0$ at $\CO(r^2)$:} At this order, we find the constraint $C_{ABCD}[q] = 0$. This implies that $q_{AB}$ is conformally flat so we can write
\begin{equation}
\begin{split}\label{app:qform}
q_{AB}(x)  = e^{2\Phi(x)} {\hat q}_{AB}(x)  , \qquad {\hat q}_{AB}(x) = \d_{CD} \p_A \chi^C(x) \p_B \chi^D(x) .
\end{split}
\end{equation}
\item \textbf{$r^2 R_{ABCr} [G] |_{\ci^+_-} =0$ at $\CO(r)$:} At this order, we find the constraint
\begin{equation}
\begin{split}
\o_{ABC} \equiv D_{[A} h_{\1B]C} + \frac{1}{3} ( D \cdot h_\1 )_{[A} q_{B]C} |_{\ci^+_-} = 0 . 
\end{split}
\end{equation}
Along with EoM, this implies
\begin{equation}
\begin{split}\label{app:h1form}
h_{\1AB} = - u R_{\{AB\}} - ( 2 D_{\{A} D_{B\}} + R_{\{AB\}} ) ( e^\Phi C  ) . 
\end{split}
\end{equation}
This can be further written in terms of the covariant derivative $\hat{D}$ associated with $\hat{q}$ as
\begin{equation}
\begin{split}
h_{\1AB} = - 2 e^{\Phi(x)} {\hat D}_{\{A} {\hat D}_{B\}}( e^{-\Phi(x)} u + C(x)).
\end{split}
\end{equation}

\item \textbf{$r^4 R_{ArBr}[G] |_{\ci^+_-} =0$ at $\CO(r^2)$ and $\CO(r^2\ln  r)$:} At this order, we find
\begin{equation}
\begin{split}\label{DABdef}
{\wt h}_{\2AB} |_{\ci^+_-} = h_{\ttt{2,1}AB}  |_{\ci^+_-}  = 0 , \qquad {\wt h}_\2 \equiv  \left[ h_\2 - \frac{1}{4} h_\1^2 \right]^\tf . 
\end{split}
\end{equation}
The conditions \eqref{app:qform} and \eqref{app:h1form} simplify the evolution equation of $h_\ttt{2,1}$ to $\p_u h_{\ttt{2,1}AB} = 0$. The boundary condition \eqref{DABdef} then implies
\begin{equation}
\begin{split}\label{app:h31du}
h_\ttt{2,1} = 0 ~ \stackrel{\text{EoM}}{\implies} ~ \b_\ttt{3,1} = \b_\ttt{4,2} = W_\ttt{3,1} = U_\ttt{2,1} = 0 , \quad \p_u h_\ttt{3,1} = \p_u h_\ttt{4,2} = 0 . 
\end{split}
\end{equation}

\item\textbf{$r^4 R_{ArBr}[G] |_{\ci^+_-} =0$ at $\CO(r \ln  r)$ and $\CO(\ln^2  r)$:} At this order, we find
\begin{equation}
\begin{split}
h_\ttt{3,1} |_{\ci^+_-} = h_\ttt{4,2} |_{\ci^+_-} = 0 \quad \stackrel{\eqref{app:h31du}}{\implies} \quad h_\ttt{3,1} = h_\ttt{4,2} = 0 .
\end{split}
\end{equation}
The EoM then automatically implies the vanishing of several more log terms in the large $r$ expansion, namely
\begin{equation}
\begin{split}\label{app:h41du}
\b_\ttt{4,1} = W_\ttt{4,1} = W_\ttt{5,3} = U_\ttt{3,2} = 0 , \qquad \p_u h_\ttt{4,1} = 0 .
\end{split}
\end{equation}

\item  \textbf{$r^4 R_{ArBr}[G] |_{\ci^+_-} =0$ at $\CO(\ln  r)$:} At this order, we find
\begin{equation}
\begin{split}
h_\ttt{4,1} |_{\ci^+_-} = 0 \quad \stackrel{\eqref{app:h41du}}{\implies} \quad h_\ttt{4,1} = 0  \quad \stackrel{\text{EoM}}{\implies} \quad W_\ttt{5,2} = 0 . 
\end{split}
\end{equation}

\item \textbf{$r^4 R_{ArBr}[G] |_{\ci^+_-} =0$ at $\CO(r)$ and $\CO(1)$:} This implies
\begin{equation}
\begin{split}\label{S0-prop-1}
{\wt h}_{\3AB}  |_{\ci^+_-}  = {\wt h}_{\4AB} |_{\ci^+_-}  = 0 ,
\end{split}
\end{equation}
where
\begin{equation}
\begin{split}
\label{app:htilde-def}
{\wt h}_\3 &\equiv \left( h_\3 -  h_\1 {\wt h}_\2 + 2 \b_\2 h_\1 \right)^\tf , \\
{\wt h}_\4 &\equiv \left( h_\4 - \frac{1}{2} {\wt h}_\2^2 - \frac{1}{2} h_\1 {\wt h}_\3  -  \frac{1}{4} h_\1 {\wt h}_\2 h_\1 + 4 \b_\2 {\wt h}_\2 +  \frac{1}{96} h_\1  \tr [ h_\1^3 ] \right)^\tf . 
\end{split}
\end{equation}
Once the constraints described so far are imposed \emph{all} the other constraints in \eqref{app:Riemann-test} turn out to be automatically satisfied!

\item \textbf{$W_\ttt{5,1}=U_\ttt{3,1}=0$:} After a large amount of simplification of the solutions derived in the previous section (equations \eqref{app:W} and \eqref{app:U}), we find that
\begin{equation}\label{eq.compare}
\begin{split}
U_\ttt{3,1} &= - \frac{1}{2} D \cdot D \cdot {\wt h}_\3 - \frac{1}{4} \tr [ {\wt h}_\3 R ]  ,   \\
W_\ttt{5,1} &= - \frac{4}{5}  \left( D \cdot \left[  {\wt h}_\4 - \frac{1}{2} h_\1 {\wt h}_\3 \right] + \frac{1}{16} \tr [ 3 h_\1 D {\wt h}_\3 - {\wt h}_\3 D h_\1 ] \right)  .  
\end{split}
\end{equation}
To prove that these quantities vanish, we take a $u$ derivative of the equation above and then use the evolution equations \eqref{hdu}. Following a long calculation, we find 
\begin{equation}
\begin{split}\label{U31-W51-eq}
\p_u U_\ttt{3,1} &= 0 , \qquad \p_u W_\ttt{5,1} = 0 .  
\end{split}
\end{equation}
Then, using the boundary conditions \eqref{S0-prop-1} we find
\begin{equation}
\begin{split}
U_\ttt{3,1} |_{\ci^+_-} = W_\ttt{5,1}  |_{\ci^+_-}  =  0 \quad \stackrel{\eqref{U31-W51-eq}}{\implies} \quad U_\ttt{3,1}  = W_\ttt{5,1}  =  0 .  
\end{split}
\end{equation}

\end{enumerate}

With this, we have completed the proof that the vanishing Riemann tensor constraints \eqref{app:Riemann-test} imply that all the log terms in the large $r$ expansion vanish identically!

\subsection*{Explicit Large $r$ Expansion}

Having shown that all the log terms vanish once the Riemann tensor constraint has been imposed, we now present explicit formulas for all the large $r$ coefficients of the metric. 

Metrics in $\CS_\text{analytic}$ admit the following large $r$ expansion
\begin{equation}
\begin{split}\label{app:metric-exp-S0}
\b &= \frac{\b_\2}{r^2} + \frac{\b_\3 }{r^3} + \frac{\b_\4 }{r^4}  + \CO(r^{-5}) ,  \\
U &= U_\0 + \frac{U_\1}{r} + \frac{U_\2  }{r^2} + \frac{U_\3 }{r^3} + \CO(r^{-4}),\\
W^A &= \frac{W^A_\2}{r^2} + \frac{W^A_\3   }{r^3} + \frac{W^A_\4 }{r^4} + \frac{W^A_\5 }{r^5}  + \CO(r^{-6}) ,   \\
h_{AB} &= q_{AB} + \frac{h_{\1AB}}{r} + \frac{h_{\2AB}  }{r^2} + \frac{h_{\3AB} }{r^3}  + \frac{h_{\4AB}}{r^4}  + \CO(r^{-5}) .
\end{split}
\end{equation}
where $q_{AB}$ is conformally flat and 
\begin{equation}
\begin{split}
\label{app:h1-sol}
h_{\1AB} = - u R_{\{AB\}} - [ 2 D_{\{A} D_{B\}} + R_{\{AB\}} ] ( e^\Phi C  ) .
\end{split}
\end{equation}
The solutions will be described in terms of the tilded fields defined in \eqref{app:htilde-def}. We recall that ${\wt h}_\3$ and ${\wt h}_\4$ satisfy the constraints
\begin{equation}
\begin{split}
D \cdot D \cdot {\wt h}_\3  &= -  \frac{1}{2} \tr [ {\wt h}_\3 R ]   ,   \\
D \cdot \bigg(  {\wt h}_\4 - \frac{1}{2} h_\1 {\wt h}_\3 \bigg)  &=  \frac{1}{16} \tr [  {\wt h}_\3 D h_\1 - 3 h_\1 D {\wt h}_\3 ] ,  
\end{split}
\end{equation}
and they vanish on $\ci^+_-$,
\begin{equation}
\begin{split}
{\wt h}_\2 |_{\ci^+_-} = {\wt h}_\3 |_{\ci^+_-} = {\wt h}_\4 |_{\ci^+_-} = 0.
\end{split}
\end{equation}
In fact, from the linearised analysis of solutions in $\CS_\text{canonical}$ and the fact that these are related to solutions in $\CS_\text{analytic}$ via BMS transformations, we can determine stronger conditions on these fields
\begin{equation}
\begin{split}\label{app:no-log-htilde-largeu}
{\wt h}_\2 = \CO(u^{-3}) , \qquad  {\wt h}_\3 = \CO(u^{-2}) , \qquad {\wt h}_\4 = \CO(u^{-1}) \quad \text{as $u \to -\infty$.}
\end{split}
\end{equation}
The large $r$ coefficients of $h_{AB}$ are then given by 
\begin{equation}
\begin{split}\label{app:nolog-h}
h_\2 &= {\wt h}_\2 + \frac{1}{4} h_\1^2 - 4 q\b_\2  ,  \\
h_\3 &= {\wt h}_\3 + \frac{1}{2} [ h_\1 {\wt h}_\2 + {\wt h}_\2 h_\1 ]  - 2 \b_\2 h_\1 - \frac{1}{48} q\,  \tr [  h_\1^3   ] ,  \\
h_\4 &=  {\wt h}_\4  + \frac{1}{2} {\wt h}_\2^2 + \frac{1}{4} [ h_\1 {\wt h}_\3 + {\wt h}_\3 h_\1  ] + \frac{1}{4} h_\1 {\wt h}_\2 h_\1 \\
&\qquad \qquad \qquad  - 4 \b_\2 {\wt h}_\2 - \frac{1}{96} h_\1  \tr [ h_\1^3 ]  + \frac{1}{8} q \, \tr \bigg[  h_\1 {\wt h}_\3  +  \b_\2 h_\1^2  + \frac{1}{16} h_\1^4  \bigg] . 
\end{split}
\end{equation}
These satisfy the trace properties
\begin{equation}
\begin{split}\label{app:nolog-htr}
\tr [h_\1] &= 0,   \\
\tr [h_\2] &=  \frac{1}{2} \tr [  h_\1^2 ] ,   \\
\tr [h_\3] &= \tr \left[ h_\1 {\wt h}_\2 - \frac{1}{12} h_\1^3  \right]  ,   \\
\tr [h_\4] &= \tr \bigg[ h_\1 {\wt h}_\3  + \frac{1}{2}  {\wt h}_\2^2 + \frac{1}{4} h_\1^2 {\wt h}_\2 + \frac{1}{2} \b_\2 h_\1^2  + \frac{1}{32} h_\1^4    \bigg]  . 
\end{split}
\end{equation}
These relations ensure that the Bondi determinant condition $\det(h_{AB}) = \det(q_{AB})$ is satisfied. The large $r$ coefficients of the inverse metric are
\begin{equation}
\begin{split}\label{app:nolog-hbAB} 
{\bar h}_\1 &= - h_\1 ,  \\
{\bar h}_\2 &=  -  {\wt h}_\2 + \frac{3}{4} h_\1^2 + 4 q\b_\2  ,   \\
{\bar h}_\3 &=  -  {\wt h}_\3 + \frac{1}{2} [ h_\1 {\wt h}_\2 + {\wt h}_\2 h_\1 ]  - 6 \b_\2  h_\1 - \frac{1}{2} h_\1^3  + \frac{1}{48} q\,  \tr [  h_\1^3   ]   ,   \\
{\bar h}_\4 &=  - {\wt h}_\4  + \frac{1}{2} {\wt h}_\2^2 + \frac{3}{4} [ h_\1 {\wt h}_\3 + {\wt h}_\3 h_\1  ]  - \frac{1}{4} [  h_\1^2 {\wt h}_\2 +  {\wt h}_\2 h_\1^2  + h_\1 {\wt h}_\2 h_\1 ] \\
&\qquad \qquad - \frac{1}{32} h_\1  \tr [ h_\1^3 ]  + \frac{5}{16} h_\1^4 - 4 \b_\2 {\wt h}_\2  + 6 \b_\2 h_\1^2  + 24 q \b_\2^2 \\
&\qquad \qquad  - \frac{1}{8} q \, \tr \bigg[  h_\1 {\wt h}_\3 + \frac{1}{16} h_\1^4  \bigg] . 
\end{split}
\end{equation}
The large $r$ coefficients of $\b$ are
\begin{equation}
\begin{split}\label{app:nolog-b}
\b_\2 &= - \frac{1}{64} \tr [ h_\1^2 ] ,  \\
\b_\3 &= - \frac{1}{24} \tr \left[ h_\1 {\wt h}_\2 - \frac{1}{4} h_\1^3 \right] ,  \\
\b_\4 &= 8 \b_\2^2 - \frac{1}{128} \tr \bigg[ 6 h_\1 {\wt h}_\3   + 4 {\wt h}_\2^2 - 2  h_\1^2 {\wt h}_\2 + \frac{3}{4} h_\1^4  \bigg]. 
\end{split}
\end{equation}
The large $r$ coefficients of $W$ are
\begin{equation}
\begin{split}\label{app:nolog-W}
W_\2 &= - \frac{1}{6} D \cdot h_\1  ,  \\
W_\3 &=  - \frac{1}{3}  D \cdot \left(  {\wt h}_\2 - \frac{3}{4}  h_\1^2 - 18 q \b_\2 \right)   ,  \\
 W_\4 &=  D \cdot \left( - \frac{3}{4}  {\wt h}_\3 + \frac{1}{8} [ h_\1 {\wt h}_\2 + {\wt h}_\2 h_\1 ] - \frac{3}{8}  h_\1^3 - 10 \b_\2 h_\1    + \frac{1}{2} q \b_\3 + \frac{7}{64}  q  \tr [  h_\1^3 ] \right) \\
&\qquad \qquad - \frac{1}{8} \tr [ {\wt h}_\2 D h_\1  ]  +  \frac{1}{6}  {\wt h}_\2  D \cdot h_\1 + \frac{4}{3} \b_\2  D \cdot h_\1  .  
\end{split}
\end{equation}

The large $r$ coefficients of $U$ are 
\begin{equation}
\begin{split}
\label{app:nolog-U}
U_\0 &= \frac{1}{12} R ,  \\
U_\1 &=   - \frac{1}{8} \tr [ h_\1 R ]  ,   \\
U_\2 &=   - \frac{1}{6} D \cdot D \cdot {\wt h}_\2 - \frac{1}{4}  \tr [  {\wt h}_\2 R ] + \frac{1}{16} \tr [ h_\1^2 R  ]  +  \frac{5}{6} \b_\2 R  +  \frac{1}{36}    (D \cdot h_\1 )^2     .
\end{split}
\end{equation}
The large $r$ coefficients of $h_{AB}$ satisfy the following evolution equations
\begin{equation}
\begin{split}
\label{app:nolog-hdu}
\p_u {\wt h}_{\1AB} &= - R_{\{AB\}} ,   \\
\p_u {\wt h}_{\3AB} &= \frac{2}{3} D_{\{A} ( D \cdot {\wt h}_\2 )_{B\}} - \frac{1}{2} D^2 {\wt h}_{\2AB}  +  ( R {\wt h}_\2 )_{\{AB\}} - \frac{1}{12} R {\wt h}_{\2AB}  ,   \\
\p_u {\wt h}_{\4AB} &= \frac{1}{2} D_{\{A} ( D \cdot {\wt h}_\3 )_{B\}} - \frac{1}{4} D^2 {\wt h}_{\3AB}    + \frac{1}{24} R {\wt h}_{\3AB}  \\
&\qquad + \frac{1}{4} h_\1^{CD}  D_C D_D {\wt h}_{\2AB} + \frac{1}{24} h_\1^{CD}   D_{\{A} D_{B\}} {\wt h}_{\2CD}   - \frac{1}{4} h_\1^{CD}  D_C D_{\{A} {\wt h}_{\2B\}D}   \\
&\qquad  - \frac{1}{6} h_{\1C\{A}  D^C ( D \cdot {\wt h}_\2 )_{B\}}  + \frac{1}{4}  h_{\1C\{A}  D_{B\}} ( D \cdot {\wt h}_\2 )^C   - \frac{1}{4} h_{\1C\{A} D^2 {\wt h}_\2^C{}_{B\}}  \\
&\qquad + \frac{1}{12} D_{\{A} h_{\1B\}C} (D \cdot {\wt h}_\2 )^C - \frac{5}{12} D_{\{A} h_\1^{CD} D_{B\}} {\wt h}_{\2CD}  + \frac{3}{4} D_C {\wt h}_{\2D\{A}  D_{B\}} h_\1^{CD} \\
&\qquad + \frac{1}{2} D_C h_{\1D\{A} D_{B\}} {\wt h}_\2^{CD} -  D_C h_{\1D\{A} D^C {\wt h}_\2^D{}_{B\}} + \frac{1}{9} ( D \cdot h_\1 )_{\{A} ( D \cdot {\wt h}_\2 )_{B\}} \\
&\qquad - \frac{1}{4} ( {\wt h}_\2 h_\1  R)_{\{AB\}} + \frac{3}{8} ( h_\1 {\wt h}_\2 R  )_{\{AB\}}  - \frac{1}{8} ( h_\1 R {\wt h}_\2 )_{\{AB\}}   \\
&\qquad + \frac{1}{24} R (h_\1 {\wt h}_\2 )_{\{AB\}}  - \frac{11}{24} D_{\{A} D_{B\}} h_{\1CD} {\wt h}_\2^{CD}  + \frac{1}{2} D_{\{A} D^C h^D_{\1B\}} {\wt h}_{\2CD}  \\
&\qquad + \frac{1}{12} \left( D \cdot D \cdot h_\1 + \frac{3}{2}  \tr [ h_\1 R ]  \right) {\wt h}_{\2AB}  - \frac{1}{24}R_{\{AB\}} \tr [ h_\1 {\wt h}_\2 ] .  
\end{split}
\end{equation}
Finally, we have evolution equations for $U_\3$ and $W_\5$ which are best described also in terms of the following tilded fields,
\begin{equation}
\begin{split}\label{app:nolog-U3W5tilde-def}
{\wt U}_\3 &\equiv U_\3 + 8 \p_u  ( \b_\2^2 ) + h_{\1AB} W_\2^A W_\2^B + 4 W_\2^A D_A  \b_\2  + \frac{R}{576}  \tr [ h_\1^3 ]  - \frac{1}{128} \p_u  \tr [    h_\1^4  ] , \\
{\wt W}_\5 &\equiv W_\5 + 10 D ( \b_\2^2 )  - \frac{1}{256}  D \tr [ h_\1^4 ]    - \frac{1}{48} \tr[ h_\1^3 ] W_\2 \\
&\qquad \qquad + 6 \b_\2 h_\1 W_\2 + \frac{1}{2} h_\1^3 W_\2 - \frac{1}{96} h_\1 D \tr[h_\1^3] + \frac{3}{2} h_\1^2 D \b_\2 .
\end{split}
\end{equation}
Then, the evolution equations take the form
\begin{equation}
\begin{split}\label{app:nolog-U3du}
\p_u {\wt U}_\3 &= \frac{1}{8} ( \p_u {\wt h}_\2 )^2 + \frac{1}{24} D^2 D_A D_B {\wt h}_\2^{AB}  - \frac{1}{96} h_{\1AB} \p_u D^2 {\wt h}_\2^{AB} + \frac{1}{8} h_{\1AB} \p_u D^A D_C {\wt h}_\2^{BC} \\
&\qquad + \frac{1}{24} D_C h_{\1AB} \p_u D^C {\wt h}_\2^{AB} + \frac{1}{6} D^C h_{\1AC} \p_u D_B {\wt h}_\2^{AB} + \frac{1}{32} \p_u {\wt h}_\2^{AB}  D^2 h_{\1AB}  \\
&\qquad - \frac{1}{72} \p_u {\wt h}_\2^{AB} D_A D^C h_{\1BC}  + \frac{1}{6} \p_u {\wt h}_\2^{AB} ( R h_\1 )_{AB} - \frac{1}{32} R \p_u {\wt h}_\2^{AB} h_{\1AB} \\
&\qquad - \frac{1}{144} R D_A D_B {\wt h}_\2^{AB} + \frac{5}{48} R_{AB} D^2 {\wt h}_\2^{AB}  + \frac{1}{12} D_C R_{AB} D^C {\wt h}_\2^{AB} - \frac{1}{72} D_A {\wt h}_\2^{AB} D_B R  \\
&\qquad + \frac{1}{24} {\wt h}_\2^{AB} D^2 R_{AB} + \frac{1}{36} R {\wt h}_\2^{AB} R_{AB} - \frac{1}{8} {\wt h}_\2^{AB} (R^2)_{AB} . 
\end{split}
\end{equation}
and
\begin{equation}
\begin{split}\label{app:nolog-W5du}
& 5 \p_u {\wt W}_{\5A} = D_A {\wt U}_\3  + \frac{3}{4} D^2 ( D \cdot {\wt h}_\3 )_A  - \frac{3}{32} D_A \tr [ {\wt h}_\3 R ]   - 3 R_{AB} D_C {\wt h}_\3^{BC} + \frac{1}{2} R ( D \cdot {\wt h}_3 )_A \\
& + \frac{3}{4} R^{BC} D_B {\wt h}_{\3CA} - \frac{1}{12} {\wt h}_{\3AB} D^B R  + \frac{3}{8}  \p_u {\wt h}_{\2BC}   D_A {\wt h}_{\2}^{BC}  - \frac{1}{8} {\wt h}_{\2BC} \p_u  D_A {\wt h}_{\2}^{BC} \\
&+ \frac{1}{2}  {\wt h}^{BC}_\2 \p_u D_C{\wt h}_{\2AB} + \frac{3}{2}  \p_u {\wt h}_{\2AB} D_C {\wt h}^{BC}_\2 - \frac{1}{2} \p_u {\wt h}^{BC}_\2 D_C {\wt h}_{\2AB}    + \frac{1}{2}  {\wt h}_{\2AB} \p_u D_C {\wt h}^{BC}_\2  \\
& + \frac{1}{32} (h_\1^2)^{BC} \p_u D_A {\wt h}_{\2BC} - \frac{1}{4} (h_\1^2)_{AB} \p_u ( D \cdot {\wt h}_\2 )^B - 6 \b_\2 \p_u ( D \cdot {\wt h}_\2 )_A \\
&- \frac{1}{4} (h_\1^2)^{BC} \p_u D_B {\wt h}_{\2CA}   + \frac{1}{12} h_{\1AB} h_{\1CD}  \p_u D^B {\wt h}_\2^{CD} - \frac{1}{4} h_{\1AB} h_{\1CD}  \p_u D^D {\wt h}_\2^{BC} \\
& + \frac{1}{16} ( h_\1 \p_u {\wt h}_\2 )_{BC} D_A h_\1^{BC} + 14 \p_u {\wt h}_{\2AB} D^B \b_\2 + \frac{7}{2} ( h_\1 \p_u {\wt h}_\2  W_\2 )_A  \\
& +  \frac{7}{2} ( \p_u {\wt h}_\2 h_\1 W_\2 )_A - \frac{2}{3} \tr [ h_\1 \p_u {\wt h}_\2 ] W_{\2A} + \frac{1}{12} h_{\1AB} \p_u {\wt h}_{\2CD} D^B h_\1^{CD}  \\
& + \frac{7}{12} h_{\1AB} D^B D \cdot D \cdot {\wt h}_\2 - \frac{1}{4} h_{\1AB} D^2 ( D \cdot {\wt h}_\2 )^B   + \frac{5}{16} h_\1^{BC} D_A D_B ( D \cdot {\wt h}_\2 )_C \\
& + \frac{1}{64} h_\1^{BC} D_A D^2 {\wt h}_{\2BC}  - \frac{1}{4} h_\1^{BC} D_B D_C ( D \cdot {\wt h}_\2 )_A - \frac{1}{4} h_\1^{BC} D_B D^2 {\wt h}_{\2CA} \\
& + \frac{5}{16} D_A h_{\1BC} D^B ( D \cdot {\wt h}_\2 )^C + \frac{1}{64} D_A h_{\1BC} D^2 {\wt h}_\2^{BC} - \frac{1}{2} W_{\2A} D \cdot D \cdot {\wt h}_\2 \\
& - 2 W_\2^B D_A ( D \cdot {\wt h}_\2 )_B + \frac{1}{3} W_\2^B D_B ( D \cdot {\wt h}_\2 )_A + 4 W_\2^B D^2 {\wt h}_{\2AB}  + \frac{1}{8} D_D h_{\1BC} D_A D^D {\wt h}_\2^{BC} \\
& - \frac{1}{4} D^B h_\1^{CD} D_C D_D {\wt h}_{\2AB} + \frac{1}{8} D_A D_B h_{\1CD} D^B {\wt h}^{CD}_\2 + \frac{11}{3} D^B W_\2^C D_C {\wt h}_{\2BA}  \\
& + \frac{1}{24} D^2 h_{\1AB} ( D \cdot {\wt h}_\2 )^B + \frac{1}{16} D^2 h_{\1BC} D_A {\wt h}_\2^{BC} + \frac{5}{24} D^2 h_\1^{BC} D_B {\wt h}_{\2CA}  \\
&+ \frac{1}{1152} R h_{\1BC} D_A {\wt h}_\2^{BC} + \frac{5}{32} ( R h_\1 )_{BC} D_A {\wt h}_\2^{BC}  + \frac{3}{8} ( R h_\1 )_{AB} ( D \cdot {\wt h}_\2 )^B  \\
&+ \frac{31}{24} ( h_\1 R )_{AB} ( D \cdot {\wt h}_\2 )^B - \frac{1}{6} R_{AB} h_{\1CD} D^B {\wt h}_\2^{CD}  + \frac{3}{4} R_{AB} h_{\1CD} D^D {\wt h}_\2^{BC}\\
& + \frac{19}{24}  R_{CD} h_{\1AB} D^B {\wt h}_\2^{CD} - \frac{1}{2} R_{CD} h_{\1AB} D^D {\wt h}_\2^{BC}  - \frac{13}{72} R ( h_\1 D \cdot {\wt h}_\2 )_A \\
&+ \frac{5}{16} \tr [ R h_\1 ] ( D \cdot {\wt h}_\2 )_A + \frac{1}{18} R h_\1^{BC} D_C {\wt h}_{\2AB} + \frac{1}{4}  (  h_\1 R )^{BC} D_C {\wt h}_{\2AB}\\
&  - \frac{5}{3} ( R h_\1 )^{BC} D_C {\wt h}_{\2AB}  - \frac{1}{2} D_B D_C W_{\2A}  {\wt h}_\2^{BC} + \frac{9}{4} {\wt h}_{\2AB} D^2 W_\2^B - \frac{7}{4} {\wt h}_{\2AB} D^B D \cdot W_\2 \\
& -\frac{59}{1152} R D_A h_{\1BC} {\wt h}_\2^{BC} + \frac{17}{32} R_C{}^D D_A h_{\1BD} {\wt h}_\2^{BC} - \frac{1}{6} R_{AB} D^B h_{\1CD} {\wt h}_\2^{CD} \\
& - \frac{33}{4}   ( R {\wt h}_\2 W_\2 )_A - 8   ( {\wt h}_\2 R W_\2 )_A + \frac{1}{6} \tr[ {\wt h}_\2 R ] W_{\2A} + \frac{61}{24} R ( {\wt h}_\2 W_\2 )_A \\
&- \frac{107}{1152} D_A R h_{\1BC} {\wt h}^{BC}_\2 + \frac{13}{32} D_A R_{BC} ( h_\1 {\wt h}_\2 )^{BC} + \frac{1}{48} h_{\1AB} {\wt h}_\2^{BC} D_C R \\
& + \frac{11}{12} h_{\1AB} {\wt h}_\2^{CD} D^B R_{CD} + \frac{1}{2} D^B R_{CD} h_\1^{CD} {\wt h}_{\2AB} . 
\end{split}
\end{equation}

\subsection{Canonical Solutions}
\label{app:mrS0-explicit-details}

Solutions in $\CS_\text{canonical}$ are obtained from those in $\CS_\text{analytic}$ by setting $\mr{q}_{AB} = \d_{AB}$ and $h_{\1AB} = 0$. The metric functions in the solution space $\CS_\text{canonical}$ are distinguished by the superscript $\circ$. The metric functions admit the following large $r$ expansion
\begin{equation}
\begin{split}\label{app:metric-exp-S0dot}
\mr\b &= \frac{\mr \b_\4 }{r^4}  + \CO(r^{-5}) ,  \qquad {\mr h}_{AB} = \d_{AB} + \frac{\mr h_{\2AB}  }{r^2} + \frac{\mr h_{\3AB} }{r^3}  + \frac{\mr h_{\4AB}}{r^4}  + \CO(r^{-5}) , \\
\mr U &= \frac{\mr U_\2  }{r^2} + \frac{ \mr  U_\3 }{r^3} + \CO(r^{-4}) , \qquad {\mr W}^A =  \frac{\mr W^A_\3   }{r^3} + \frac{\mr W^A_\4 }{r^4} + \frac{\mr W^A_\5 }{r^5}  + \CO(r^{-6}) ,   \\
\end{split}
\end{equation}
The EoM constraints derived in the previous section significantly simplify in this case so it is quite useful to reproduce them here,
\begin{equation}
\label{Bondi-frame-sol} 
\begin{split}
& \mr{\b}_\4 = - \frac{1}{32} \tr  [ \mr{h}_\2^2 ]  ,\qquad \mr{U}_\2 = - \frac{1}{6} \p \cdot \p \cdot h_\2 ,\qquad \mr{W}_\3 = - \frac{1}{3} \p \cdot  \mr{h}_\2 ,   \\
&\mr{W}_\4 = - \frac{3}{4} \p \cdot \mr{h}_\3 , \qquad \p_u \mr{h}_{\3AB} = \frac{2}{3} \p_{\{A} ( \p \cdot \mr{h}_\2 )_{B\}} - \frac{1}{2} \p^2 \mr{h}_{\2AB} , \qquad \p \cdot \p \cdot \mr{h}_{\3AB} = 0 , \\
& \p_u \left( \mr{h}_{\4AB} - \frac{1}{2} ( \mr{h}_\2^2 )_{AB} \right) = \frac{1}{2} \p_{\{A} ( \p \cdot \mr{h}_\3 )_{B\}} - \frac{1}{4} \p^2 \mr{h}_{\3AB}  , \qquad \p \cdot \left( \mr{h}_\4 - \frac{1}{2} \mr{h}_\2^2  \right) = 0  .  
\end{split}
\end{equation}
The evolution equations for $\mr{U}_\3$ and $\mr{W}_\5$ are
\begin{equation}
\begin{split}\label{U3-evolution-eq}
\p_u \mr{U}_\3 &= \frac{1}{8} ( \p_u \mr{h}_\2 )^2 + \frac{1}{24} \p^2 \p \cdot \p \cdot \mr{h}_\2  , \\
\p_u \mr{W}_\5 &= \frac{1}{5} \left[ \p \mr{U}_\3  + \frac{3}{4} \p^2 ( \p \cdot \mr{h}_\3 )  + \frac{1}{2} \tr [ \p_u \mr{h}_\2 \p \mr{h}_\2 ] \right. \\
&\left. \qquad \qquad \qquad + \frac{1}{2} \p \cdot (   \mr{h}_\2 \overleftrightarrow{\p_u} \mr{h}_\2  )  + \p_u \left( \mr{h}_\2 \p \cdot \mr{h}_\2   -  \frac{1}{16} \p \tr [  \mr{h}_\2^2  ] \right) \right] . 
\end{split}
\end{equation}

\subsection{Comments on the News Tensor}
\label{sec:news-tensor}
For comparison purposes, it is useful to spell out the relationship between $h_{(2)AB}$ in the general case and $\mr{h}_{(2)AB}$ (modulo the trivial change from $\d_{AB}$ to $q_{AB}$). They are as follows
\begin{equation}
h_{(2)AB}=h^\tf_{\2 AB}+\frac{q_{AB}}{8}h_\1^2=\mr{h}_{\2 AB}+\frac{q_{AB}}{8}h_\1^2,
\end{equation}
\begin{equation}
\wt{h}_{\2 AB}=h_{(2)AB}-\frac{1}{4}h_{\1 A}\cdot h_{\1 B}-\frac{q_{AB}}{16}h_\1^2=\mr{h}_{\2 AB}-\frac{1}{4}h_{\1 A}\cdot h_{\1 B}+\frac{q_{AB}}{16}h_\1^2.
\end{equation}
In the canonical solution space $\CS_\text{canonical}$, the news tensor appears naturally as 
\begin{equation}\label{eq.radnews}
\mr{N}_{AB}:=\partial_u \mr{h}_{\2 AB}.
\end{equation}
and is conjugate to $\mr{h}_{\2 AB}$ in the symplectic form. This also stems from the integration function of the fourth main equation upon choosing a particular normalization. In \cite{Capone:2021ouo} it was observed that the fourth main equation gives 
\begin{equation}
\frac{N_{AB}}{c}=\partial_u\left(\frac{1}{2}h_{\2AB}-\frac{1}{4}h_{\1 A}\cdot h_{\1 B}\right),
\end{equation}
(this is a reduction of (3.7) in \cite{Capone:2021ouo} to our b.c.) where $c$ is the normalization factor chosen as $c=2$ to match the standard normalization \eqref{eq.radnews} - so that 
\begin{equation}
N_{AB}=\mr{N}_{AB}-\frac{1}{2}\partial_u\left(h_{\1 A}\cdot h_{\1 B}-\frac{q_{AB}}{4}h_\1^2\right).
\end{equation}
It was argued, correctly, that the additional part with respect to $\mr{N}_{AB}$ carries the information of the "Geroch tensor", but the identification with the true degrees of freedom in the symplectic form was not made. Indeed, the correctly gauge-invariant quantity was later found in \cite{Chowdhury:2022nus} (denoted there with $\partial_u\wt{D}_{AB}$)
\begin{equation}\label{eq.news}
\wt{N}_{AB}:=\partial_u \wt{h}_{\2 AB}=\mr{N}_{AB}-\frac{1}{4}\partial_u\left(h_{\1 A}\cdot h_{\1 B}-\frac{q_{AB}}{4}h_\1^2\right),
\end{equation}
which enters the symplectic form as the variable conjugate to $\wt{h}_{\2 AB}$. We observe that $\wt{N}_{AB}$ is nothing but the only non-trivial contribution to the component $C_{\cdot A \cdot B}$ of the Weyl tensor at leading order. Although not explicit there, this should perhaps be expected both from \cite{Geroch:1977} and on physical grounds. This discussion carries over to any number of even dimensions and unifies our understanding of the radiative degrees of freedom across all even dimensions: the invariant notion of news is given by the asymptotic form of the Weyl tensor and is related to the integration function of the fourth main equation by a gauge adjustment of the latter.

\section{Finite BMS Transformations from \texorpdfstring{$\CS_\text{canonical} \to \CS_\text{analytic}$}{Sdot0toS0}: Details}
\label{app:finite-BMS-transforms}

In this Appendix, we present the details for the finite BMS transformation which maps solutions from $\CS_\text{canonical}$ to those in $\CS_\text{analytic}$. Given a metric $\mr{G}$ in $\CS_\text{canonical}$, the new metric $G$ obtained via a diffeomorphism is given by 
\begin{equation}
\begin{split}\label{app:BMStransform}
G_{\mu\nu} &=  - e^{2\mr{\b}(\mr X)} \mr{U}(\mr X) \p_\mu \mru \p_\nu \mru  - e^{2\mr{\b}(\mr X)} \left[ \p_\mu \mru \p_\nu \mr{r} + \p_\mu \mr{r} \p_\nu \mru \right]  \\
&\qquad \qquad \qquad \qquad \qquad \qquad + {\mr g}_{AB}(\mr X) \left[ \p_\mu \mr{x}^A   -  \mr{W}^A(\mr X) \p_\mu \mru \right] \left[ \p_\nu \mr{x}^B - \mr{W}^B(\mr X)  \p_\nu \mru  \right] .
\end{split}
\end{equation}
We are interested in diffeomorphisms which preserve the Bondi-Sachs gauge conditions,
\begin{equation}
\begin{split}\label{app:Bondi-gauge-cond}
G_{rr} = G_{rA} = \p_r \det ( r^{-2} G_{AB} ) = 0 .
\end{split}
\end{equation}
We start by analysing the equations \eqref{app:BMStransform} at leading order in $r$ where $\mr X^\mu( X)$ has the form
\begin{equation}
\begin{split}
\mru = \mru_\0 + \CO(r^{-1}) , \qquad \mr{r} = r \mr{r}_\m + \CO(1) , \qquad \mrx^A = \mrx_\0^A + \CO(r^{-1}) . 
\end{split}
\end{equation}
Substituting this into \eqref{app:BMStransform} and expanding to leading order, we find
\begin{equation}
\begin{split}
G_{uu} &= r^2 [ \p_u \mrx_\0 ]^2 + \CO(r) , \qquad G_{ur} =  - \p_u \mru_\0 \mr{r}_\m + \CO(r^{-1}) .
\end{split}
\end{equation}
Since we need $G_{uu} = \CO(1)$ and $G_{ur} = - 1 + \CO(r^{-2})$ we must have $\mrx_\0^A = \chi^A(x)$, $\p_u \mru_\0 \mr{r}_\m = 1$. Using this and moving to subleading order, we find
\begin{equation}
\begin{split}
G_{uu} =  - 2 r \p_u \mr{r}_\m \p_u \mru_\0 + \CO(1) \qquad \implies \qquad \p_u \mr{r}_\m \p_u \mru_\0 = 0 . 
\end{split}
\end{equation}
It follows from this that
\begin{equation}
\begin{split}
\mrx_\0^A = \chi^A(x) , \qquad \mru_\0 = \k(u,x) \equiv e^{-\Phi(x)} u + C(x) , \qquad \mr{r}_\1 = e^{\Phi(x)} .
\end{split}
\end{equation}

Having completed the leading order analysis, we move on to the full large $r$ expansion of $\mr X^\mu( X)$ which takes the form
\begin{equation}
\begin{split}\label{app:finite-BMS-transform}
\mr{u} &= \k + e^{-\Phi} \frac{\mr{u}_\1}{r} + e^{-2\Phi} \frac{\mr{u}_\2}{r^2} + e^{-3\Phi}  \frac{\mr{u}_\3}{r^3} + e^{-4\Phi} \frac{ \mr{u}_\4}{r^4} + \CO(r^{-5}) , \\
\mr{r} &=  r e^{\Phi} + \mr{r}_\0 + e^{-\Phi} \frac{ \mr{r}_\1}{r} + e^{-2\Phi} \frac{ \mr{r}_\2}{r^2} + e^{-3\Phi} \frac{\mr{r}_\3}{r^3} + \CO(r^{-4}) , \\
\mr{x}^A &= \chi^A + e^{-\Phi} \frac{ \mr{x}^A_\1}{r} + e^{-2\Phi} \frac{ \mr{x}^A_\2}{r^2} + e^{-3\Phi}  \frac{\mr{x}^A_\3}{r^3} + e^{-4\Phi} \frac{\mr{x}^A_\4}{r^4} + \CO(r^{-5}) . \\
\end{split}
\end{equation}
Substituting this expansion into \eqref{app:BMStransform} and imposing the conditions \eqref{app:Bondi-gauge-cond}, we can obtain each of the coefficients shown above. The general procedure is as follows. We first analyse $G_{rA} = 0$ at $\CO(r^{-n})$ which fixes $\hat{x}_\ttt{n\!+\!1}^A  = \p_{\chi^B} x^A \mr{x}_\ttt{n\!+\!1}^B$. We next analyse $G_{rr} = 0$ at $\CO(r^{-n-2})$ which fixes $\mr{u}_\ttt{n\!+\!1}$. We then analyse the final constraint $\p_r\det(G_{AB}/r^2)=0$ at $\CO(r^{-n-2})$ to fix $\mr{r}_\ttt{n}$. This procedure is completed iteratively for each $n$ (starting at $n=0$).

Here, we shall simply present the results of this analysis. It will be convenient to introduce the following quantities
\begin{equation}
\begin{split}
\mu_A \equiv \p_A \k , \qquad \Psi_{AB} \equiv {\hat D}_A {\hat D}_B \k . 
\end{split}
\end{equation}
The coefficients are then given by 
\begin{equation}
\begin{split}
{\hat x}_\1^A &=  - \mu^A ,  \\
{\hat x}_\2^A &= \frac{1}{4} \tr[\Psi] \mu^A ,  \\
{\hat x}_\3^A &=  \frac{1}{3} {\hat h}_\2^{AB} \mu_B  +  \left( \frac{1}{8}  \tr [ \Psi^2 ] -  \frac{3}{32} \tr[\Psi]^2 \right) \mu^A ,    \\
{\hat x}_\4^A &= \frac{1}{4} {\hat h}_\3^{AB} \mu_B + \frac{1}{24} {\hat D}^A {\hat h}_\2^{BC} \mu_B \mu_C  + \frac{1}{4} \msD_\1 {\hat h}_\2^{AB} \mu_B - \frac{1}{4} \tr[\Psi]  {\hat h}_\2^{AB}  \mu_B - \frac{1}{8} \mu^2  {\hat W}_\3^A  \\
&\qquad \qquad \qquad  + \left( - \frac{1}{12} {\hat D}_B [ {\hat h}_\2^{BC} \mu_C ] + \frac{1}{12}\tr [ \Psi^3 ]  - \frac{1}{8}  \tr[\Psi] \tr [ \Psi^2 ] + \frac{1}{24}  \tr[\Psi]^3 \right) \mu^A ,   \\
\mru_\1 &=  - \frac{1}{2} \mu^2  ,  \\
\mr{u}_\2 &= \frac{1}{8} \tr[\Psi]  \mu^2 ,    \\
\mr{u}_\3 &= \frac{1}{6} {\hat h}_\2^{AB} \mu_A \mu_B + \left( \frac{1}{16} \tr [ \Psi^2 ]  - \frac{3}{64} \tr[\Psi]^2  \right) \mu^2 ,   \\
\mr{u}_\4 &= \left( \frac{1}{8} {\hat h}_\3^{AB} + \frac{1}{8} \msD_\1 {\hat h}_\2^{AB}  + \frac{1}{24} {\hat D}^A {\hat h}_\2^{BC}  \mu_C - \frac{1}{8}  \tr[\Psi] {\hat h}_\2^{AB} \right) \mu_A   \mu_B \\
&\qquad + \left( - \frac{1}{8} \mu_A  {\hat W}_\3^A  - \frac{1}{24} {\hat D}_B [ {\hat h}_\2^{BC} \mu_C ] + \frac{1}{24}  \tr [ \Psi^3 ] - \frac{1}{16}  \tr[\Psi] \tr [ \Psi^2 ] + \frac{1}{48} \tr[\Psi]^3  \right) \mu^2 ,  \\
\mr{r}_\0 &= \frac{1}{4} \tr[\Psi] ,    \\
\mr{r}_\1 &= \frac{1}{8} \tr [ \Psi^2 ] -  \frac{1}{32} \tr[\Psi]^2 ,  \\
\mr{r}_\2 &=  \frac{1}{4} {\hat W}_\3^A \mu_A - \frac{1}{12} {\hat D}_A [ {\hat h}_\2^{AB} \mu_B ] +  \frac{1}{12}  \tr [ \Psi^3 ] -  \frac{1}{16}  \tr[\Psi] \tr [ \Psi^2 ]  + \frac{1}{96} \tr[\Psi]^3 ,  \\
\mr{r}_\3 &= \bigg[  \frac{3}{16} \SD_u {\hat h}_\3^{AB}   - \frac{1}{16} \mu^C  {\hat D}_C \SD_u {\hat h}_\2^{AB}  + \frac{1}{48} \mu^2 \SD_u^2 {\hat h}_\2^{AB}  - \frac{1}{24} \tr[\Psi] \SD_u {\hat h}_\2^{AB} + \frac{1}{8} {\hat D}^A {\hat D}_C {\hat h}_\2^{BC} \\
&\qquad - \frac{1}{12} \SD_u {\hat h}_\2^{AC} \Psi_C^B  - \frac{1}{96}  {\hat D}^2 {\hat h}_\2^{AB} \bigg] \mu_A \mu_B + \bigg[ - \frac{1}{4} {\hat D}_B {\hat h}_\3^{AB}  + \frac{5}{48} \tr[\Psi]   {\hat D}_B {\hat h}_\2^{AB}  \\
&\qquad - \frac{1}{48}  \mu^2  {\hat D}_B \SD_u {\hat h}_\2^{AB}  - \frac{1}{24} {\hat D}_B {\hat h}_\2^{BC} \Psi_C^A   + \frac{1}{16} {\hat D}^A {\hat h}_\2^{BC} \Psi_{BC}  - \frac{1}{24}  {\hat D}^B {\hat h}_{\2CA} \Psi_{BC}  \bigg] \mu_A   \\
&\qquad - \frac{1}{32} {\hat D}_A {\hat D}_B {\hat h}_\3^{AB} \mu^2 - \frac{1}{16} \tr[ {\hat h}_\3 \Psi ]  - \frac{1}{12} \tr [ {\hat h}_\2 \Psi^2 ]  + \frac{1}{16} \tr [ \Psi^4 ]  - \frac{3}{128} \tr [ \Psi^2 ]^2  \\
&\qquad  - \frac{1}{16} \tr[\Psi] \tr[\Psi^3 ]  + \frac{9}{256} \tr[\Psi]^2 \tr [ \Psi^2 ] - \frac{9}{2048} \tr[\Psi]^4   ,  \\
\end{split}
\end{equation}
We can extract the large $r$ coefficients of $h_{AB}$ as
\begin{equation}
\begin{split}
h_{\1AB} &= -  2 e^\Phi \left( \Psi_{AB} - \frac{1}{4}  {\hat q}_{AB} \tr[\Psi]  \right) , \\
h_{\2AB} &= {\hat h}_{\2AB}  +  (\Psi^2)_{AB} - \frac{1}{2} \tr[\Psi] \Psi_{AB} + \frac{1}{4}  {\hat q}_{AB} \tr [ \Psi^2 ] ,  \\
h_{\3AB} &= e^{-\Phi} \left[  {\hat h}_{\3AB} + \msD_\1 {\hat h}_{\2AB}  - 2  {\hat W}_{\3\{A} \mu_{B\}} + \frac{2}{3} {\hat D}_{\{A}  [ {\hat h}_{\2B\}C} \mu^C ]   - 2 ( {\hat h}_\2 \Psi )_{(AB)}  \right. \\
&\left. - \frac{1}{4} \left( \tr [ \Psi^2 ] - \frac{1}{4} \tr[\Psi]^2 \right) \Psi_{AB} + \left( \frac{1}{6}  \tr [ \Psi^3 ]  - \frac{1}{16} \tr[\Psi] \tr [ \Psi^2 ] + \frac{1}{192} \tr[\Psi]^3 \right)  {\hat q}_{AB} \right] ,  \\
h_{\4AB} &=  e^{-2\Phi} \bigg[ \hat{h}_{\4AB} -  \mr{r}_\0 \hat{h}_{\3AB} -  \mr{u}_\1 \SD_u \hat{h}_{\3AB}   +  \CL_{\hx_\1} \hat{h}_{\3AB} \\
&\qquad + [ - 2 {\hat D} \k {\hat W}_\4  - 2 [ \CL_{{\mr x}_\1} {\hat W}_\3 ] {\hat D} \k  - 2  {\hat W}_\3 {\hat D}  \mr{u}_\1+  2  \mr{u}_\1  \SD_u {\hat W}_\3 {\hat D} \k  \\
&\qquad +  2   \mr{r}_\0 {\hat W}_\3 {\hat D} \k -  {\hat U}_\2 {\hat D} \k {\hat D} \k  - 2 {\hat D}  \mr{u}_\2 {\hat D} \mr{r}_\0  - 2 {\hat D} \mr{u}_\1 {\hat D}  \mr{r}_\1  - 2 {\hat D} \k {\hat D} \mr{r}_\2 ]_{(AB)}  \\
&\qquad + \frac{1}{2} \left[  \mr{u}^2_\1 \SD_u  -  2 \mr{u}_\1 \hx_\1^C   {\hat D}_C  - \left( 2   \mr{u}_\2 +  \hx^C_\1 \hx^D_\1 {\hat D}_C {\hat D}_D \k \right) \right]  \SD_u \hat{h}_{\2AB}  \\
&\qquad + \CL_{\hx_\2} {\hat h}_{\2AB}  + \frac{1}{2} \hx^C_\1 \hx^D_\1 {\hat D}_C {\hat D}_D \hat{h}_{\2AB}  +  {\hat D}_A \hx_\1^C {\hat D}_B \hx_\1^D \hat{h}_{\2CD}  \\
&\qquad  + 2 \hx^D_\1 {\hat D}_D \hat{h}_{\2C(A}  {\hat D}_{B)} {\mr x}_\1^C  -  2   \mr{u}_\1  \SD_u \hat{h}_{\2C(A}  {\hat D}_{B)} {\mr x}_\1^C \\
&\qquad +  4 \mr{r}_\0  \left[   {\hat D}_{(A} \hx_{\3B)} + {\hat D}_{(A} \hx^C_\1 {\hat D}_{B)} \hx_{\2C}  \right]  +  [ 2   {\hat D} \hx_\4 + ( {\hat D} \hx_\2 )^2 ]_{(AB)}   \\
&\qquad + 2 {\hat D}_{(A}  \hx^C_\3 {\hat D}_{B)} \hx_{\1C}  + 4  [ \mr{r}_\2 + \mr{r}_\0 \mr{r}_\1 ]  [ {\hat D} \hx_\1 ]_{(AB)}  \\
&\qquad  +  [ 2 \mr{r}_\1 + \mr{r}_\0^2  ]   [ 2 {\hat D} \hx_\2 +  ( {\hat D} \hx_\1 )^2  ]_{(AB)}  +  [ 2  \mr{r}_\3 + 2 \mr{r}_\0 \mr{r}_\2 +  \mr{r}^2_\1  ] {\hat q}_{AB} \bigg] . 
\end{split}
\end{equation}
From this, we can extract that
\begin{equation}
\begin{split}
{\wt h}_{\2AB} &= {\hat h}_{\2AB} ,  \\
{\wt h}_{\3AB} &= e^{-\Phi} \left[ {\hat h}_{\3AB} + \msD_\1 {\hat h}_{\2AB}  - 2  {\hat W}_{\3\{A} \mu_{B\}} + \frac{2}{3} {\hat D}_{\{A}  [ {\hat h}_{\2B\}C} \mu^C ]  - \frac{1}{2}\tr[\Psi] {\hat h}_{\2AB} \right] .  
\end{split}
\end{equation}

\bibliography{srbib}
\bibliographystyle{utphys}

\end{document}